\documentclass[english,english, reprint, notitlepage, superscriptaddress, longbibliography]{revtex4-1}
\usepackage[T1]{fontenc}
\usepackage[latin9]{inputenc}
\usepackage{mathtools}
\usepackage{amsmath}
\usepackage{amssymb}
\usepackage{graphicx}

\usepackage[breaklinks=true]{hyperref}

\makeatletter

\usepackage{babel}
\usepackage{xcolor}

\makeatother

\begin{document}

\title{Impact of Pre-symptomatic Transmission on Epidemic Spreading in Contact Networks: \\ A Dynamic Message-passing Analysis}

\author{Bo Li}
\email{b.li10@aston.ac.uk}
\affiliation{Non-linearity and Complexity Research Group, Aston University, Birmingham, B4 7ET, United Kingdom}

\author{David Saad}
\email{d.saad@aston.ac.uk}
\affiliation{Non-linearity and Complexity Research Group, Aston University, Birmingham, B4 7ET, United Kingdom}

\begin{abstract}	 
Infectious diseases that incorporate pre-symptomatic transmission are challenging to monitor, model, predict and contain. We address this scenario by studying a variant of a stochastic susceptible-exposed-infected-recovered model on arbitrary network instances using an analytical framework based on the method of dynamic message-passing. This framework provides a good estimate of the probabilistic evolution of the spread on both static and contact networks, offering a significantly improved accuracy with respect to  individual-based mean-field approaches while requiring a much lower computational cost compared to numerical simulations. It facilitates the derivation of epidemic thresholds, which are phase boundaries separating parameter regimes where infections can be effectively contained from those where they cannot. These have clear implications on different containment strategies through topological (reducing contacts) and infection parameter changes (e.g., social distancing and wearing face masks), with relevance to the recent COVID-19 pandemic. 
\end{abstract}

\maketitle

\section{Introduction}

Rapid spreading of infectious diseases has had a devastating impact on societies throughout human history but has become more critical in modern society due to dense population in urban areas and the increase in human mobility facilitated by global transportation networks. A recent threat is the spread of the COVID-19 disease caused by the SARS-CoV-2 virus, which has led to a pandemic with severe impact on public health and the global economy. 
One prominent feature of this disease is that the presymptomatic and asymptomatic viral carriers can spread the disease as well, which poses a big challenge to contact tracing and disease containment~\cite{WangYixuan2020,RuiyunLi2020,He2020,Moghadas2020}. Therefore, it is crucial to understand the significance of these undetected transmissions and estimate their impact. Of particular importance are parameter regimes where pre- and asymptomatic infections result in a complete breakdown of our ability to identify infected individuals and contain the spread.

Numerous studies that investigate the spread of the COVID-19 disease aim at predicting the causes of the spreading processes and examine the effectiveness of non-pharmaceutical intervention strategies~\cite{Gatto2020,Chinazzi2020,Hao2020}. It is common to model the evolution of the population mass of each group (e.g.,  susceptible, exposed, infected, and recovered) by deterministic differential equations~\cite{Moghadas2020Apr,Gatto2020,Moghadas2020}.
While being simplistic and tractable, such a method assumes homogeneous mixing of the population (in a city or within an age group) and neglects the social contact network structures of the specific instance investigated~\cite{Satorras2015}. Large scale agent-based simulations are also widely used, which provide a more detailed picture of the spreading processes but are very computationally demanding and suffer from a lack of principled understanding~\cite{Ferguson2020,Adam2020,Reich2020,Davies2020}. To obtain a reliable statistical description of the process, one has to increase the number of samples significantly as the system size increases, which makes the computation prohibitive for large systems.
Analytical treatments to the epidemic spreading processes on heterogeneous contact networks are valuable both in providing solutions in specific instances and in exploring the typical macroscopic behavior of an ensemble of systems with similar characteristics; the latter also results in generic and intuitive understanding.

In this paper, we analyze diseases spreading with presymptomatic transmission such as COVID-19 by studying a variant of stochastic susceptible-exposed-infected-recovered (SEIR) model on contact networks, in which nodes in exposed states can also spread the disease without showing symptoms. For simplicity, the contact networks are viewed as static, serving as substrates on which the disease spreads. We derive the dynamic message-passing (DMP) equations for this model, which provide a good approximation to the complex stochastic spreading dynamics on general networks and facilitates theoretical analyses~\cite{Karrer2010,Lokhov2014,Lokhov2015,Shrestha2015,Wang2017}.
Based on this framework, we derive the epidemic thresholds and their dependence on different intervention methods. The emphasis of this paper is to provide a more accurate description of the complex spreading dynamics through DMP to obtain a more intuitive physical picture and to clarify the effects of some containment strategies.

The remainder of the paper is organized as follows. We introduce the SEIR model in Sec.~\ref{sec:model}, and derive the dynamical equations in Sec.~\ref{sec:theory}. We then perform a linear stability analysis of the dynamical equations in Sec.~\ref{sec:linearized_dynamics}, based on which the epidemic thresholds are obtained and analyzed in Sec.~\ref{sec:threshold}. In Sec.~\ref{sec:NB_centrality}, we show that nonbacktracking centrality can be used to predict the outbreak profile. In Sec.~\ref{sec:dilution}, we investigate the effects of reducing contacts on slowing down the spread of the disease. Finally, we summarize our findings and discuss some limitations and outlook.

\section{The Model\label{sec:model}}

The contact network is represented by a graph $G(V,\mathcal{E})$, where $V$ is the set of nodes and $\mathcal{E}$ is the set of edges. We assume that the network has only one connected component. Each individual resides on a node, assuming one of four states, susceptible ($S$), exposed ($E$), infected ($I$) and recovered ($R$) at any particular time step. We assume that a node in the exposed state has contracted the disease but has not developed symptoms yet. Unlike the usual SEIR model~\cite{Keeling2008}, the exposed nodes can also spread the disease. The dynamical process of the modified SEIR model in discrete time is defined in the form of transition probabilities of states of neighboring nodes (say $i$ and $j$) and the state evolution of an individual node $i$,

\begin{align}
S(i)+E(j) & \xrightarrow{\alpha_{ji}}E(i)+E(j),\label{eq:SEIR_def1}\\
S(i)+I(j) & \xrightarrow{\beta_{ji}}E(i)+I(j),\label{eq:SEIR_def2}\\
E(i) & \xrightarrow{\nu_{i}}I(i),\label{eq:SEIR_def3}\\
I(i) & \xrightarrow{\mu_{i}}R(i),\label{eq:SEIR_def4}
\end{align}
where $\alpha_{ji}$($\beta_{ji}$) is the probability that node $j$ being in the exposed(infected) state transmits the disease to its susceptible neighboring node $i$ at a certain time step. We assume that each time step corresponds to one day, keeping in mind that a finer time scale can also be considered. At each time step, an existing exposed node $i$ becomes infected (i.e., develops symptoms) with probability $\nu_{i}$, while an existing infected node $i$ recovers with probability $\mu_{i}$. Therefore, the average periods of incubation and recovery are $1/\nu_{i}$ and $1/\mu_{i}$, respectively. At a certain time step, the symptom-development and recovery processes are assumed to occur after possible transmission activities. Since we will contrast the properties of the SEIR and SIR models, we also introduce the transition probabilities of the latter,

\begin{align}
S(i)+I(j) & \xrightarrow{\beta_{ji}}I(i)+I(j),\label{eq:SIR_def1}\\
I(i) & \xrightarrow{\mu_{i}}R(i),\label{eq:SIR_def2}
\end{align}
which has been widely studied in the literature~\cite{Satorras2015}. We remark that both models are Markovian processes, implying an exponential distribution for both symptom-development and recovery times, which may not be fully realistic for many diseases including COVID-19~\cite{Satorras2015,Tian2020}. Nevertheless, they both represent relevant models that provide insights, offer an approximate and effective description of the spreading process and are amenable to analysis.

The epidemiological parameters depend on the nature of the disease and the intervention strategies being imposed; they are usually estimated based on observations and can be subject to a high degree of uncertainty. As for COVID-19, the average incubation period is about 5.2 days~\cite{QunLi2020,He2020}. Infectiousness is estimated to start from 2.3 days before the onset of symptoms~\cite{He2020}, while it is argued~\cite{Ashcroft2020} that infectiousness can start much earlier (one needs to look back at 5 days to catch 97\% of presymptomatic infections). The time needed for the symptoms to disappear depends on disease severity of the patient. In Ref.~\cite{He2020}, it is inferred that infectiousness declines rapidly within 7 days. In~\cite{Perera2020}, it is found (from patients with mostly mild COVID-19) that viral subgenomic RNA, which provide evidence of replicative intermediates of the virus, were detectable up to 8 days after the onset of symptoms. In this paper, we define the recovered ($R$) state where the exposed and/or infected individual effectively looses infectiousness, irrespective of whether the symptoms persist or not. Therefore, an infected individual who has been put into isolation and can no longer infect others is also categorized to be in state $R$. To address the COVID-19 disease, we set $\nu_{i}=1/5,\mu_{i}=1/8$ according to these previous findings.

The transmission probabilities $\alpha_{ji},\beta_{ji}$ are more difficult to estimate. Based on estimates from previous studies~\cite{RuiyunLi2020}, we set $\alpha_{ji}=\beta_{ji}/2$ in some experiments but will also consider other parameter combinations in establishing the epidemic thresholds phase diagram.
For simplicity, we also assume that the parameters are homogeneous, i.e., $\alpha_{ij}=\alpha,\beta_{ij}=\beta,\nu_{i}=\nu,\mu_{i}=\mu$, while any infection and/or recovery parameter distributions can also be accommodated.
Various intervention strategies have different impacts on these epidemiological parameters.

Our model can be easily extended to accommodate other aspects of disease modeling.
Some studies report cases where infected individuals remain asymptomatic throughout the course of the infection; however, these cases seem to have a much lower secondary attack rate~\cite{Koh2020,Byambasuren2020}. 
To keep the analysis simple, we do not consider asymptomatic individuals who do not become infected prior to recovery but briefly discuss how the frameworks used could accommodate such cases in Appendix~\ref{app:AsympState}. We also discuss the extension to a model with an additional compartment where the exposed individual is non-contagious for a period of time, and derive the corresponding DMP equations in Appendix~\ref{app:SCEIRmodel}.  Despite the simplicity assumptions made, the proposed SEIR model captures the essential  characteristics of presymptomatic transmission and constitutes an effective approximation of the realistic spreading dynamics.

\section{Theoretical Frameworks\label{sec:theory}}

\subsection{Individual-Based Mean-Field Approach\label{subsec:IBMF}}

Since the exact solutions of the stochastic spreading processes Eqs.~(\ref{eq:SEIR_def1})-(\ref{eq:SEIR_def4}) are difficult to obtain, various approximation methods have been developed to tackle such complex dynamics~\cite{Satorras2015}. A simple method is the individual-based mean-field (IBMF) approach~\cite{Youssef2011,Satorras2015}, which expresses the evolution of the marginal distribution $P_{\sigma}^{i}(t)$ that each node $i$ belongs to state $\sigma$ by assuming the independence on the probabilities of neighboring nodes. 

Consider the SEIR model defined in Sec.~\ref{sec:model}. For node $i$ being in the susceptible state $S$, it will remain in state $S$ in the next time step if none of its neighbors transmits an infection signal to node $i$. In the IBMF framework, this occurs with probability $\prod_{k\in\partial i}\big[1-\alpha_{ki}P_{E}^{k}(t)-\beta_{ki}P_{I}^{k}(t)\big]$, where $\partial i$ denotes the set of nodes adjacent to node $i$. Therefore the evolution of the marginal probability node $i$ in state $S$ is given by

\begin{equation}
P_{S}^{i}(t+1)=P_{S}^{i}(t)\prod_{k\in\partial i}\big[1-\alpha_{ki}P_{E}^{k}(t)-\beta_{ki}P_{I}^{k}(t)\big].
\end{equation}
The probability of node $i$ in the exposed state $E$ increases if there is an infection signal from its neighbors, while it decreases
with rate $\nu_{i}$ (probability of transforming into state $I$) as infection symptoms appear. The corresponding IBMF dynamical equation is

\begin{align}
P_{E}^{i}(t+1)= & (1-\nu_{i})P_{E}^{i}(t)\nonumber \\
+P_{S}^{i}(t) & \bigg\{1-\prod_{k\in\partial i}\big[1-\alpha_{ki}P_{E}^{k}(t)-\beta_{ki}P_{I}^{k}(t)\big]\bigg\},\label{eq:Emarginal}
\end{align}
Similarly, the evolution of $P_{I}^{i}(t)$ and $P_{R}^{i}(t)$ are given by
\begin{align}
P_{I}^{i}(t+1)= & (1-\mu_{i})P_{I}^{i}(t)+\nu_{i}P_{E}^{i}(t),\label{eq:Imarginal}\\
P_{R}^{i}(t+1)= & P_{R}^{i}(t)+\mu_{i}P_{I}^{i}(t).
\end{align}

This approach has been used to investigate similar models addressing the COVID-19 pandemic~\cite{MoBaichuan2020,Basnarkov2020}. However, the drastic simplification based on the independence assumption of probabilities may lead to large approximation errors~\cite{Youssef2011}. One source of errors comes from the mutual infection effect due to this decorrelation assumption~\cite{Shrestha2015,Koher2019}. For instance, suppose that a node $i$, having probability $P_{E}^{i}(t)$ in the exposed state, infects its susceptible neighboring node $k$ at time $t$, then node $k$ can also reinfect node $i$ at time $t+1$ with some probability, which is an artifact of neglecting the correlation between nodes $i$ and $k$. Such effects need to be correctly accounted for to improve accuracy.

\subsection{Dynamic Message-passing Approach}
\label{subsec:DMP}

The dynamic message-passing approach, an algorithm that originates from the statistical physics literature~\cite{Karrer2010,Lokhov2014,Lokhov2015} avoids the mutual infection effect by considering the irreversible complete trajectories of the system. Formally, the DMP equations can be derived from the belief propagation equations of dynamical trajectories, which is especially useful when the correct set of dynamical variables are difficult to determine straightforwardly~\cite{Lokhov2015,HanlinSun2019}. In this section, we provide an intuitive derivation of the DMP equations of the SEIR model, while we give the more formal derivation based on belief propagation in Appendix~\ref{app:deriveDMP}.

Similar to the IBMF approach, the DMP method aims at deriving the evolution of the marginal distributions. Consider the marginal probability $P_{S}^{i}(t)$ of node $i$ being at state $S$ at time $t$, it is given by
\begin{equation}
P_{S}^{i}(t)=P_{S}^{i}(0)\prod_{k\in\partial i}\theta^{k\to i}(t),\label{eq:PS_i_t_DMP}
\end{equation}
where $\theta^{k\to i}(t)$ is the probability that node $i$ has not contracted the disease from node $k$ up to time $t$. We have made the assumption that the probability which node $i$ has not contracted the disease from its neighbors up to time $t$ factorizes as $\prod_{k\in\partial i}\theta^{k\to i}(t)$. This assumption is valid in tree graphs but constitutes a good approximation in many loopy networks~\cite{Lokhov2015}. 

The message $\theta^{k\to i}$ decreases if node $k$ transmits the infection signal to node $i$, which occurs with probability $\alpha_{ki}$
if node $k$ is in state $E$ or with probability $\beta_{ki}$ if node $k$ is in state $I$. Therefore, it follows the update rule

\begin{align}
\theta^{k\to i}(t+1)= & \theta^{k\to i}(t)-\alpha_{ki}\psi^{k\to i}(t)-\beta_{ki}\phi^{k\to i}(t),\label{eq:theta_ki_t_iter}
\end{align}
where $\psi^{k\to i}(t)$ is the probability that $k$ is in state $E$ but has not transmitted the infection signal to node $i$, and $\phi^{k\to i}(t)$ is the cavity probability (on a graph where node $i$ is absent - a cavity) that $k$ is in state $I$ but has not transmitted the infection signal to node $i$ up to time $t$. 

The message $\phi^{k\to i}$ decreases if node $k$ transmits the infection signal to node $i$ or changes from state $I$ into state $R$; note that the two processes can occur at the same time step. On the other hand, it increases if node $k$ changes from state $E$ into $I$. Therefore, it is updated according to 

\begin{align}
\phi^{k\to i}(t+1)= & \big(1-\beta_{ki}\big)\big(1-\mu_{k}\big)\phi^{k\to i}(t)\nonumber \\
 & +\big(1-\alpha_{ki}\big)\nu_{k}\psi^{k\to i}(t).\label{eq:phi_ki_t_iter}
\end{align}

Similarly, the message $\psi^{k\to i}(t)$ decreases if node $k$ transmits the infection signal to node $i$ or changes from state $E$ into state $I$, while it increases if node $k$ changes from state $S$ into $E$. In computing the probability increment due to the latter case, one needs to exclude the contribution from node $i$ to node $k$ in the previous time steps to avoid the effect of mutual infection. This is achieved through defining
\begin{equation}
P_{S}^{k\to i}(t)=P_{S}^{k}(0)\prod_{l\in\partial k\backslash i}\theta^{l\to k}(t),\label{eq:PS_ki_t_DMP}
\end{equation}
which is computed in the cavity graph assuming that node $i$ has been removed. Then the message $\psi^{k\to i}(t)$ follows the update
rule

\begin{align}
\psi^{k\to i}(t+1)= & (1-\alpha_{ki})\big(1-\nu_{k}\big)\psi^{k\to i}(t)\nonumber \\
 & +\big[P_{S}^{k\to i}(t)-P_{S}^{k\to i}(t+1)\big].\label{eq:psi_ki_t_iter}
\end{align}

Upon computing the messages $\{\theta^{k\to i}(t),\phi^{k\to i}(t),\psi^{k\to i}(t),P_{S}^{i\to j}(t)\}$ from the update rules, the marginal probability $P_{S}^{i}(t)$ can be obtained by Eq.~(\ref{eq:PS_i_t_DMP}). The marginal probabilities of other states can also be determined as
\begin{align}
P_{R}^{i}(t+1)= & P_{R}^{i}(t)+\mu_{i}P_{I}^{i}(t),\label{eq:PR_DMP_iter}\\
P_{I}^{i}(t+1)= & (1-\mu_{i})P_{I}^{i}(t)+\nu_{i}P_{E}^{i}(t),\label{eq:PI_DMP_iter}\\
P_{E}^{i}(t+1)= & 1-P_{S}^{i}(t+1)-P_{I}^{i}(t+1)-P_{R}^{i}(t+1),\label{eq:PE_DMP_iter}
\end{align}
We assume that the nodes are either in susceptible or exposed states at time $t=0$, such that the initial conditions are solely determined by $P_{S}^{i}(0)$ as
\begin{align}
\psi^{i\to j}(0) & =1-P_{S}^{i}(0),\\
\theta^{i\to j}(0) & =\phi^{i\to j}(0)=0,\\
P_{E}^{i}(0) & =1-P_{S}^{i}(0),\\
P_{I}^{i}(0) & =P_{R}^{i}(0)=0.
\end{align}
If node $i$ is selected as the initial exposed node, the initial condition is simply set as $P_{E}^{i}(0)=1,P_{S}^{i}(0)=P_{I}^{i}(0)=P_{R}^{i}(0)=0$. Based on the initial data, the messages are solved by updating the DMP Equations~(\ref{eq:theta_ki_t_iter})-(\ref{eq:psi_ki_t_iter}) forward in time, after which the marginal probabilities are determined by Equations~(\ref{eq:PS_i_t_DMP}), (\ref{eq:PR_DMP_iter}), (\ref{eq:PI_DMP_iter})and (\ref{eq:PE_DMP_iter}). 
The computational complexity of the DMP algorithm is linear in the number of time steps and in number of edges as $\mathcal{O}(|\mathcal{E}|T)$. Therefore, the DMP approach saves a significant amount of computational resources compared to Monte Carlo (MC) simulations, which requires many realizations to obtain reliable results. 
On the other hand, it is more demanding than the IBMF approach which deals with node-wise variables rather than edge-wise variables. Nevertheless, if the network is sparse, i.e., the average degree $\langle d\rangle=2|\mathcal{E}|/N\ll N$, then the DMP approach has the same order of computational complexity as the IBMF approach. This is relevant to the case of disease spreading as the number of close contacts each person has is limited~\cite{Mossong2008}, except for super-spreaders.

\subsection{Evaluation on Contact Networks}

Here we evaluate the effectiveness of the developed theories on contact networks, which are either artificially generated or adapted from some realistic human contact data. The realistic contact networks are taken from data sets obtained in the SocioPatterns collaboration~\cite{sociopatterns}, where the temporal face-to-face human contacts are projected to static contact networks as described in Appendix~\ref{app:contactnetworks}.

\begin{figure}
\includegraphics[scale=1.05]{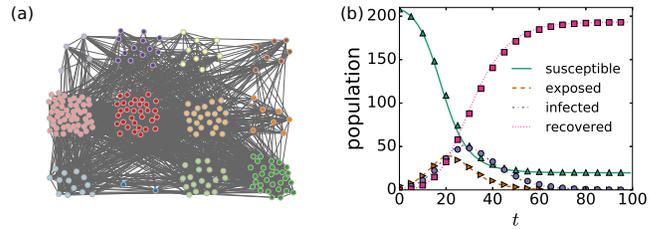}
\caption{(a) Structure of a contact network in the workplace taken from the SocioPatterns data (WP2015). The color of a node represents the department that the individual belongs to. (b) Evolution of the average population of each state. The parameters used are $\nu=0.2,\mu=0.125,\beta=0.016,\alpha=\beta/2$. 
At time $t=0$, there are 5 exposed nodes. The trajectories predicted by the DMP approach (represented by lines) match well with those from MC simulations (averaged over $10^{4}$ realizations, represented by dots). \label{fig:WP2015_net_and_simulation}}
\end{figure}

Many realistic contact networks exhibit community structures. For instance, in workplaces, people usually interacts more frequently with other people from the same department compared to those from other departments; in schools, students from the same class also interacts more frequently. An exemplar contact network in the workplace (WP2015) from the SocioPatterns data is depicted in Fig.~\ref{fig:WP2015_net_and_simulation}(a). We run the DMP algorithm for this contact network by randomly selecting 5 nodes as the initial exposed nodes and iterating for $T=100$ time steps. The evolution of the population of each compartment is shown in Fig.~\ref{fig:WP2015_net_and_simulation}(b). The number of exposed or infected cases first rises and then decreases, and eventually dies out when herd immunity is reached. The trajectories predicted by the DMP approach match well with those from MC simulations in this case. 

\begin{figure}
\includegraphics[scale=1.05]{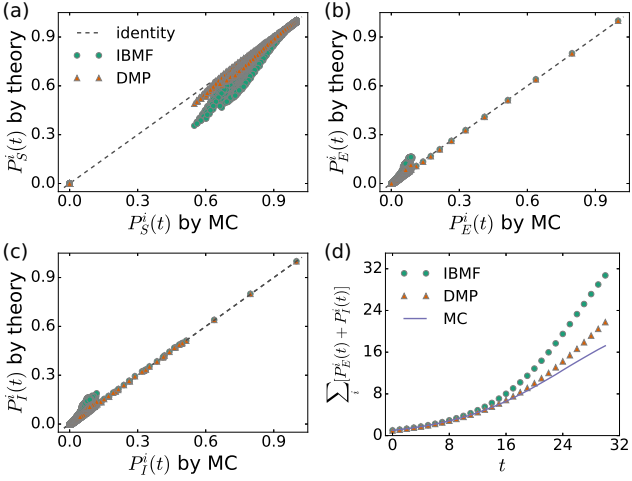}
\caption{Comparison between theory and MC simulation (averaged over $10^{4}$ realizations). The underlying network is a random regular graph with $N=100,d=10$. The parameters used are $T=30,\nu=0.2,\mu=0.125,\beta=0.03,\alpha=\beta/2$. At time $t=0$, there is only one exposed node. The accuracy of the DMP approximation is much better than the IBMF approach.\label{fig:error_detailed_rrg100}}
\end{figure}

As another example, we evaluate the efficacy of our framework on random regular graphs, where all nodes have the same degree and are connected randomly. Only one node is selected as the initial exposed node, and the system is simulated for $T=30$ time steps. The results are shown in Fig.~\ref{fig:error_detailed_rrg100}, exhibiting a much better approximation accuracy of the DMP approach compared to IBMF.
Due to the effect of mutual infection, both IBMF and DMP tend to overestimate the outbreak.

\begin{figure}
\includegraphics[scale=1.05]{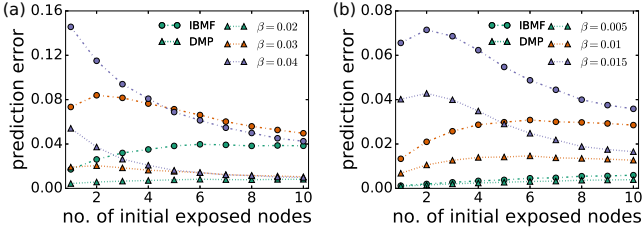}
\caption{Comparison between theory and MC simulation for different parameters, where the prediction error is calculated as $\frac{1}{NT} \sum_{\sigma\in\{S,E,I\}}\sum_{i,t}|P_{\text{simulation},\sigma}^{i}(t)-P_{\text{theory},\sigma}^{i}(t)|$. The parameters used are $T=30,\nu=0.2,\mu=0.125,\alpha=\beta/2$. Each data point is averaged over 5 instances with different sets of randomly selected initial exposed nodes. (a) Random regular graph with $N=100,d=10$. (b) Contact network in the workplace taken from the SocioPatterns data (WP2015). The prediction errors depend on the number of initial exposed nodes and the epidemiological parameter $\beta$.\label{fig:error_vs_no_init_seeds_and_beta}}
\end{figure}

In Fig.~\ref{fig:error_vs_no_init_seeds_and_beta}, we systematically compare the results between theories and MC simulation, showing that DMP provides a much better approximation than the IBMF approach. It is found that the prediction errors of both theoretical approaches generally increase with the epidemiological parameter $\beta$, and also depend on the number of initial exposed nodes. Intuitively, large values of $\beta$ lead to larger growth rates of the infections, in which case a small approximation error in early time steps could be amplified in late times. For large $\beta$, the prediction errors typically decrease as the number of initial exposed nodes increases. 
One possible reason is that the infections spread out from a unique source are correlated, so that the independence assumption in the IBMF and DMP approaches deteriorates~\cite{Altarelli2014}, e.g., the assumption that the messages in Eq.~(\ref{eq:PS_i_t_DMP}) factorize does not hold strictly. On the other hand, if there are multiple initial seeds that trigger the outbreak, the infection signals into a node will be less correlated, which partly preserves the decorrelation assumption. 

The network topology also impacts on the approximation accuracy of the theories as shown in Appendix~\ref{app:experiments}. As mentioned before, DMP avoids the mutual infection by excluding one step backtracking interaction, it does not take into account mutual infections due to backtracking of multiple steps, which can be non-negligible in networks with many short loops. In Appendix~\ref{app:experiments}, we find that the approximation accuracy deteriorates significantly when the localization of nonbacktracking centrality is present, an effect we will discuss below.

\section{Linearized Dynamics and Stability\label{sec:linearized_dynamics}}

\subsection{Linearized Dynamics of the DMP Equations}

The fate of the spreading processes depends on the epidemiological parameters. For large transmission probabilities $\beta$ and $\alpha$, the disease can spread out to a large fraction of the network, while it tends to die out quickly for small transmission probabilities. There exist thresholds for these parameters, above which the epidemic outbreaks occur. One commonly used method to determine epidemic thresholds is to examine whether the disease-free state is linearly stable to small perturbations~\cite{Boguna2002,Satorras2015}. 

Specifically, the initial disease-free state is perturbed infinitesimally as $P_{S}^{i}(0)=1-\epsilon^{i}$; if such perturbation diverges,
then the outbreak tend to spread out globally. At the initial stage, the message $\theta^{k\to i}(t)$, which denotes the probability node $k$ has not passed the infection signal to node $i$, can also be expressed as $\theta^{k\to i}(t)=1-\delta^{k\to i}(t)$. At time $t+1$, we have

\begin{align}
\theta^{k\to i}(t+1)= & 1-\delta^{k\to i}(t+1)\\
= & 1-\delta^{k\to i}(t)-\alpha_{ki}\psi^{k\to i}(t)-\beta_{ki}\phi^{k\to i}(t),\nonumber 
\end{align}
which implies

\begin{equation}
\delta^{k\to i}(t+1)=\delta^{k\to i}(t)+\alpha_{ki}\psi^{k\to i}(t)+\beta_{ki}\phi^{k\to i}(t),
\end{equation}
where $\delta^{k\to i}(t)$, $\psi^{k\to i}(t)$ and $\phi^{k\to i}(t)$ have small values. Expanding Eq.~(\ref{eq:PS_ki_t_DMP}) and keeping first order of $\psi^{k\to i}(t)$ and $\phi^{k\to i}(t)$ leads to

\begin{align}
P_{S}^{i\to j}(t+1)\approx & P_{S}^{i\to j}(t)-\sum_{k\in\partial i\backslash j}\big[\alpha_{ki}\psi^{k\to i}(t)+\beta_{ki}\phi^{k\to i}(t)\big].
\end{align}
Then $\psi^{k\to i}(t+1)$ in Eq.~(\ref{eq:psi_ki_t_iter}) can be approximated as

\begin{align}
\psi^{i\to j}(t+1) & \approx(1-\alpha_{ij})(1-\nu_{i})\psi^{i\to j}(t)\nonumber \\
 & +\sum_{k\in\partial i\backslash j}\big[\alpha_{ki}\psi^{k\to i}(t)+\beta_{ki}\phi^{k\to i}(t)\big].\label{eq:psi_ij_t_DMP_approx}
\end{align}
Equations~(\ref{eq:psi_ij_t_DMP_approx}) and (\ref{eq:phi_ki_t_iter}) constitute a linear dynamical system of the messages $\{\phi^{i\to j}(t),\psi^{i\to j}(t)\}$. 

In the following, we use homogeneous parameters $\alpha_{ij}=\alpha,\beta_{ij}=\beta,\nu_{i}=\nu,\mu_{i}=\mu$. To make the linearized dynamical equations more compact, we introduce the $2|\mathcal{E}|\times2|\mathcal{E}|$ nonbacktracking (NB) matrix
with elements 
\begin{equation}
B_{i\to j,k\to l}=\delta_{il}(1-\delta_{jk}),\label{eq:NB_matrix_def}
\end{equation}
which are non-zero if and only if the directed edge $i\to j$ follows right after edge $k\to l$, i.e., in a configuration like $k\to l(=i)\to j$, but edge $i\to j$ does not backtrack to node $k$ \footnote{We remark that there are other conventions used when defining the NB matrix in the literature; a common convention corresponds to the transpose of $B$ defined in Eq.~(\ref{eq:NB_matrix_def})~\cite{Krzakala2013}.}. Then Eqs.~(\ref{eq:psi_ij_t_DMP_approx}) and (\ref{eq:phi_ki_t_iter}) can be written in the matrix form as 
\begin{align}
\left(\begin{array}{c}
\boldsymbol{\psi}(t+1)\\
\boldsymbol{\phi}(t+1)
\end{array}\right)= & \mathcal{J}\left(\begin{array}{c}
\boldsymbol{\psi}(t)\\
\boldsymbol{\phi}(t)
\end{array}\right),\label{eq:DMP_linear_dyn_matrix_form}
\end{align}
where the $\mathcal{J}$ is the Jacobian matrix of the dynamical system defined as

\begin{equation}
\mathcal{J}=\left(\begin{array}{cc}
(1-\alpha)(1-\nu)I+\alpha B & \beta B\\
(1-\alpha)\nu I & (1-\beta)(1-\mu)I
\end{array}\right),
\end{equation}
where $I$ is the $2|\mathcal{E}|$-dimensional identity matrix. The spectral radius $\rho(\mathcal{J})$ of the Jacobian $\mathcal{J}$ determines the growth rate of the fastest mode of the linearized dynamics. The appearance of the NB matrix $B$ in the linearized dynamical equation is rooted in the fact that the one-step backtracking infection is excluded in the DMP equations (e.g., through Eq.~(\ref{eq:PS_ki_t_DMP})), therefore it also appears in other algorithms for complex networks based on linearizing belief propagation, such as the applications in community detection~\cite{Krzakala2013} and percolation~\cite{Karrer2014}.

\subsection{Spectral Properties of the Jacobian}

The condition that $\rho(\mathcal{J})\ge1$ corresponds to an exponential growth of the linearized dynamics in Eq.~(\ref{eq:DMP_linear_dyn_matrix_form}), which implies that the disease is likely to spread out globally. The solution of $\rho(\mathcal{J})=1$ marks the phase boundary of the epidemiological parameters. Since the matrix elements of $\mathcal{J}$ are non-negative, the Perron-Frobenius theorem asserts that (i) its leading eigenvalue $\lambda_{\mathcal{J}}^{\text{max}}$ (defined as the eigenvalue having the largest real part) equals its spectral radius $\rho(\mathcal{J})$ and therefore is real and non-negative and (ii) there exists an eigenvector with non-negative and nonzero elements corresponding to $\lambda_{\mathcal{J}}^{\text{max}}$~\cite{Horn2012}. 

Consider the eigenvalue equation of the Jacobian matrix $\mathcal{J}$,
\begin{equation}
\mathcal{J}\left(\begin{array}{c}
\boldsymbol{u}\\
\boldsymbol{v}
\end{array}\right)=\lambda_{\mathcal{J}}\left(\begin{array}{c}
\boldsymbol{u}\\
\boldsymbol{v}
\end{array}\right),
\end{equation}
which can be simplified to

\begin{align}
\boldsymbol{v} & =\frac{(1-\alpha)\nu}{\lambda_{\mathcal{J}}-(1-\beta)(1-\mu)}\boldsymbol{u},\label{eq:relation_v_and_u}\\
B\boldsymbol{u} & =\frac{\lambda_{\mathcal{J}}-(1-\alpha)(1-\nu)}{\alpha+\beta\frac{(1-\alpha)\nu}{\lambda_{\mathcal{J}}-(1-\beta)(1-\mu)}}\boldsymbol{u}.\label{eq:u_eigvector_of_B}
\end{align}
It implies that $\big[\lambda_{\mathcal{J}}-(1-\alpha)(1-\nu)\big]/\big[\alpha+\beta\frac{(1-\alpha)\nu}{\lambda_{\mathcal{J}}-(1-\beta)(1-\mu)}\big]$ is an eigenvalue of the NB matrix $B$ with eigenvector $\boldsymbol{u}$, denoted as $\lambda_{B}$. The Perron-Frobenius theorem also guarantees that the leading eigenvalue $\lambda_{B}^{\max}$ of $B$ is real and non-negative. It can be shown that the leading eigenvalue $\lambda_{\mathcal{J}}^{\max}$ is related to $\lambda_{B}^{\max}$ as shown in Appendix~\ref{app:EpidemicThreshold}
\begin{align}
\lambda_{\mathcal{J}}^{\max}= & \frac{1}{2}\big[(1-\alpha)(1-\nu)+(1-\beta)(1-\mu)+\alpha\lambda_{B}^{\max}\big]\nonumber \\
 & +\frac{1}{2}\bigg[\big((1-\alpha)(1-\nu)-(1-\beta)(1-\mu)+\alpha\lambda_{B}^{\max}\big)^{2}\nonumber \\
 & \qquad+4(1-\alpha)\nu\beta\lambda_{B}^{\max}\bigg]^{1/2}.\label{eq:lamJmax_of_lamBmax}
\end{align}
In this way, we relate the dynamical properties of the SEIR model to the epidemiological parameters and network structure properties, where the latter is subtly conveyed through the eigenvalue $\lambda_{B}^{\max}$. This is in contrast to the growth rate given by the commonly used basic reproduction number $R_{0}$, defined as the expected number of secondary cases caused by a single randomly-selected exposed individual when the rest of the population are susceptible. In the SEIR model considered here, the $R_{0}$ is estimated to be (see Appendix~\ref{app:ReproductionNo})
\begin{equation}
R_{0}=\langle d\rangle\left(\frac{\alpha}{\nu}+\frac{\beta}{\mu}\right),\label{eq:R0_SEIR}
\end{equation}
which only depends on the averaged degree $\langle d\rangle$, but neglects possible higher order structures of the contact networks. It has long been recognized that the $R_{0}$ measure is deficient in network epidemiology~\cite{Youssef2011,Satorras2015}.

Another useful measure is the effective reproduction number $R(t)$, which has a similar definition to $R_{0}$ but is based on the expected secondary infections to the remaining susceptible population at time $t$~\cite{Keeling2008}. The network structure and dynamical model play a role in determining $R(t)$, as the real-time susceptible population needs to be estimated, which is difficult to carry out analytically. In general, applying $R(t)$ requires solving the dynamics and it is more suitable as an indicator for monitoring the spread (e.g., as in Ref.~\cite{Adriana2021}), which differs from the role of $R_{0}$ or the epidemic threshold as predictors.

The computation of the leading eigenvalue $\lambda_{B}^{\max}$ can be demanding as the NB matrix is of size $2|\mathcal{E}|\times2|\mathcal{E}|$, which is a large matrix if the underlying network is relatively dense. It has been observed that the spectrum of $B$ can be obtained from a much smaller matrix of size $2N\times2N$~\cite{Watanabe2009,Krzakala2013}
\begin{equation}
M=\begin{pmatrix}0 & D-I_{N}\\
-I_{N} & A
\end{pmatrix},\label{eq:M_matrix}
\end{equation}
where $I_{N}$ is the $N$-dimensional identity matrix, $D$ is the diagonal matrix of node degrees with elements $D_{ij}=d_{i}\delta_{ij}$, and $A$ is the adjacency matrix with elements satisfying $A_{ij}=1$ if $(i,j)\in\mathcal{E}$ and $A_{ij}=0$ otherwise. Intuitively speaking, the reduction of complexity comes from compressing the edge-based data, e.g., $\phi^{i\to j}(t)$ and $\psi^{i\to j}(t)$, to node-based data~\cite{Krzakala2013} as shown in Appendix~\ref{app:Eigenvalue}. This allows us to work with networks of relatively large sizes.

\section{Epidemic Threshold\label{sec:threshold}}

\subsection{Determining the Critical Points}

Equation~(\ref{eq:lamJmax_of_lamBmax}) gives rise to the epidemic threshold as predicted by the DMP approach through the solution of $\lambda_{\mathcal{J}}^{\max}(\beta,\alpha,\nu,\mu,\lambda_{B}^{\max})=1$, keeping in mind that $\lambda_{\mathcal{J}}^{\max}=\rho(\mathcal{J})$. 

The same derivation can also be applied to the IBMF equations, as shown in Appendix~\ref{app:Eigenvalue}, leading to

\begin{align}
\lambda_{\mathcal{J}^{\text{MF}}}^{\max}= & \frac{1}{2}\big[(1-\nu)+(1-\mu)+\alpha\lambda_{A}^{\max}\big]\\
+ & \frac{1}{2}\bigg[\big((1-\nu)-(1-\mu)+\alpha\lambda_{A}^{\max}\big)^{2}+4\nu\beta\lambda_{A}^{\max}\bigg]^{1/2},\nonumber 
\end{align}
which relates the maximal eigenvalue $\lambda_{\mathcal{J}^{\text{MF}}}^{\max}$ of the Jacobian matrix $\mathcal{J}^{\text{MF}}$ of the IBMF equations in Sec.~\ref{subsec:IBMF} and the adjacency matrix $A$ of the network. Solving $\lambda_{\mathcal{J}^{\text{MF}}}^{\max}(\beta,\alpha,\nu,\mu,\lambda_{A}^{\max})=1$ yields the epidemic threshold as predicted by the IBMF approach. 

The epidemic thresholds obtained by the two theoretical approaches are to be contrasted with those obtained from numerical simulations, where the large time limit is taken such that the outbreaks saturate and the final state of each node $i$ is either susceptible or recovered. The fraction of nodes that have been infected is an order parameter $r=\sum_{i}P_{R}^{i}(\infty)/N$ defining the phase transition from localized infections to global epidemics. Since statistical fluctuation is large near criticality, one can estimate the critical point through the variability measure~\cite{Shu2015}

\begin{equation}
C_{r}=\frac{\sqrt{\langle r^{2}\rangle-\langle r\rangle^{2}}}{\langle r\rangle},
\end{equation}
which peaks at the critical point.

\subsection{Phase Transition in Random Regular Graphs}

\begin{figure}
\includegraphics[scale=1.05]{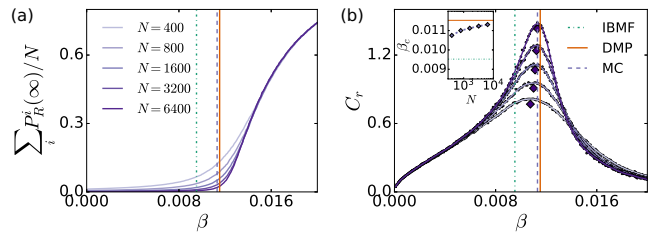}

\caption{Phase transition behavior of the SEIR model in random regular graphs of degree $d=10$. The parameters are $\nu=0.2,\mu=0.125,\alpha=\beta/2$, while $\lambda_{A}^{\max}=10,\lambda_{B}^{\max}=9$. The systems are initiated with 5 exposed nodes and are simulated for a sufficiently long time such that every node that has contracted the disease has recovered. In both (a) and (b), the three vertical lines correspond to the critical points $\beta_{c}$ obtained via different approaches, and different lines correspond to networks of different sizes. In both the IBMF and the DMP approaches, $\beta_{c}$ is the same for networks of different sizes as the degree is fixed. In MC simulation, $\beta_{c}$ is obtained through the largest network ($N=6400$) considered. (a) Order parameter $r=\sum_{i}P_{R}^{i}(\infty)/N$ as a function of $\beta$, obtained in MC simulations. (b) Variability of the order parameter, i.e., $C_{r}=\sqrt{\langle r^{2}\rangle-\langle r\rangle^{2}}/\langle r\rangle$, as a function of $\beta$. Since the variability $C_{r}$ obtained in simulations (data points in circle markers) is subject to fluctuations, we apply curve fitting to the experimental data using Gaussian processes with a radial basis function kernel (the fitted curves are shown in solid lines)~\cite{Rasmussen2006}. The maximum point of $C_{r}$ in the fitted curve (diamond-shape marker) is marked as the critical point obtained from MC simulations. Inset of (b) shows that the critical point $\beta_{c}$ obtained in MC simulations approaches the one obtained by the DMP approach as the network size increases. \label{fig:phase_transition_rrg_very_beta}}
\end{figure}

As an example, we consider random regular graphs of degree $d=10$, where the leading eigenvalues of the matrices $A$ and $B$ have exact expressions as $\lambda_{A}^{\max}=d=10,\lambda_{B}^{\max}=d-1=9$, irrespective of the network size, as described in Appendix~\ref{app:Eigenvalue}. We also fix the values of $\nu$ and $\mu$, let $\alpha=\beta/2$ and consider the phase transition by varying $\beta$. It is shown in Fig.~\ref{fig:phase_transition_rrg_very_beta}(a) that a significant fraction of the systems nodes are affected by the epidemic outbreak above the critical point $\beta_{c}$. 
In Fig.~\ref{fig:phase_transition_rrg_very_beta}(b), we pinpoint the critical point $\beta_{c}$ from MC simulations through the variability measure $C_{r}$, and compare them to those obtained via the IBMF and the DMP approaches, where it is observed that the DMP approach provides a much better estimation. In Appendix~\ref{app:Eigenvalue} we observed that the epidemic has a small probability to die out even for large $\beta$, a behavior that also appears in the SIR model~\cite{Shu2015} and may impact on the estimation of $\beta_{c}$ through simulations~\cite{Koher2019}. This is not captured by the theories which only consider averaged quantities. It would be an interesting future direction to study the deviations from the mean behaviors~\cite{Bianconi2018} and possible heterogeneous structures~\cite{Kuehn2017}.

\section{Exploring Phase Diagram and the Impact of Intervention Strategies}

In this section, we explore phase diagrams in different scenarios via the DMP approach and examine the impact of some intervention strategies.
Any specific intervention strategy will have an effect on the system parameters, and therefore influence the dynamics. Here, we primarily examine their impact on epidemic thresholds in the asymptotic limit.
In particular, we expect that various social distancing measures effectively reduce $\beta$ and $\alpha$.
Some intervention strategies may influence the transmission probabilities $\beta$ and $\alpha$ in different manners, so we consider the phase diagrams in parameter subspace spanned by $\beta$ and $\alpha$, rather than keeping a fixed ratio between them. 

The critical line separating the parameter regions of localized infections and global outbreaks, obtained by solving $\lambda_{\mathcal{J}}^{\max}(\beta,\alpha,\nu,\mu,\lambda_{B}^{\max})=1$ for $\alpha$, has the following expression
\begin{align}
\alpha_{c}(\beta)= & \nu\big[\beta(\lambda_{B}^{\max}+\mu-1)-\mu\big]/\nonumber \\
 & \big[\beta(1-\mu)(1-\nu)+\beta\lambda_{B}^{\max}(\mu+\nu-1)\nonumber \\
 & \,-\mu(\lambda_{B}^{\max}+\nu-1)\big].\label{eq:alpha_c_vs_beta_SEIR}
\end{align}

While there is no presymptomatic transmission in the SIR model defined by Eqs.~(\ref{eq:SIR_def1}) and (\ref{eq:SIR_def2}), we can compare with the expression obtained for the critical transmission probability~\cite{Koher2019}
\begin{equation}
\beta_{c}^{\text{SIR}}=\frac{\mu}{\lambda_{B}^{\max}+\mu-1}.\label{eq:beta_c_SIR}
\end{equation}
As a comparison, we sketch the phase boundaries $\alpha_{c}(\beta)$ of the SEIR model and $\beta=\beta_{c}^{\text{SIR}}$ of the SIR model for certain $\nu,\mu$ and $\lambda_{B}^{\max}$ in Fig.~\ref{fig:phase_diagram_beta_alpha_plane}(a). The epidemic will not spread globally when the parameters are in both regions (I) and (II) in the SIR model. In contrast, the disease will die out only in region (I) in the SEIR model. Particular caution is needed in region (II), where $\beta$ is small enough so that it is safe for the SIR model, but the presymptomatic transmission is sufficiently significant (i.e., $\alpha>\alpha_{c}(\beta)$) to cause a global epidemic. 

\subsection{Cases For $\alpha=0$ and $\beta=0$}

Consider the special case $\alpha=0$, solving $\alpha_{c}(\beta)=0$ gives 
\begin{equation}
\beta_{c}^{\text{SEIR}}|_{\alpha=0}=\frac{\mu}{\lambda_{B}^{\max}+\mu-1},\label{eq:beta_c_at_alpha0_SEIR}
\end{equation}
which coincides with $\beta_{c}^{\text{SIR}}$. It indicates that the intersection of the two critical lines occurs at $\alpha=0$ (as shown in Fig.~\ref{fig:phase_diagram_beta_alpha_plane}(a)) is a general phenomenon. Physically, this special case corresponds to the traditional SEIR model where the exposed nodes are not infectious, which has the same epidemic thresholds with the SIR model, irrespective of the incubation period $1/\nu$.

Similarly, consider the special case $\beta=0$, then
\begin{equation}
\alpha_{c}^{\text{SEIR}}|_{\beta=0}=\frac{\nu}{\lambda_{B}^{\max}+\nu-1},\label{eq:alpha_c_at_beta0_SEIR}
\end{equation}
which is effectively the critical point of an SIR model viewing the state $I$ as ``recovered'', as expected. Equations~(\ref{eq:beta_c_at_alpha0_SEIR}) and (\ref{eq:alpha_c_at_beta0_SEIR}) explain some of the behaviors in Fig.~\ref{fig:phase_diagram_beta_alpha_plane}(b)(c)(d), as explained below. 

\subsection{Effect of Increasing $\mu$ and Quick Isolation of Symptomatic Individuals}

When $\nu$ and $\lambda_{B}^{\max}$ are fixed, varying $\mu$ will only affect the intersects of the phase boundary on the $\beta$-axis ($\beta_{c}^{\text{SEIR}}|_{\alpha=0}$ depends on $\mu$) but not the $\alpha$-axis ($\alpha_{c}^{\text{SEIR}}|_{\beta=0}$ is independent of $\mu$), as seen in Fig.~\ref{fig:phase_diagram_beta_alpha_plane}(b). Physically, since different values of $\mu$ represent different recovery rates, increasing $\mu$ can be effectively realized by isolating all nodes in state $I$ before they recover. If such a policy can be executed strictly and timely, it corresponds to a large $\mu$ that can significantly expand the parameter region of the epidemic-free phase, leaving a lot of flexibility in implementing social distancing measures (i.e., many usual social interactions can still be allowed). Nevertheless, if the presymptomatic transmission probability $\alpha$ is large enough, e.g., $\alpha>\alpha_{c}^{\text{SEIR}}|_{\beta=0}$, then isolating the patients in state $I$ alone is insufficient to slow down the spread; in this case, identifying nodes in state $E$ through contact tracing or mass testing, and/or implementing stricter social distancing measure become necessary.

\subsection{Diseases with Different Incubation Periods}

Similarly, when $\mu$ and $\lambda_{B}^{\max}$ are fixed, varying $\nu$ will only impact on the intersects of the phase boundaries with the $\alpha$-axis but not those with the $\beta$-axis, as seen in Fig.~\ref{fig:phase_diagram_beta_alpha_plane}(d). Physically, different $\nu$ values correspond to diseases with different incubation periods. For smaller $\nu$, the exposed nodes have a longer time to infect their neighbors, which makes it more difficult to combat the epidemic spreading.

\subsection{Approximation of the Phase Boundary}

\begin{figure}
\includegraphics[scale=1.05]{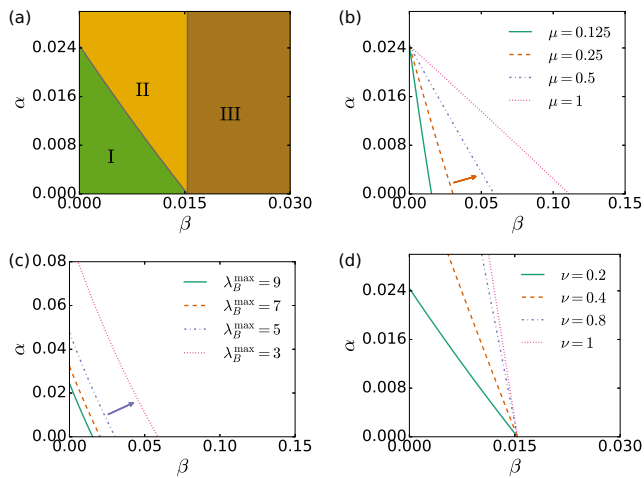}

\caption{Phase diagrams of the SEIR model in the $(\beta,\alpha)$-plane. In panels (b)(c)(d), the critical lines are obtained by solving Eq.~(\ref{eq:alpha_c_vs_beta_SEIR}). 
(a) The parameters used are $\nu=0.2,\mu=0.125,\lambda_{B}^{\max}=9$. The phase boundary separating region (I) from the others is given by Eq.~(\ref{eq:alpha_c_vs_beta_SEIR}), while the line separating region (II) and region (III) is $\beta=\beta_{c}^{\text{SIR}}$, where $\beta_{c}^{\text{SIR}}$ is given by Eq.~(\ref{eq:beta_c_SIR}). The transmission probabilities $\beta, \alpha$ can be reduced by imposing measures such as maintaining social distance and wearing face mask.
(b) The parameters used are $\nu=0.2,\lambda_{B}^{\max}=9$.  The arrow points to the direction of change in phase boundary due to increasing $\mu$, which enlarges the disease-free region and can be realized by identifying and isolating nodes in state $I$ more effectively.
(c) The parameters used are $\nu=0.2,\mu=0.125$. The arrow points to the direction of change in phase boundary due to increasing $\lambda_{B}^{\max}$, which can be realized by reducing contacts (self-isolation, lock-down).
(d) The parameters used are $\mu=0.125,\lambda_{B}^{\max}=9$. Different $\nu$ values correspond to diseases with different incubation periods.
\label{fig:phase_diagram_beta_alpha_plane}}

\end{figure}

Finally, although the phase boundary $\alpha_{c}(\beta)$ of the SEIR model is in general nonlinear, the cases considered in Fig.~\ref{fig:phase_diagram_beta_alpha_plane} exhibit an almost linear relation, except for a very small $\lambda_{B}^{\max}$. This can be seen more explicitly in the limit of large $\lambda_{B}^{\max}$, where one of the sufficient conditions is that the network has a large average degree (e.g., $\lambda_{B}^{\max}=d-1$ in random regular graphs), and requires the transmission probability $\beta$ and $\alpha$ to be small enough for the disease to die out. Under the condition of large $\lambda_{B}^{\max}$ and small $\beta$, Eq.~(\ref{eq:alpha_c_vs_beta_SEIR})
can be approximated as
\begin{align}
\alpha_{c}(\beta) & \approx-\frac{\nu(\lambda_{B}^{\max}+\mu-1)}{\mu(\lambda_{B}^{\max}+\nu-1)}\beta+\frac{\nu}{\lambda_{B}^{\max}+\nu-1},\nonumber \\
 & \approx-\frac{\nu}{\mu}\beta+\frac{\nu}{\lambda_{B}^{\max}}.\label{eq:alpha_c_vs_beta_SEIR_approx}
\end{align}
Equation~(\ref{eq:alpha_c_vs_beta_SEIR_approx}) explains the phenomena shown in Fig.~\ref{fig:phase_diagram_beta_alpha_plane}(c) that when $\nu$ and $\mu$ are fixed but $\lambda_{B}^{\max}$ is reduced, the epidemic-free region expands, but the slope of the phase boundary remains roughly unchanged. Physically, reducing $\lambda_{B}^{\max}$ can be achieved by limiting the number of contacts between nodes, as will be shown in Sec.~\ref{sec:dilution}.

A similar linear relation of the phase boundary can be obtained by the condition $R_{0}=1$, where $R_{0}$ satisfies Eq.~(\ref{eq:R0_SEIR}),
yielding
\begin{equation}
\alpha_{c}^{R_{0}}(\beta)=-\frac{\nu}{\mu}\beta+\frac{\nu}{\langle d\rangle},\label{eq:alpha_c_vs_beta_by_R0}
\end{equation}
which coincides with Eq.~(\ref{eq:alpha_c_vs_beta_SEIR_approx}) if we identify $\lambda_{B}^{\max}$ as $\langle d\rangle$. It turns out that the condition $\lambda_{B}^{\max}=\langle d\rangle$ holds approximately for Poisson random graphs~\cite{Martin2014}, as will also be shown in Sec.~\ref{sec:dilution}. Therefore, \emph{the critical line obtained via the basic reproduction number is a good estimation in dense Poisson random graphs, but may become a poor approximation otherwise}.

\section{Prediction of Outbreak Profile by the Nonbacktracking Centrality\label{sec:NB_centrality}}

Similar to the eigenvalues, the eigenvectors of the Jacobian matrix also provide valuable information on the dynamics. Consider the eigen-decomposition of the Jacobian as $\mathcal{J}=\sum_{a}\lambda_{\mathcal{J}}^{a}\boldsymbol{\xi}^{a}(\boldsymbol{\xi}^{-1})_{a}$, and note that the messages can be decomposed using the eigenvectors as bases, i.e., $\big(\boldsymbol{\psi}(t),\boldsymbol{\phi}(t)\big)=\sum_{a}c^{a}(t)\boldsymbol{\xi}^{a}$. In light of the linearized dynamics, the component $c^{\max}(t)$ corresponding to the leading eigenvalue $\lambda_{\mathcal{J}}^{\max}$, will dominate when $\lambda_{\mathcal{J}}^{\max}>1$ as the system evolves. Therefore, we can use the leading eigenvector $\boldsymbol{\xi}^{\max}$ of the Jacobian $\mathcal{J}$ to predict the outcome $\big(\boldsymbol{\psi}(T),\boldsymbol{\phi}(T)\big)$ of the dynamics.

According to Eq.~(\ref{eq:relation_v_and_u}), the two components of $\boldsymbol{\xi}^{\max}=(\tilde{\boldsymbol{u}},\tilde{\boldsymbol{v}})$ are proportional to each other, i.e., $\tilde{\boldsymbol{v}}\propto\tilde{\boldsymbol{u}}$. Thus, it is sufficient to examine one component only. In what follows, we consider $\tilde{\boldsymbol{u}}$, which is the leading eigenvector of the NB matrix $B$ according to Eqs.~(\ref{eq:u_eigvector_of_B}) and (\ref{eq:lamJmax_of_lamBmax}). Furthermore, we are ultimately interested in the marginal probabilities, which relate to the incoming messages to each node as seen in Eq.~(\ref{eq:PS_i_t_DMP}). In the linearized dynamics, the probability of newly infection of node $i$ is given by

\begin{equation}
P_{S}^{i}(t)-P_{S}^{i}(t+1)\approx\sum_{k\in\partial i}\big[\alpha\psi^{k\to i}(t)+\beta\phi^{k\to i}(t)\big].\label{eq:dPSi_DMP}
\end{equation}
Therefore, we need to consider the incoming vector of the leading eigenmode $\tilde{\boldsymbol{u}}$
\begin{align}
\tilde{u}_{i}^{\text{in}} & :=\sum_{k\in\partial i}\tilde{u}_{k\to i},
\end{align}
which is known as the nonbacktracking centrality~\cite{Martin2014}. Interestingly, in addition to the leading eigenvalue $\lambda_{B}^{\max}$, the NB centrality $\tilde{\boldsymbol{u}}^{\text{in}}$ can be obtained through the much smaller matrix $M$~\cite{Krzakala2013} as shown in Appendix~\ref{app:Eigenvalue}. The NB centrality has been shown to play an important role in percolation and SIR model in networks~\cite{Karrer2014, Rogers2015, Kuehn2017}.

\begin{figure}
\includegraphics[scale=1.05]{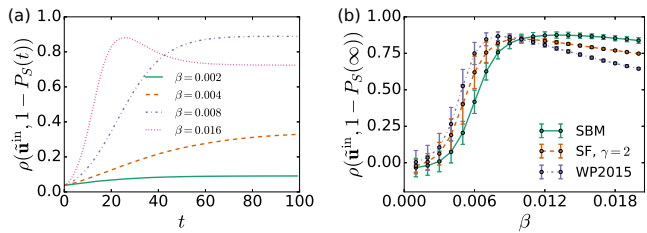}

\caption{Correlation coefficient $\rho$ between the nonbacktracking centrality $u_{i}^{\text{in}}$ and the outbreak profile, measured by the probability that each node $i$ has been infected, i.e., $1-P_{S}^{i}(t)$. The parameters are $\nu=0.2,\mu=0.125$. At time $t=0$, there are 5 initial exposed nodes. (a) Correlation coefficient $\rho$ of the spread on the WP2015 network as a function of time. (b) Correlation coefficient $\rho$ as a function of $\beta$ in different networks in the large time limit, where the spreading processes have saturated. Each data point is averaged over 5 instances with different sets of randomly selected initial exposed nodes.\label{fig:rho_vs_t_and_beta}}
\end{figure}

Based on Eq.~(\ref{eq:dPSi_DMP}), the probability of newly infection $\sum_{k\in\partial i}\big[\alpha\psi^{k\to i}(t)+\beta\phi^{k\to i}(t)\big]$ can be identified as the incoming vectors of the messages as $\alpha\psi_{i}^{\text{in}}(t)+\beta\phi_{i}^{\text{in}}(t)$, which will be increasingly more aligned with $\tilde{u}_{i}^{\text{in}}$ as time evolves. Thus, we can use $\tilde{\boldsymbol{u}}^{\text{in}}$ to predict the relative strengths of the outbreak, indicated by $\{1-P_{S}^{i}(t)\}$. Figure~\ref{fig:rho_vs_t_and_beta}(a) demonstrates that, for $\beta$ large enough, the evolution of correlation coefficient $\rho$ between $\tilde{\boldsymbol{u}}^{\text{in}}$ and the profile of the outbreak on the WP2015 network generally increases with time. For small $\beta=0.002$, the correlation remains low as the disease will die out. For a rather large $\beta=0.016$, the correlation coefficient $\rho$ increases rapidly in the initial stage of the development of the spreading, and then decreases to a lower level. This is because for a very large $\beta$, most nodes are likely to be infected eventually, irrespective of the spatial structure of the network. As shown in Fig.~\ref{fig:rho_vs_t_and_beta}(b), such relation between $\rho$ and $\beta$ is also observed in other networks, including a graph generated by a stochastic block model (SBM) and a scale-free network (SF) with degree exponent  $\gamma=2$ (details of the networks will be discussed in Sec.~\ref{sec:dilution}).

Similarly, the IBMF approach dictates that the leading eigenvector of the adjacency matrix $A$ (known as the eigenvector centrality~\cite{Martin2014}) is a predictor of the outbreak profile~\cite{Basnarkov2020}. Comparison between different centrality measures is briefly discussed in Appendix~\ref{app:experiments}, where the NB centrality generally provides a better prediction than the eigenvector centrality. As was mentioned before, the DMP approach only avoids the effect of mutual infection due to one-step backtracking, and the approximation accuracy can deteriorate if counteracting the effect of one-step backtracking is insufficient. This is of particular concern when the NB centrality $\tilde{\boldsymbol{u}}^{\text{in}}$ displays the localization phenomenon, where the centrality values of a few nodes are much larger than the others~\cite{Martin2014,Satorras2020}. In Appendix~\ref{app:experiments} we demonstrated the degradation of the approximation power of the DMP equations and the NB centrality in random networks with a relatively large planted clique (i.e., a complete subgraph), which possesses the localization property~\cite{Martin2014}; we also observed a poor approximation accuracy in one of the contact networks from the SocioPatterns data, recorded in a high school in 2013 (HS2013)~\cite{Mastrandrea2015}, where the NB centrality $\tilde{\boldsymbol{u}}^{\text{in}}$ is more localized on a few communities. 

\section{Effect of Reducing Contacts\label{sec:dilution}}

In addition to the social distancing rules that lower the transmission probabilities $\beta$ and $\alpha$, reducing social contacts between individuals is also an effective measure to slow down the spread of the disease. We examine its effects in the static contact networks considered here, by removing edges between nodes. Such a measure will change the network structure and reduce the leading eigenvalue $\lambda_{B}^{\max}$ of the matrix $B$, which can enlarge the epidemic-free region in the parameter space as shown in Fig.~\ref{fig:phase_diagram_beta_alpha_plane}(c). In general, the leading eigenvalue $\lambda_{B}^{\max}$ can depend intricately on the network structure. In the special case of configuration model where the node degrees follow a given distribution $P(d)$ and the nodes are wired randomly, it is found in Refs.~\cite{Krzakala2013,Martin2014} that $\lambda_{B}^{\max}$ can be approximated as
\begin{equation}
\lambda_{B}^{\max}\approx\frac{\langle d^{2}\rangle-\langle d\rangle}{\langle d\rangle},\label{eq:lamB_max_approx}
\end{equation}
which already indicates that even in the high degree limit, the epidemic thresholds obtained by $R_{0}$ in Eq.~(\ref{eq:alpha_c_vs_beta_by_R0}) do not generally coincide with those obtained in the DMP approach in Eq.~(\ref{eq:alpha_c_vs_beta_SEIR_approx}). Specifically, the second moment of the degree distribution is also relevant for epidemic thresholds in uncorrelated random networks~\cite{Satorras2015}, which is not captured by Eq.~(\ref{eq:alpha_c_vs_beta_by_R0}). A more refined approximation taking into account the relation of degrees of neighboring nodes is given in Ref.~\cite{Satorras2020}. The accuracy of these approximations depends on the validity of the uncorrelated random network assumption and/or the presence of localization of NB centrality~\cite{Satorras2020} (see also Appendix~\ref{app:approximation_accuracy_lambdaB}).

\begin{figure}
\includegraphics[scale=1.05]{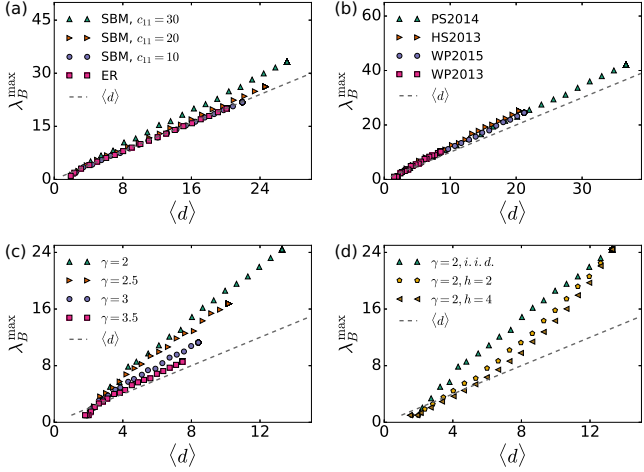}

\caption{Leading eigenvalue $\lambda_{B}^{\max}$ of the nonbacktracking matrix $B$ vs average degree $\langle d\rangle$ of the network. The rightmost dot of a curve represents the original network, while other dots correspond to the networks obtained by removing existing edges while keeping the largest connected component. Edges are removed randomly in panels (a)(b)(c), while edges adjacent to high-degree nodes are removed preferentially in panel (d). (a) ER random graph and networks generated through SBM with $N=200$. (b) Contact networks extracted from data in the SocioPatterns collaboration, including a network in a primary school (PS2014)~\cite{Gemmetto2014}, in a high school (HS2013)~\cite{Mastrandrea2015} and workplaces (WP2013, WP2015)~\cite{GNOIS2015,Genois2018}. (c) Scale-free networks generated from the configuration model. The network size is $N=400$ and the node degree is limited with $d<d^{\max}=100$. (d) Scale-free network with $\gamma=2$ (same as the one in panel (c)). In the edge removal process, a node $i$ is firstly selected according to the probability $p_{i}=\frac{d_{i}^{h}}{\sum_{i}d_{i}^{h}},h>0$, which is biased toward high-degree nodes; then one of the edges adjacent to node $i$ is randomly selected and removed. Different fractions of removed edges result in networks of different average degrees $\langle d\rangle$.\label{fig:dilution_experiments}}
\end{figure}

Due to the dilution of connections between nodes, a network may become disconnected, resulting in fragmented components~\cite{Barabasi2016}; if such cases happen, we keep the largest connected component in the following experiment.
In Fig.~\ref{fig:dilution_experiments}(a), Erd{\H o}s--R{\' e}nyi (ER) random graphs and networks generated via a stochastic block model (SBM) are considered. A network generated by SBM has 4 communities, where each community comprises 50 nodes; node $i$ in community $a$ is connected to $c_{ab}$ nodes of community $b$ on average. Here, we consider $c_{ab}=4,\forall a\neq b$ and $c_{aa}=10,\forall a\neq1$, while different values of $c_{11}$ are considered. When $c_{11}=10$, all four communities are statistically equivalent, and the node degrees follow a Poisson distribution, similar to the ER random graphs. For such Poisson random graphs, $\lambda_{B}^{\max}\approx\langle d\rangle$ according to Eq.~(\ref{eq:lamB_max_approx}), which is justified experimentally by the numerical results of Fig.~\ref{fig:dilution_experiments}(a). On the other hand, $\lambda_{B}^{\max}$ deviates from $\langle d\rangle$ in SBM networks with $c_{11}>0$. The networks become sparser when edges are removed randomly, and $\lambda_{B}^{\max}$ decreases linearly with $\langle d\rangle$. A quasi-linear decreasing trend is also observed in the random edge removal experiments in contact networks from the SocioPatterns collaboration as shown in Fig.~\ref{fig:dilution_experiments}(b), as well as in scale-free networks where the node degrees follow the power-law distribution $P(d)=d^{-\gamma}/\sum_{d=d^{\min}}^{d^{\max}}d^{-\gamma}$, as shown in Fig.~\ref{fig:dilution_experiments}(c). 
	
In Fig.~\ref{fig:dilution_experiments}(c), the leading eigenvalues $\lambda_{B}^{\max}$ of scale-free networks can deviate significantly from $\langle d\rangle$, especially for more heterogeneous networks with a small $\gamma$ value. Therefore, predicting the cause of the spread through $R_{0}$ becomes very unreliable in scale-free networks, an effect which has been observed in network epidemiology studies~\cite{Satorras2015}. Physically, there exist a small number of hubs (i.e., nodes with very large degrees) in scale-free networks, which can be viewed as super-spreaders that significantly facilitate the spread of the disease. In light of this, restricting contacts of these high-degree nodes preferentially can effectively reduce $\lambda_{B}^{\max}$ as shown in Fig.~\ref{fig:dilution_experiments}(d), and consequently lower the epidemic thresholds.

\section{Discussion and Outlook}

In this paper, we studied the SEIR model with presymptomatic transmissions, a feature of the COVID-19 disease, in both artificial and realistic contact networks through the dynamic message-passing method. The DMP approach provides a much better approximation compared to the IBMF approach while being much less computationally demanding than MC simulations, which is the most prominent feature of this method. The linear stability analysis of the DMP equations gives rise to the epidemic thresholds and phase diagrams of the models, where their dependence on epidemiological parameters and the network structure are elucidated. A larger presymptomatic transmission probability value $\alpha$ leads to a lower critical point $\beta_{c}$, which makes the strategy of blocking only symptomatic transmission less effective. We also show that different intervention strategies impact on the epidemic thresholds in different manners. The influence of network structure on the epidemic thresholds is represented by the leading eigenvalue $\lambda_{B}^{\max}$ of the nonbacktracking matrix $B$, which encodes more subtle structural information in contrast to the average number of contacts $\langle d\rangle$ appearing in the basic reproduction number $R_{0}$. Additionally, we demonstrated that the nonbacktracking centrality $\tilde{\boldsymbol{u}}^{\text{in}}$ related to the leading eigenvector of the matrix $B$ can effectively predict the relative strength of the outbreak.

On the other hand, it is worthwhile mentioning some limitations of the DMP method. Firstly, as the DMP approach is based on the decorrelation assumption of the infection signals, it may become a less accurate approximation when the correlations between trajectories are non-negligible, which can happen when there is a single initial exposed node seeding the dynamics and/or there are many short loops in the network. Second, the approximation accuracy of the DMP approach also deteriorates when counteracting the one-step backtracking reaction is insufficient to avoid the effect of mutual infection; this effect has been observed in networks where the nonbacktracking centrality exhibits a localization phenomenon as shown in Appendix~\ref{app:experiments}. It is an interesting future direction to further characterize the condition of nonbacktracking centrality localization, its impact on spreading processes and possible improvements~\cite{Satorras2020}.

`The theoretical frameworks were mostly applied to contact networks in some specific scenarios or those exhibiting particular characteristics, such as the presence of community structure or high-degree hubs. The applications on a wider scale (e.g., in a city) require considering additional network characteristics, such as the mixing pattern of different age groups~\cite{Mossong2008}, the household structure, and so on~\cite{Prem2017}. Additional states, such as hospitalized and dead, can also be considered in order to model the pressure on public-health services and social cost. Since presymptomatic transmissions make it more difficult to contain the disease by dealing with the symptomatic cases only, an extension of particular interest is to examine the effectiveness and limitation of (manual or digital) contact tracing~\cite{Bianconi2021, Baker2020, Adriana2021}, mass testing and other strategies which can identify exposed individuals that have not shown symptoms. The DMP equations developed here will also benefit future works which aim at optimal deployment of resources (e.g., vaccines) to contain the spread of epidemics with presymptomatic transmissions~\cite{Altarelli2014, Lokhov2017, HanlinSun2019}.

\begin{acknowledgments}
B.L. and D.S. acknowledge support from European Union's Horizon 2020 research and innovation program under the Marie Sk{\l}odowska-Curie grant agreement No.~835913. D.S. acknowledges support from the EPSRC program grant TRANSNET (EP/R035342/1). 
\end{acknowledgments}

\begin{widetext}
\appendix

\section{Deriving DMP Equations From Dynamic Belief Propagation}
\label{app:deriveDMP}

\subsection{Belief Propagation Equations of Trajectories}

In this Appendix, we derive the DMP equations from the principled
dynamic belief propagation established in Ref.~\cite{Lokhov2015}. It
is based on the message $m^{i\to j}(\vec{\sigma}_{i}|\vec{\sigma}_{j})$,
which is the cavity probability of the dynamical trajectory $\vec{\sigma}_{i}=[\sigma_{i}^{0},...,\sigma_{i}^{T}]$
(where $\sigma_{i}^{t}\in\{S,E,I,R\}$) of node $i$ in the cavity
graph in which node $j$ has been removed. Since the transition between
states is irreversible (only permitted in the order $S\to E\to I\to R$),
the trajectory $\vec{\sigma}_{i}$ can be parametrized by three transition
times $(\tau_{i},\omega_{i},\varepsilon_{i})$ as $\vec{\sigma}_{i}=|S_{0}SSE_{\tau_{i}}EEI_{\omega_{i}}IIIR_{\varepsilon_{i}}RR_{T}\rangle$.

The dynamic belief propagation for the modified SEIR model takes the
following form

\begin{equation}
m^{i\to j}(\tau_{i},\omega_{i},\varepsilon_{i}|\tau_{j},\omega_{j},\varepsilon_{j})=\sum_{\{\tau_{k},\omega_{k},\varepsilon_{k}\}_{k\in\partial i\backslash j}}W_{SEIR}\prod_{k\in\partial i\backslash j}m^{k\to i}(\tau_{k},\omega_{k},\varepsilon_{k}|\tau_{i},\omega_{i},\varepsilon_{i}),
\end{equation}
where $W_{SEIR}$ is the transition kernel

\begin{align}
W_{SEIR}= & \bigg\{ P_{E}^{i}(0)\mathbb{I}(\tau_{i}=0)+P_{S}^{i}(0)\mathbb{I}(\tau_{i}>0)\nonumber \\
& \quad\times\prod_{t'=0}^{\tau_{i}-2}\prod_{k\in\partial i}\bigg(1-\alpha_{ki}\mathbb{I}(\tau_{k}\leq t')\mathbb{I}(\omega_{k}\geq t'+1)-\beta_{ki}\mathbb{I}(\omega_{k}\leq t')\mathbb{I}(\varepsilon_{k}\geq t'+1)\bigg)\nonumber \\
& \quad\times\bigg[1-\prod_{k\in\partial i}\bigg(1-\alpha_{ki}\mathbb{I}(\tau_{k}\leq\tau_{i}-1)\mathbb{I}(\omega_{k}\geq\tau_{i})-\beta_{ki}\mathbb{I}(\omega_{k}\leq\tau_{i}-1)\mathbb{I}(\varepsilon_{k}\geq\tau_{i})\bigg)\bigg]\bigg\}\nonumber \\
& \times\bigg[\prod_{t''=\tau_{i}}^{\omega_{i}-2}(1-\nu_{i})\bigg]\nu_{i}\bigg[\prod_{t'''=\omega_{i}}^{\varepsilon_{i}-2}(1-\mu_{i})\bigg]\mu_{i}\mathbb{I}(\tau_{i}<\omega_{i}<\varepsilon_{i})\prod_{k\in\partial i}\mathbb{I}(\tau_{k}<\omega_{k}<\varepsilon_{k}).
\end{align}
The marginal of a trajectory of node $i$ is computed as

\begin{equation}
m^{i}(\tau_{i},\omega_{i},\varepsilon_{i})=\sum_{\{\tau_{k},\omega_{k},\varepsilon_{k}\}_{k\in\partial i}}W_{SEIR}\prod_{k\in\partial i}m^{k\to i}(\tau_{k},\omega_{k},\varepsilon_{k}|\tau_{i},\omega_{i},\varepsilon_{i}),
\end{equation}

The cavity probability of a trajectory has the following properties
\begin{align}
m^{i\to j}(\tau_{i},\omega_{i},\varepsilon_{i}+1|\cdot) & =(1-\mu_{i})m^{i\to j}(\tau_{i},\omega_{i},\varepsilon_{i}|\cdot),\label{eq:m_property_1}\\
m^{i\to j}(\tau_{i},\omega_{i}+1,\varepsilon_{i}|\cdot) & =\frac{1-\nu_{i}}{1-\mu_{i}}m^{i\to j}(\tau_{i},\omega_{i},\varepsilon_{i}|\cdot)\mathbb{I}(\varepsilon_{i}>\omega_{i}+1),\label{eq:m_property_2}\\
m^{i\to j}(\tau_{i},\omega_{i}+1,\varepsilon_{i}+1|\cdot) & =(1-\nu_{i})m^{i\to j}(\tau_{i},\omega_{i},\varepsilon_{i}|\cdot),\label{eq:m_property_3}
\end{align}
where similar relations hold for $m^{i}(\tau_{i},\omega_{i},\varepsilon_{i})$. In addition, if $\tau_{j} \geq \tau_{i}$, we have
\begin{equation}
m^{i\to j}(\tau_{i},\omega_{i},\varepsilon_{i}|\tau_{j}, \cdot, \cdot) = m^{i\to j}(\tau_{i},\omega_{i},\varepsilon_{i}|T, \omega_{j}, \varepsilon_{j}), \quad \forall T \geq \tau_{i}. \label{eq:m_property_4}
\end{equation}

\subsection{Deriving the Messages and Probability of Being in State $S$}

The cavity probability of node $i$ being in state $S$ at time $t$
is obtained by tracing over the probability of trajectories $m^{i \to j}$ in the cavity graph (assuming node $j$ is absent by setting $\tau_{j} = T$),

\begin{align}
P_{S}^{i\to j}(t) & =\sum_{\tau_{i},\omega_{i},\varepsilon_{i}}\mathbb{I}(t<\tau_{i}<\omega_{i}<\varepsilon_{i})m^{i\to j}(\tau_{i},\omega_{i},\varepsilon_{i}|T, \cdot, \cdot),
\end{align}
where a similar relation holds between $P_{S}^{i}(t)$
and $m^{i}(\tau_{i},\omega_{i},\varepsilon_{i})$.

Using the above definition, we compute the cavity probability $P_{S}^{i\to j}(t+1)$
as

\begin{align}
P_{S}^{i\to j}(t+1)= & \sum_{\tau_{i}>t+1}\sum_{\omega_{i}>\tau_{i}}\sum_{\varepsilon_{i}>\omega_{i}}m^{i\to j}(\tau_{i},\omega_{i},\varepsilon_{i}|T, \cdot, \cdot)\nonumber \\
= & \sum_{\tau_{i}>t+1}\sum_{\omega_{i}>\tau_{i}}\sum_{\varepsilon_{i}>\omega_{i}}\sum_{\{\tau_{k},\omega_{k},\varepsilon_{k}\}_{k\in\partial i\backslash j}}W_{SEIR}\prod_{k\in\partial i\backslash j}m^{k\to i}(\tau_{k},\omega_{k},\varepsilon_{k}|\tau_{i},\omega_{i},\varepsilon_{i})\nonumber \\
= & P_{S}^{i}(0)\prod_{k\in\partial i\backslash j}\bigg\{\sum_{\tau_{k},\omega_{k},\varepsilon_{k}}\mathbb{I}(\tau_{k}<\omega_{k}<\varepsilon_{k})m^{k\to i}(\tau_{k},\omega_{k},\varepsilon_{k}|T, \cdot, \cdot)\nonumber \\
& \quad\times\prod_{t'=0}^{t}\bigg[1-\alpha_{ki}\mathbb{I}(\tau_{k}\leq t')\mathbb{I}(\omega_{k}\geq t'+1)-\beta_{ki}\mathbb{I}(\omega_{k}\leq t')\mathbb{I}(\varepsilon_{k}\geq t'+1)\bigg]\bigg\}\nonumber \\
=: & P_{S}^{i}(0)\prod_{k\in\partial i\backslash j}\theta^{k\to i}(t+1),
\end{align}
where we have replaced $\prod_{k\in\partial i\backslash j}m^{k\to i}(\tau_{k},\omega_{k},\varepsilon_{k}|\tau_{i},\omega_{i},\varepsilon_{i})$ in the second line by $\prod_{k\in\partial i\backslash j}m^{k\to i}(\tau_{k},\omega_{k},\varepsilon_{k}|T,\cdot,\cdot)$ (a property imposed by the transition kernel and Eq.~(\ref{eq:m_property_4})) and traced
over the $\tau_{i},\omega_{i},\varepsilon_{i}$~\cite{Lokhov2015};
the message $\theta^{k\to i}(t+1)$ is defined as

\begin{align}
\theta^{k\to i}(t+1):= & \sum_{\tau_{k},\omega_{k},\varepsilon_{k}}\mathbb{I}(\tau_{k}<\omega_{k}<\varepsilon_{k})m^{k\to i}(\tau_{k},\omega_{k},\varepsilon_{k}|T, \cdot, \cdot)\nonumber \\
& \times\prod_{t'=0}^{t}\bigg(1-\alpha_{ki}\mathbb{I}(\tau_{k}\leq t')\mathbb{I}(\omega_{k}\geq t'+1)-\beta_{ki}\mathbb{I}(\omega_{k}\leq t')\mathbb{I}(\varepsilon_{k}\geq t'+1)\bigg),
\end{align}
which has the physical meaning of the cavity probability that node
$k$ (in either exposed or infected state) has not transmitted the
infection signal to node $i$ up to time $t+1$. In order to eliminate
the explicit dependence on the microscopic trajectories, we compute
the iteration scheme of $\theta^{k\to i}(t+1)$ as

\begin{align}
\theta^{k\to i}(t+1)-\theta^{k\to i}(t)= & \sum_{\tau_{k},\omega_{k},\varepsilon_{k}}\mathbb{I}(\tau_{k}<\omega_{k}<\varepsilon_{k})m^{k\to i}(\tau_{k},\omega_{k},\varepsilon_{k}|\cdot)\nonumber \\
& \times\prod_{t'=0}^{t-1}\bigg(1-\alpha_{ki}\mathbb{I}(\tau_{k}\leq t')\mathbb{I}(\omega_{k}\geq t'+1)-\beta_{ki}\mathbb{I}(\omega_{k}\leq t')\mathbb{I}(\varepsilon_{k}\geq t'+1)\bigg)\nonumber \\
& \times\bigg(-\alpha_{ki}\mathbb{I}(\tau_{k}\leq t)\mathbb{I}(\omega_{k}\geq t+1)-\beta_{ki}\mathbb{I}(\omega_{k}\leq t)\mathbb{I}(\varepsilon_{k}\geq t+1)\bigg)\nonumber \\
=: & -\alpha_{ki}\psi^{k\to i}(t)-\beta_{ki}\phi^{k\to i}(t),\label{eq:theta_update_rule}\\
\phi^{k\to i}(t):= & \sum_{\tau_{k},\omega_{k},\varepsilon_{k}}\mathbb{I}(\tau_{k}<\omega_{k}<\varepsilon_{k})\prod_{t'=0}^{t-1}\bigg(1-\alpha_{ki}\mathbb{I}(\tau_{k}\leq t')\mathbb{I}(\omega_{k}\geq t'+1)-\beta_{ki}\mathbb{I}(\omega_{k}\leq t')\mathbb{I}(\varepsilon_{k}\geq t'+1)\bigg)\nonumber \\
& \times m^{k\to i}(\tau_{k},\omega_{k},\varepsilon_{k}|\cdot)\mathbb{I}(\omega_{k}\leq t)\mathbb{I}(\varepsilon_{k}\geq t+1),\\
\psi^{k\to i}(t):= & \sum_{\tau_{k},\omega_{k},\varepsilon_{k}}\mathbb{I}(\tau_{k}<\omega_{k}<\varepsilon_{k})\prod_{t'=0}^{t-1}\bigg(1-\alpha_{ki}\mathbb{I}(\tau_{k}\leq t')\mathbb{I}(\omega_{k}\geq t'+1)-\beta_{ki}\mathbb{I}(\omega_{k}\leq t')\mathbb{I}(\varepsilon_{k}\geq t'+1)\bigg)\nonumber \\
& \times m^{k\to i}(\tau_{k},\omega_{k},\varepsilon_{k}|\cdot)\mathbb{I}(\tau_{k}\leq t)\mathbb{I}(\omega_{k}\geq t+1)\nonumber \\
= & \sum_{\tau_{k},\omega_{k},\varepsilon_{k}}\mathbb{I}(\tau_{k}<\omega_{k}<\varepsilon_{k})\prod_{t'=0}^{t-1}\bigg(1-\alpha_{ki}\mathbb{I}(\tau_{k}\leq t')\bigg)m^{k\to i}(\tau_{k},\omega_{k},\varepsilon_{k}|\cdot)\mathbb{I}(\tau_{k}\leq t)\mathbb{I}(\omega_{k}\geq t+1),
\end{align}
where we have introduced the message $\phi^{k\to i}(t)$ (the cavity
probability that $k$ is in state $I$ but has not transmitted the
infection signal) and $\psi^{k\to i}(t)$ (the cavity probability
that $k$ is in state $E$ but has not transmitted the infection signal).

The message $\phi^{k\to i}(t)$ is computed as

\begin{align}
\phi^{k\to i}(t) 
= & \sum_{\tau_{k},\omega_{k},\varepsilon_{k}}\mathbb{I}(\tau_{k}<\omega_{k}<\varepsilon_{k})\prod_{t'=0}^{t-2}\bigg(1-\alpha_{ki}\mathbb{I}(\tau_{k}\leq t')\mathbb{I}(\omega_{k}\geq t'+1)-\beta_{ki}\mathbb{I}(\omega_{k}\leq t')\mathbb{I}(\varepsilon_{k}\geq t'+1)\bigg)\nonumber \\
& \times\bigg(1-\alpha_{ki}\mathbb{I}(\tau_{k}\leq t-1)\mathbb{I}(\omega_{k}\geq t)-\beta_{ki}\mathbb{I}(\omega_{k}\leq t-1)\mathbb{I}(\varepsilon_{k}\geq t)\bigg)\nonumber \\
& \times m^{k\to i}(\tau_{k},\omega_{k},\varepsilon_{k}|\cdot)\bigg[\mathbb{I}(\omega_{k}\leq t-1)\mathbb{I}(\varepsilon_{k}\geq t+1)+\mathbb{I}(\varepsilon_{k}\geq t+1)\delta_{\omega_{k},t}\bigg]\nonumber \\
= & \sum_{\tau_{k},\omega_{k},\varepsilon_{k}}\mathbb{I}(\tau_{k}<\omega_{k}<\varepsilon_{k})\prod_{t'=0}^{t-2}\bigg(1-\alpha_{ki}\mathbb{I}(\tau_{k}\leq t')\mathbb{I}(\omega_{k}\geq t'+1)-\beta_{ki}\mathbb{I}(\omega_{k}\leq t')\mathbb{I}(\varepsilon_{k}\geq t'+1)\bigg)\nonumber \\
& \times m^{k\to i}(\tau_{k},\omega_{k},\varepsilon_{k}|\cdot)\big(1-\beta_{ki}\big)\mathbb{I}(\omega_{k}\leq t-1)\mathbb{I}(\varepsilon_{k}\geq t+1)\nonumber \\
& +\sum_{\tau_{k},\varepsilon_{k}}\mathbb{I}(\tau_{k}<t<\varepsilon_{k})\prod_{t'=0}^{t-2}\bigg(1-\alpha_{ki}\mathbb{I}(\tau_{k}\leq t')\big)m^{k\to i}(\tau_{k},t,\varepsilon_{k}|\cdot)\big(1-\alpha_{ki}\mathbb{I}(\tau_{k}\leq t-1)\bigg)\nonumber \\
= & (1-\beta_{ki})(1-\mu_{k})\phi^{k\to i}(t-1)+\sum_{\tau_{k},\varepsilon_{k}}\mathbb{I}(\tau_{k}<t<\varepsilon_{k})\prod_{t'=0}^{t-1}\bigg(1-\alpha_{ki}\mathbb{I}(\tau_{k}\leq t')\bigg)m^{k\to i}(\tau_{k},t,\varepsilon_{k}|\cdot),\label{eq:phi_k_i_t_v1}
\end{align}
where the second term in the last line needs to be simplified.

To proceed further, we first compute the update rule of $\psi^{k\to i}(t)$
as 

\begin{align}
\psi^{k\to i}(t) & =\sum_{\tau_{k},\omega_{k},\varepsilon_{k}}\mathbb{I}(\tau_{k}<\omega_{k}<\varepsilon_{k})\prod_{t'=0}^{t-2}\bigg(1-\alpha_{ki}\mathbb{I}(\tau_{k}\leq t')\bigg)\nonumber \\
& \quad\times\bigg(1-\alpha_{ki}\mathbb{I}(\tau_{k}\leq t-1)\bigg)m^{k\to i}(\tau_{k},\omega_{k},\varepsilon_{k}|\cdot)\bigg[\mathbb{I}(\tau_{k}\leq t-1)\mathbb{I}(\omega_{k}\geq t+1)+\delta_{\tau_{k},t}\mathbb{I}(\omega_{k}\geq t+1)\bigg]\nonumber \\
& =(1-\alpha_{ki})\sum_{\tau_{k},\omega_{k},\varepsilon_{k}}\mathbb{I}(\tau_{k}<\omega_{k}<\varepsilon_{k})\prod_{t'=0}^{t-2}\bigg(1-\alpha_{ki}\mathbb{I}(\tau_{k}\leq t')\bigg)\nonumber \\
& \quad\times m^{k\to i}(\tau_{k},\omega_{k},\varepsilon_{k}|\cdot)\mathbb{I}(\tau_{k}\leq t-1)\mathbb{I}(\omega_{k}\geq t+1)+\sum_{\omega_{k},\varepsilon_{k}}\mathbb{I}(t<\omega_{k}<\varepsilon_{k})m^{k\to i}(t,\omega_{k},\varepsilon_{k}|\cdot)\nonumber \\
& \xRightarrow{\text{let }\omega_{k}=\omega_{k}'+1,\varepsilon_{k}=\varepsilon_{k}'+1}\nonumber \\
& =(1-\alpha_{ki})\sum_{\tau_{k},\omega_{k}',\varepsilon_{k}'}\mathbb{I}(\tau_{k}<\omega_{k}'<\varepsilon_{k}')\prod_{t'=0}^{t-2}\bigg(1-\alpha_{ki}\mathbb{I}(\tau_{k}\leq t')\bigg)\nonumber \\
& \quad\times(1-\nu_{k})m^{k\to i}(\tau_{k},\omega_{k},\varepsilon_{k}|\cdot)\mathbb{I}(\tau_{k}\leq t-1)\mathbb{I}(\omega_{k}'\geq t)+\sum_{\omega_{k},\varepsilon_{k}}\mathbb{I}(t<\omega_{k}<\varepsilon_{k})m^{k\to i}(t,\omega_{k},\varepsilon_{k}|\cdot)\nonumber \\
& =(1-\alpha_{ki})(1-\nu_{k})\psi^{k\to i}(t-1)+\sum_{\omega_{k},\varepsilon_{k}}\mathbb{I}(t<\omega_{k}<\varepsilon_{k})m^{k\to i}(t,\omega_{k},\varepsilon_{k}|\cdot)\nonumber \\
& =(1-\alpha_{ki})(1-\nu_{k})\psi^{k\to i}(t-1)-\big(P_{S}^{k\to i}(t)-P_{S}^{k\to i}(t-1)\big).\label{eq:psi_k_i_t_v1}
\end{align}
We observed that the message $\psi^{k\to i}(t)$ can also be expressed
in another form

\begin{align}
\psi^{k\to i}(t)= & \sum_{\tau_{k},\omega_{k},\varepsilon_{k}}\mathbb{I}(\tau_{k}<\omega_{k}<\varepsilon_{k})\prod_{t'=0}^{t-2}\bigg(1-\alpha_{ki}\mathbb{I}(\tau_{k}\leq t')\bigg)\nonumber \\
& \times\bigg(1-\alpha_{ki}\mathbb{I}(\tau_{k}\leq t-1)\bigg)m^{k\to i}(\tau_{k},\omega_{k},\varepsilon_{k}|\cdot)\nonumber \\
& \times\bigg[\mathbb{I}(\tau_{k}\leq t-1)\mathbb{I}(\omega_{k}\geq t)+\delta_{\tau_{k},t}\mathbb{I}(\omega_{k}\geq t)-\delta_{\omega_{k},t}\mathbb{I}(\tau_{k}\leq t-1)\bigg]\nonumber \\
= & \big(1-\alpha_{ki}\big)\psi^{k\to i}(t-1)+\sum_{\omega_{k},\varepsilon_{k}}\mathbb{I}(t<\omega_{k}<\varepsilon_{k})m^{k\to i}(t,\omega_{k},\varepsilon_{k}|\cdot)\nonumber \\
& -\sum_{\tau_{k},\varepsilon_{k}}\mathbb{I}(\tau_{k}<t<\varepsilon_{k})\prod_{t'=0}^{t-1}\bigg(1-\alpha_{ki}\mathbb{I}(\tau_{k}\leq t')\bigg)m^{k\to i}(\tau_{k},t,\varepsilon_{k}|\cdot)\nonumber \\
= & \big(1-\alpha_{ki}\big)\psi^{k\to i}(t-1)-\big(P_{S}^{k\to i}(t)-P_{S}^{k\to i}(t-1)\big)\nonumber \\
& -\sum_{\tau_{k},\varepsilon_{k}}\mathbb{I}(\tau_{k}<t<\varepsilon_{k})\prod_{t'=0}^{t-1}\bigg(1-\alpha_{ki}\mathbb{I}(\tau_{k}\leq t')\bigg)m^{k\to i}(\tau_{k},t,\varepsilon_{k}|\cdot),\label{eq:psi_k_i_t_v2}
\end{align}

Comparing Eq.~(\ref{eq:psi_k_i_t_v1}) and Eq.~(\ref{eq:psi_k_i_t_v2})
yields

\begin{align}
\sum_{\tau_{k},\varepsilon_{k}}\mathbb{I}(\tau_{k}<t<\varepsilon_{k})\prod_{t'=0}^{t-1}\big(1-\alpha_{ki}\mathbb{I}(\tau_{k}\leq t')\big)m^{k\to i}(\tau_{k},t,\varepsilon_{k}|\cdot) & =(1-\alpha_{ki})\nu_{k}\psi^{k\to i}(t-1),\label{eq:term_to_close_DMP_1}
\end{align}
Inserting Eq.~(\ref{eq:term_to_close_DMP_1}) into Eq.~(\ref{eq:phi_k_i_t_v1})
gives the update rule of $\phi^{k\to i}(t)$

\begin{equation}
\phi^{k\to i}(t)=(1-\beta_{ki})(1-\mu_{k})\phi^{k\to i}(t-1)+(1-\alpha_{ki})\nu_{k}\psi^{k\to i}(t-1),
\end{equation}
which closes the update rules of the messages $\theta^{i\to j},\phi^{i\to j},\psi^{i\to j}$
and $P_{S}^{i\to j}$. Collecting the incoming messages to node $i$
gives rise to the marginal probability of node $i$ being in state
$S$

\begin{equation}
P_{S}^{i}(t)=P_{S}^{i}(0)\prod_{k\in\partial i}\theta^{k\to i}(t).
\end{equation}

\subsection{Probabilities of Being in State $E,I,R$}

Finally, we consider the marginal probabilities of node $i$ in state
$E,I,R$, defined as 

\begin{align}
P_{E}^{i}(t) & =\sum_{\tau_{i},\omega_{i},\varepsilon_{i}}\mathbb{I}(\tau_{i}\leq t<\omega_{i}<\varepsilon_{i})m^{i}(\tau_{i},\omega_{i},\varepsilon_{i}),\\
P_{I}^{i}(t) & =\sum_{\tau_{i},\omega_{i},\varepsilon_{i}}\mathbb{I}(\tau_{i}<\omega_{i}\leq t<\varepsilon_{i})m^{i}(\tau_{i},\omega_{i},\varepsilon_{i}),\\
P_{R}^{i}(t) & =\sum_{\tau_{i},\omega_{i},\varepsilon_{i}}\mathbb{I}(\tau_{i}<\omega_{i}<\varepsilon_{i}\leq t)m^{i}(\tau_{i},\omega_{i},\varepsilon_{i}),
\end{align}
We first compute 
\begin{align}
P_{R}^{i}(t+1) & =\sum_{\tau_{i},\omega_{i},\varepsilon_{i}}\mathbb{I}(\tau_{i}<\omega_{i}<\varepsilon_{i}\leq t+1)m^{i}(\tau_{i},\omega_{i},\varepsilon_{i})\nonumber \\
 & =P_{R}^{i}(t)+\sum_{\tau_{i},\omega_{i}}\mathbb{I}(\tau_{i}<\omega_{i}\leq t)m^{i}(\tau_{i},\omega_{i},t+1),\label{eq:PR_v1}
\end{align}
and notice that

\begin{align}
(1-\mu_{i})P_{I}^{i}(t) & =(1-\mu_{i})\sum_{\tau_{i},\omega_{i},\varepsilon_{i}}\mathbb{I}(\tau_{i}<\omega_{i}\leq t<\varepsilon_{i})m^{i}(\tau_{i},\omega_{i},\varepsilon_{i})\nonumber \\
 & =\sum_{\tau_{i},\omega_{i},\varepsilon_{i}}\mathbb{I}(\tau_{i}<\omega_{i}\leq t<\varepsilon_{i})m^{i}(\tau_{i},\omega_{i},\varepsilon_{i}+1)\nonumber \\
 & \xRightarrow{\text{let }\varepsilon_{i}'=\varepsilon_{i}+1}\nonumber \\
 & =\sum_{\tau_{i},\omega_{i},\varepsilon_{i}'}\mathbb{I}(\tau_{i}<\omega_{i}\leq t<\varepsilon_{i}'-1)m^{i}(\tau_{i},\omega_{i},\varepsilon_{i}')\nonumber \\
 & =\sum_{\tau_{i},\omega_{i},\varepsilon_{i}'}\bigg[\mathbb{I}(\tau_{i}<\omega_{i}\leq t<\varepsilon_{i}')-\delta_{\varepsilon_{i}',t+1}\mathbb{I}(\tau_{i}<\omega_{i}\leq t)\bigg]m^{i}(\tau_{i},\omega_{i},\varepsilon_{i}')\nonumber \\
 & =P_{I}^{i}(t)-\sum_{\tau_{i},\omega_{i}}\mathbb{I}(\tau_{i}<\omega_{i}\leq t)m^{i}(\tau_{i},\omega_{i},t+1),\\
\mu_{i}P_{I}^{i}(t) & =\sum_{\tau_{i},\omega_{i}}\mathbb{I}(\tau_{i}<\omega_{i}\leq t)m^{i}(\tau_{i},\omega_{i},t+1),\label{eq:muPI_expression}
\end{align}
where we have made use of the property Eq.~(\ref{eq:m_property_1}).
Inserting Eq.~(\ref{eq:muPI_expression}) into Eq.~(\ref{eq:PR_v1})
gives rise to 
\begin{equation}
P_{R}^{i}(t+1)=P_{R}^{i}(t)+\mu_{i}P_{I}^{i}(t).
\end{equation}

Similarly, we compute 
\begin{align}
P_{I}^{i}(t+1) & =\sum_{\tau_{i},\omega_{i},\varepsilon_{i}}\mathbb{I}(\tau_{i}<\omega_{i}\leq t+1<\varepsilon_{i})m^{i}(\tau_{i},\omega_{i},\varepsilon_{i}),\nonumber \\
 & =(1-\mu_{i})P_{I}^{i}(t)+\sum_{\tau_{i},\varepsilon_{i}}\mathbb{I}(\tau_{i}\leq t)\mathbb{I}(\varepsilon_{i}>t+1)m^{i}(\tau_{i},t+1,\varepsilon_{i}),\label{eq:PI_v1}
\end{align}
and make use of the property Eq.~(\ref{eq:m_property_3}) to compute

\begin{align}
(1-\nu_{i})P_{E}^{i}(t) & =(1-\nu_{i})\sum_{\tau_{i},\omega_{i},\varepsilon_{i}}\mathbb{I}(\tau_{i}\leq t<\omega_{i}<\varepsilon_{i})m^{i}(\tau_{i},\omega_{i},\varepsilon_{i})\nonumber \\
 & =\sum_{\tau_{i},\omega_{i},\varepsilon_{i}}\mathbb{I}(\tau_{i}\leq t<\omega_{i}<\varepsilon_{i})m^{i}(\tau_{i},\omega_{i}+1,\varepsilon_{i}+1)\nonumber \\
 & \xRightarrow{\text{let }\omega_{i}'=\omega_{i}+1,\varepsilon_{i}'=\varepsilon_{i}+1}\nonumber \\
 & =\sum_{\tau_{i},\omega_{i}',\varepsilon_{i}'}\mathbb{I}(\tau_{i}\leq t<\omega_{i}'-1)\mathbb{I}(\omega_{i}'<\varepsilon_{i}')m^{i}(\tau_{i},\omega_{i}',\varepsilon_{i}')\nonumber \\
 & =\sum_{\tau_{i},\omega_{i}',\varepsilon_{i}}\bigg[\mathbb{I}(\tau_{i}\leq t<\omega_{i}')-\delta_{\omega_{i}',t+1}\mathbb{I}(\tau_{i}\leq t)\bigg]\mathbb{I}(\omega_{i}'<\varepsilon_{i}')m^{i}(\tau_{i},\omega_{i}',\varepsilon_{i}')\nonumber \\
 & =P_{E}^{i}(t)-\sum_{\tau_{i},\varepsilon_{i}'}\mathbb{I}(\tau_{i}\leq t)\mathbb{I}(\varepsilon_{i}>t+1)m^{i}(\tau_{i},t+1,\varepsilon_{i}'),\\
\nu_{i}P_{E}^{i}(t) & =\sum_{\tau_{i},\varepsilon_{i}}\mathbb{I}(\tau_{i}\leq t)\mathbb{I}(\varepsilon_{i}>t+1)m^{i}(\tau_{i},t+1,\varepsilon_{i}).\label{eq:nuPE_expression}
\end{align}
Inserting Eq.~(\ref{eq:nuPE_expression}) into Eq.~(\ref{eq:PI_v1})
gives rise to 
\begin{equation}
P_{I}^{i}(t+1)=(1-\mu_{i})P_{I}^{i}(t)+\nu_{i}P_{E}^{i}(t).
\end{equation}

Upon obtaining $P_{S}^{i}(t+1),P_{I}^{i}(t+1),P_{R}^{i}(t+1)$, the
probability $P_{E}^{i}(t+1)$ is given by the normalization condition
\begin{equation}
P_{E}^{i}(t+1)=1-P_{S}^{i}(t+1)-P_{I}^{i}(t+1)-P_{R}^{i}(t+1),
\end{equation}
which closes the DMP equations.

\section{Modeling Asymptomatic Transmission}
\label{app:AsympState}
As for the COVID-19 epidemic, there are patients who remain \emph{asymptomatic} prior to recovery. Here, we discuss how to accommodate potential disease transmissions due to having asymptomatic individuals in our framework. Patients who do not exhibit symptoms at the time of exposure but developed symptoms at a later time are termed \emph{presymptomatic} as described in the main text. There are many possible scenarios to model asymptomatic infections. One example is that each individual has a certain probability to become either asymptomatic or presymptomatic upon contracting the virus, in which case an additional asymptomatic state is required to model such stochastic transitions. The DMP equations can be derived accordingly, but will differ from those in this study.

Another perspective is based on the observation that whether an exposed individual will develop symptoms  or not seems to vary from person to person, based on age and pre-existing medical condition, e.g., children are more likely to have mild or no symptoms~\cite{Koh2020}. In light of this, one can assign each node $i$ a label $\ell_i$ (either probabilistically or according to additional existing information information), such that $\ell_i = 1$ if the individual is expected to develop symptoms when exposed and $\ell_i = 0$ otherwise. Nodes with different labels may have different epidemiological parameters. The disease transmission dynamics still obeys the transition rules of Sec.~\ref{sec:model}, and the DMP equations in Sec.~\ref{subsec:DMP} also apply. The only difference is that when a node $i$ is in state $I$, one should interpret it as infected with symptoms if $\ell_i=1$, or asymptomatic if $\ell_i=0$. In this way, our theoretical framework can readily accommodate asymptomatic transmission, except that a weighted-version of nonbacktracking is needed to accommodate different characteristics of symptomatic and asymptomatic individuals.

\section{Modeling Non-contagious presymptomatic Period in SCEIR Model}
\label{app:SCEIRmodel}
Another extension of the model is to take into account a possible non-contagious presymptomatic state when the viral load is too small to infect others. To accommodate this effect, we introduce an additional compartment called contracted ($C$) and define the corresponding SCEIR model. In this model, an individual in state $S$ (say node $i$) who contracts the virus will firstly assume state $C$ (non-contagious and non-symptomatic), after which that person will turn into state $E$ (contagious and non-symptomatic) with rate $\lambda_{i}$, and subsequently to states $I$ and $R$ with rates $\nu_{i}$ and $\mu_{i}$, respectively. Similar to the SEIR model, we also denote the transmission probabilities from node $j$ in state $E$ (state $I$) to a susceptible node $i$ as $\alpha_{ji}$ ($\beta_{ji}$).

Following a similar derivation from dynamic belief propagation as above, the DMP equations for the SCEIR model admit the following form
\begin{eqnarray}
	P_S^{i \to j}(t+1) & = & P_S^{i}(0) \sum_{k \in \partial i \backslash j} \theta^{k \to i}(t+1), \\
	\theta^{k \to i}(t+1) & = & \theta^{k \to i}(t) - \alpha_{ki} \psi^{k \to i}(t) - \beta_{ki} \phi^{k \to i}(t), \\
	\psi^{k \to i}(t+1) & = & (1-\alpha_{ki})(1-\nu_{k}) \psi^{k \to i}(t) + \lambda_{k} P_C^{k \to i}(t), \\
	\phi^{k \to i}(t+1) & = & (1-\beta_{ki})(1-\mu_{k}) \phi^{k \to i}(t) + (1-\alpha_{ki}) \nu_{k} \psi^{k \to i}(t), \\
	P_C^{k \to i}(t+1) & = & (1-\lambda_{k}) P_C^{k \to i}(t) - \big[ P_S^{k \to i}(t+1) - P_S^{k \to i}(t) \big],
\end{eqnarray}
where the messages $\{ P_S^{i \to j}, \theta^{k \to i}, \psi^{k \to i}, \phi^{k \to i} \}$ bear the same physical meanings as those in the SEIR model, while $P_C^{k \to i}(t)$ is the cavity probability that node $k$ is in state $C$ in the absence of node $i$. Upon obtaining these messages, the marginal probabilities of a node in each state can be computed as
\begin{eqnarray}
	P_S^{i}(t+1) & = & P_S^{i}(0) \sum_{k \in \partial i} \theta^{k \to i}(t+1), \\
	P_C^{i}(t+1) & = & (1-\lambda_{i}) P_C^{i}(t) - \big[ P_S^{i}(t+1) - P_S^{i}(t) \big], \\
	P_R^{i}(t+1) & = & P_R^{i}(t) + \mu_{i} P_I^{i}(t), \\
	P_I^{i}(t+1) & = & (1-\mu_{i}) P_I^{i}(t) + \nu_{i} P_E^{i}(t), \\
	P_E^{i}(t+1) & = & 1 - P_S^{i}(t+1) - P_C^{i}(t+1) - P_I^{i}(t+1) - P_R^{i}(t+1),
\end{eqnarray}
which starts from certain initial condition $P_{\sigma}^{i}(0), \sigma \in \{ S,C,E,I,R \}$.

\section{Contact Networks}
\label{app:contactnetworks}

\subsection{Realistic Networks}

The realistic contact networks are taken from data sets obtained from
the SocioPatterns collaboration website~\cite{sociopatterns}, where the
face-to-face contacts were recorded through wearable sensors over a
certain period. Each data set contains a lists of active contacts
between two individuals lasting for 20 seconds and the membership
information of each individual (belonging to a class or department).
To build the contact network, we first aggregate the contacts between
any two individuals $i$ and $j$, and consider the link between node
$i$ and node $j$ as active if the cumulative contact duration between them
in the recording period is not less than 60 seconds. A more precise
treatment is to retain the information of contact duration in the
transmission probabilities $\beta_{ji},\alpha_{ji}$. However, this
results in a graph with a weighted nonbacktracking (NB)
matrix, which complicates the analysis. For simplicity, we only preserve
the topological information of the resulting networks, keeping in
mind that the contact duration information can also be incorporated
in our framework. In this paper, we consider contact networks in a
primary school (PS2014)~\cite{Gemmetto2014}, a high school (HS2013)~\cite{Mastrandrea2015} and workplaces (WP2013, WP2015)~\cite{GNOIS2015,Genois2018}.

\subsection{Artificial Networks}

Artificially generated networks are also considered in this paper,
including Erd{\H o}s--R{\' e}nyi (ER) random graphs, random networks with
community structures, scale-free networks and random networks with
planted subgraph structures.

\subsubsection{Random Networks with Community Structure}

Random networks with community structures are generated through
the stochastic block model. It is specified by a list of population
of each block $(n_{1},n_{2},...,n_{r})$ and a $r\times r$ symmetric
probability matrix $P_{ab}$, where $r$ is the number of blocks.
Suppose node $i$ is assigned to block $a$ and node $j$ is assigned
to block $b$, then node $i$ and node $j$ are connected with probability
$P_{ab}$. The average number of neighbors of node $i$ (assigned
to block $a$) belonging to block $b$ is given by 
\begin{equation}
c_{ab}=\begin{cases}
P_{aa}(n_{a}-1) & \text{if }a=b,\\
P_{ab}n_{b} & \text{otherwise}.
\end{cases}
\end{equation}

\subsubsection{Scale-free Networks}

The scale-free networks are generated through the configuration model.
We first generate a degree sequence $(d_{1},d_{2},...,d_{N})$ of
size $N$ where each element follows the power law distribution
\begin{equation}
P(d_{i})=\frac{d_{i}^{-\gamma}}{\sum_{d_{j}=d^{\min}}^{d^{\max}}d_{j}^{-\gamma}},
\end{equation}
where $d^{\min}$ and $d^{\max}$ are the minimum and maximum of the
admissible degrees. In this study, we set $d^{\min}=5$ and $d^{\max}=100$,
considering the maximal number of people that one contacts can not
be arbitrarily large. After assigning a degree to each node, we randomly
connect different nodes such that each node $i$ has $d_{i}$ connections.
The resulting graph may have a few self-loops and multiple edges between
two nodes, which are simply removed to form a simple graph.

\subsubsection{Random Networks with Planted Subgraph Structures}

We also look at random networks with planted subgraph structures,
primarily for the purpose of examining the effect of localization.
Following~\cite{Martin2014}, we consider the ER random graph (with
average degree $\langle d\rangle$) with a planted hub (of degree
$d_{h}$), which is constructed by adding a hub to the existing ER
network through creating connections from the hub to $d_{h}$ randomly
selected nodes. When $d_{h}$ is large enough, i.e., $d_{h}>\langle d\rangle(\langle d\rangle+1)$,
it is argued that the eigenvector centrality (i.e., the leading eigenvector
of the adjacency matrix) is localized at the hub, while the nonbacktracking
centrality does not suffer from localization~\cite{Martin2014}.
The intuition behind the localization of eigenvector centrality is
that the hub and its neighbors are reinforcing each other, similar
to the mutual infection effect of the IBMF approach to epidemic spreading.
The NB centrality avoids this problem due to the hub by forbidding
one-step backtracking.

On the other hand, it is noticed in~\cite{Martin2014} that a relatively
large clique (i.e., a complete subgraph) can cause the NB centrality
to localize as well. Therefore, we consider the ER random graph (with
average degree $\langle d\rangle$) with a planted clique (of size
$N_{\text{clique}}$), which is constructed by randomly selecting
$N_{\text{clique}}$ nodes of the existing ER network to form a complete
subgraph. For the original ER graph, the leading eigenvalue of the
NB matrix $B$ is given by $\lambda_{B}^{\max}\approx\langle d\rangle$.
In the presence of the clique, the leading eigenvalue satisfies $\lambda_{B}^{\max}>N_{\text{clique}}-2$,
which can easily exceed $\langle d\rangle$ for large $N_{\text{clique}}$,
so that the NB centrality is dictated by the clique~\cite{Martin2014}.
There are also other subgraph structures that can cause the NB centrality
to localize, such as dense subgraphs and overlapping hubs~\cite{Satorras2020}.
The intuition behind the localization of the NB centrality is that
there is a subgraph sharing many neighbors, and avoiding one-step
backtracking is insufficient to counteract the self-reinforcement
among them.

\section{Additional Experiments}
\label{app:experiments}
\subsection{Approximation Accuracy of the Theory on Different Networks\label{subsec:approx_accuracy}}

In this Appendix, we give more examples of spreading
processes experiments to support the findings in the main text. In Fig.~\ref{fig:rrg_PEI_vs_t_vary_no_of_seeds},
it is shown that the approximation accuracy of the theories improves
as the number of initial exposed nodes increases; we expect that it
also depends on the locations of the initial seeds. 

\begin{figure}
	\includegraphics[scale=1.6]{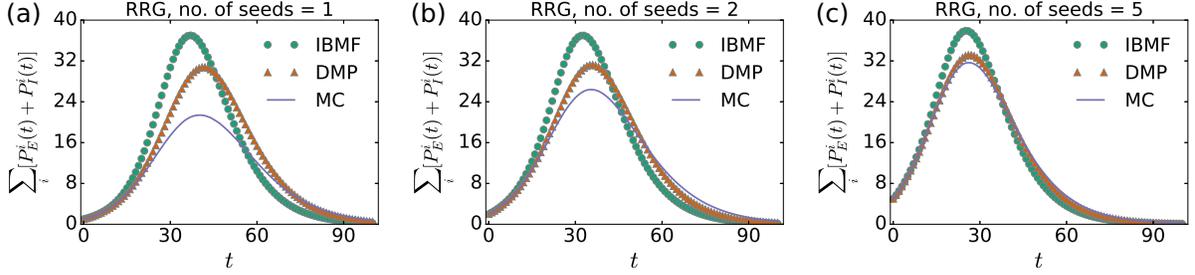}
	
	\caption{Evolution of the average number of individuals in state $E$ or $I$,
		i.e., $\sum_{i} [P_{E}^{i}(t)+P_{I}^{i}(t)]$ vs $t$. The underlying
		network is a random regular graph with $N=100,d=10$. The parameters
		are $T=100,\nu=0.2,\mu=0.125,\beta=0.03,\alpha=\beta/2$. The initial exposed nodes
		are randomly selected. (a) One initial exposed node. (b) Two initial exposed nodes. (a) Five initial exposed nodes. In general, when there are more initial seeds, the approximation accuracy of the theory becomes better. 
		\label{fig:rrg_PEI_vs_t_vary_no_of_seeds}}
\end{figure}

In Fig.~\ref{fig:ER_PEI_vs_t}, we examine the theoretical results  on ER random
networks with different average degrees and/or different planted subgraph
structures. We remark that there is no absolute fair comparison among
different networks, as they have different dependencies on the epidemiological
parameters (e.g., as in Fig.~\ref{fig:error_vs_no_init_seeds_and_beta} of the main text). Here, we fix the
values of $\nu,\mu$ and let $\alpha=\beta/2$, and choose $\beta$
such that approximately 70\% of the population have
contracted the disease in the final time $T=100$. Fig.~\ref{fig:ER_PEI_vs_t}
demonstrates that the IBMF approach becomes a better approximation
when the network is denser. This implies that the DMP approach is
superior to the IBMF approach, especially in sparsely connected
networks. When the average degree becomes higher, the trajectories
obtained by the IBMF equations are approaching those by the DMP equations.
In the limit of very dense random networks, the mass-action approximation
becomes a good approximation~\cite{Satorras2015}. As mentioned before,
a planted hub of relatively large degree can cause the leading eigenvector
of the adjacency matrix $A$ to localize~\cite{Martin2014}, which
impairs the accuracy of the IBMF approach. We show in Fig.~\ref{fig:ER_PEI_vs_t}(b)
that such a planted hub does not have a noticeable effect on the approximation
accuracy of the DMP approach. However, a planted clique which can
cause the nonbacktracking centrality to localize~\cite{Martin2014,Satorras2020},
does impair the approximation accuracy of the DMP approach, as shown in Fig.~\ref{fig:ER_PEI_vs_t}(c).

\begin{figure}
	\includegraphics[scale=1.6]{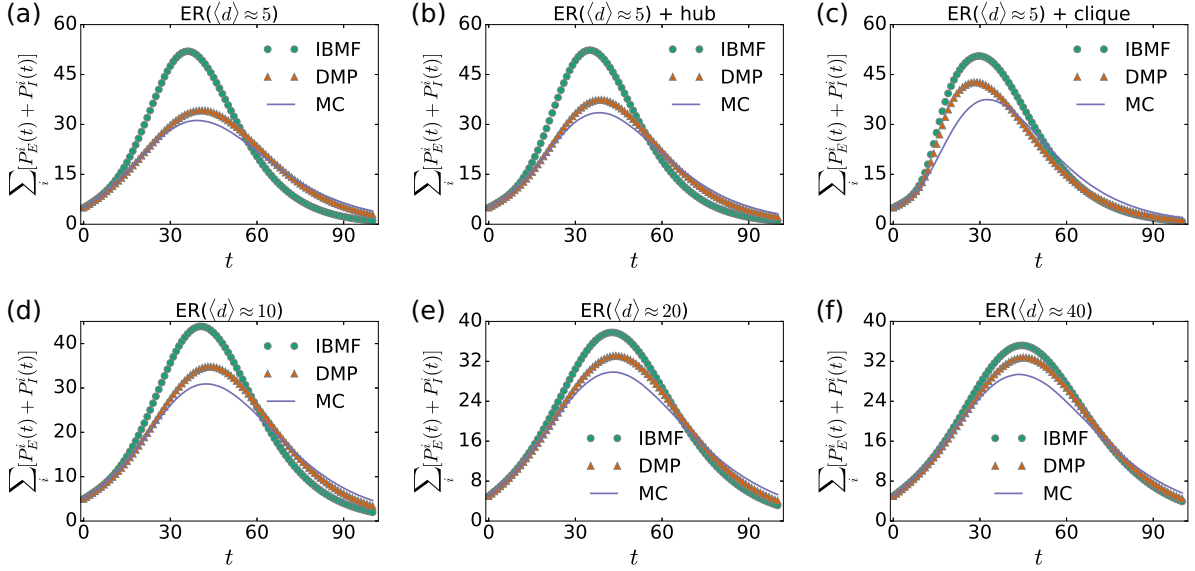}
	
	\caption{$\sum_{i}[P_{E}^{i}(t)+P_{I}^{i}(t)]$ vs $t$. The networks are ER
		random graphs of size $N=200$ with different average degrees and/or
		different planted subgraph structures. The systems start with 5 initial
		exposed nodes. The parameters are $T=100,\nu=0.2,\mu=0.125,\alpha=\beta/2$,
		while $\beta$ is selected such that in the final time $T=100$, approximately 70\% of the population has contracted the
		disease. Comparing panels (a),(d),(e),(f), it is observed that the IBMF
		approach becomes a better approximation when the network is denser
		(i.e., $\langle d\rangle$ is larger). In panel (b), a planted hub
		of degree $d_{h}=40$ is created in the networks; the DMP approach
		is still a rather good approximation in this case. In panel (c), a
		planted clique (or complete subgraph) of size $N_{\text{clique}}=20$
		is created in the networks; the approximation accuracy of the DMP
		approach is comparable poorer than other networks. \label{fig:ER_PEI_vs_t}}
\end{figure}

In Fig.~\ref{fig:ER_PEI_vs_t}, we examine the theories on the networks
extracted from contact data obtained in the SocioPatterns collaboration.
The approximation accuracy of the HS2013 network is much poorer than
the other networks, which may be attributed to the weakly localization
of the nonbacktracking centrality as shown in Fig.~\ref{fig:NB_centrality_on_SP_nets}.

\subsection{Evolution and Distribution of the Epidemic Outbreak}

In Fig.~\ref{fig:phase_transition_rrg_very_beta} of the main text, the normalized outbreak size at large
time $r=\sum_{i}P_{R}^{i}(\infty)/N$ was used to identify the critical
point $\beta_{c}$ of the SEIR model. In Fig.~\ref{fig:outbreak_evolution_and_distribution}(a),
we show the transient evolution of the normalized outbreak size, which
is defined by the fraction of nodes that has contracted the disease
as $1-\sum_{i}P_{S}^{i}(t)/N$. The normalized outbreak size increases
only gently near criticality, while it grows rapidly above the critical
point $\beta_{c}$.

As the epidemic outbreaks are triggered by a few initial seeds, there
is always a small probability that the disease will die out before
spreading out further (e.g., the infected node transforming into state
$R$ much faster than average, or the infection signal not being transmitted
to neighbors). This can also happen when $\beta>\beta_{c}$, indicating
that there are some instances where the outbreak size is small although
the system is in the global-epidemic phase, as shown in Fig.~\ref{fig:outbreak_evolution_and_distribution}(b).
This phenomenon has been observed in the SIR model~\cite{Shu2015,Koher2019}.
While the theoretical frameworks only concern the average behaviors,
they do not capture such variability of trajectories due to stochastic
fluctuations. A detailed theoretical investigation into these aspects
will be an interesting topic for future studies.

\begin{figure}
	\includegraphics[scale=1.2]{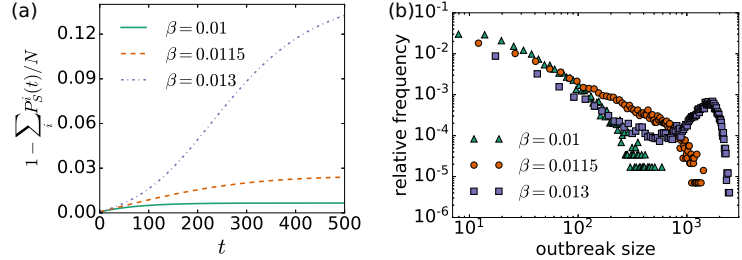}
	
	\caption{Evolution and distribution of the epidemic outbreak on a random regular
		graph with $N=6400,d=10$ in MC simulation. The systems start with
		5 initial exposed nodes. The parameters are $\nu=0.2,\mu=0.125,\alpha=\beta/2$.
		The critical point is given by $\beta_{c}\approx0.0115$. (a) Fraction
		of nodes that has contracted the disease, defined as $1-\sum_{i}P_{S}^{i}(t)/N$,
		as a function of time. (b) Distribution of outbreak size at large
		time, i.e., $\sum_{i}P_{R}^{i}(\infty)$. \label{fig:outbreak_evolution_and_distribution}}
\end{figure}

\subsection{Nonbacktracking Centrality}

As mentioned in the main text and in Sec.~\ref{subsec:approx_accuracy},
the degradation in accuracy of the approximation of the theory (IBMF or
DMP method) is correlated with the localization phenomenon of the
corresponding centrality measure, where the centrality values of a
few nodes are much larger than the others. We have showcased this
in Fig.~\ref{fig:ER_PEI_vs_t} by considering some planted subgraph
structures in a random network. We further examine the prediction
of the centrality measures on the outbreak profiles in these planted
random networks. In Fig.~\ref{fig:rho_vs_t_on_ER}(a), it is shown
that in the ER graph, both the eigenvector centrality and the NB centrality
(coming from the linear approximation of the IBMF and DMP approaches),
as well as the degree, are good predictors of the outbreak
profile, despite the poor approximation of the full nonlinear IBMF
approach in Fig.~\ref{fig:ER_PEI_vs_t}(a). On the other hand, a planted
hub causing the eigenvecter centrality to localized, degrades its prediction
accuracy as shown in Fig.~\ref{fig:rho_vs_t_on_ER}(b), while the
NB centrality appears to be a much better predictor. In the presence
of a clique, the NB centrality also predicts the outbreak poorly as
shown in Fig.~\ref{fig:rho_vs_t_on_ER}(c) due to the localization
phenomenon. 

In all three cases, the degree appears to be a good predictor
of the outbreak profile; in random networks, having more neighbors
usually implies a higher chance to contract the disease. However,
this does not hold in general, e.g., consider a hub connected to many
dangling nodes but linked to the bulk of the network through only
a few edges, such a hub node is not likely to be at high risk as indicated
by its degree. 

\begin{figure}
	\includegraphics[scale=1.6]{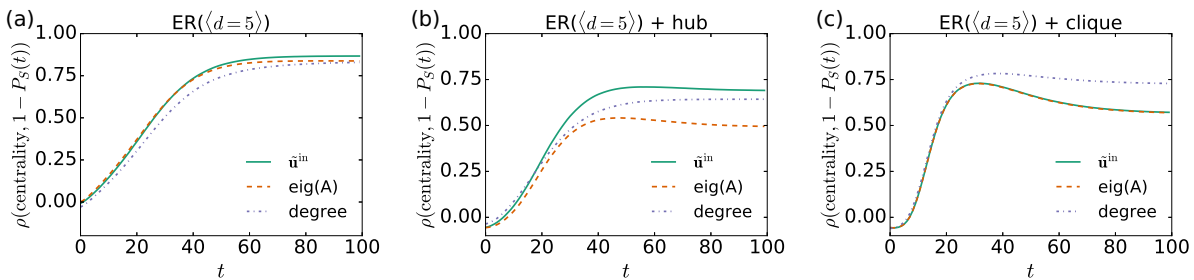}
	
	\caption{Correlation coefficient $\rho$ between various centrality measures
		and the outbreak profile $1-P_{S}(t)$, as a function of time. Here,
		eig($A$) stands for the eigenvector centrality and $\tilde{\boldsymbol{u}}^{\text{in}}$
		is the NB centrality. The networks and the parameter settings correspond
		to those in Figs.~\ref{fig:ER_PEI_vs_t}(a)(b)(c). \label{fig:rho_vs_t_on_ER}}
\end{figure}

It has been shown in Fig.~\ref{fig:SP_net_PEI_vs_t} that the approximation
accuracy of the HS2013 network is much poorer than the other networks
from the SocioPatterns data sets. In Fig.~\ref{fig:NB_centrality_on_SP_nets},
we show the NB centralities on the corresponding networks. Comparing
to other networks, the NB centrality $\tilde{\boldsymbol{u}}^{\text{in}}$
of HS2013 concentrates on two communities, which may cause
the approximation of the DMP equations to be less accurate.

\begin{figure}
	\includegraphics[scale=1.2]{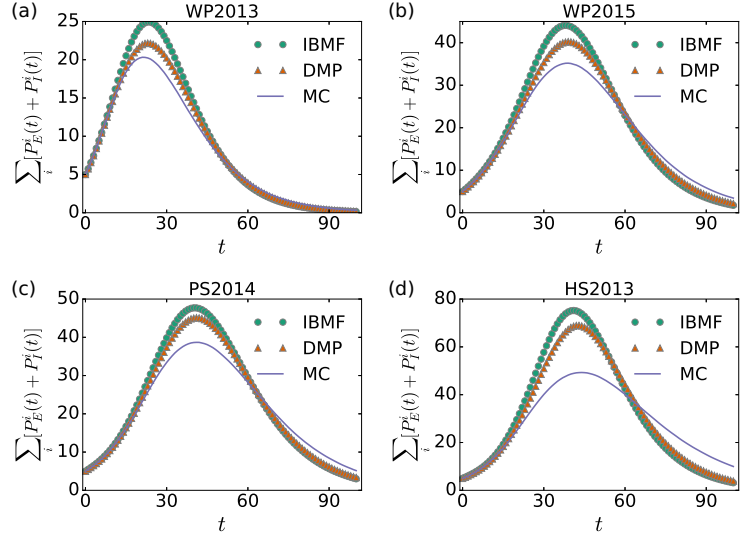}
	
	\caption{$\sum_{i}P_{E}^{i}(t)+P_{I}^{i}(t)$ vs $t$. The networks are extracted from contact data obtained in the SocioPatterns collaboration. The
		systems start with 5 initial exposed nodes. The parameters are $T=100,\nu=0.2,\mu=0.125,\alpha=\beta/2$,
		while $\beta$ is selected such that in the final time $T=100$, approximately 70\% of the population has contracted the
		disease. The approximation accuracy of the HS2013 network is much
		poorer than the other networks.\label{fig:SP_net_PEI_vs_t}}
\end{figure}

\begin{figure}
	\includegraphics[scale=1.5]{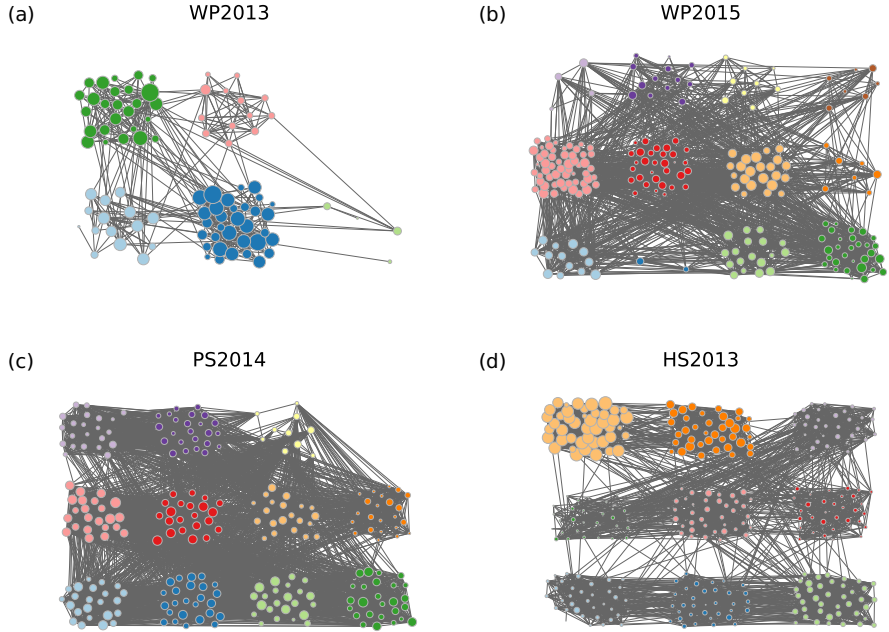}
	
	\caption{NB centrality $\tilde{\boldsymbol{u}}^{\text{in}}$ of contact networks
		from the SocioPatterns data. The marker size of node $i$ is proportional
		to the NB centrality value $\tilde{u}_{i}^{\text{in}}$. \label{fig:NB_centrality_on_SP_nets}}
\end{figure}

\section{Additional Details on The Derivation of Epidemic Threshold}
\label{app:EpidemicThreshold}

\subsection{Perron-Frobenius Theorem}

For a non-negative matrix $X$ which satisfies $X_{ij}\geq 0$,
the Perron-Frobenius (PF) theorem asserts that (i) the spectral
radius $\rho(X)$ is an eigenvalue of $X$, which implies that the
leading eigenvalue of $X$ (defined as the eigenvalue
having the largest real part) satisfies $\lambda_{X}^{\text{max}}=\rho(X)$, which is real and non-negative and (ii) there is a nonnegative and nonzero vector $\boldsymbol{u}$ (satisfying $u_{i}\geq0, \boldsymbol{u}\neq 0$)
such that $X\boldsymbol{u}=\rho(X)\boldsymbol{u}$; more properties of the leading eigenvalue and eigenvector can be deduced if the matrix $X$ is irreducible~\cite{Horn2012}.

\subsection{Leading Eigenvalue of the Jacobian of the DMP Equations}

In the main text, we have shown that the eigenvalue of the Jacobian
$\mathcal{J}$ and the eigenvalue of the nonbacktracking matrix has
the relation
\begin{equation}
\lambda_{B}=\frac{\lambda_{\mathcal{J}}-(1-\alpha)(1-\nu)}{\alpha+\beta\frac{(1-\alpha)\nu}{\lambda_{\mathcal{J}}-(1-\beta)(1-\mu)}},
\end{equation}
which can be solved for $\lambda_{\mathcal{J}}$, leading to
\begin{align}
\lambda_{\mathcal{J}}^{\pm}= & \frac{1}{2}\bigg[(1-\alpha)(1-\nu)+(1-\beta)(1-\mu)+\alpha\lambda_{B}\nonumber \\
& \pm\sqrt{\big[(1-\alpha)(1-\nu)-(1-\beta)(1-\mu)+\alpha\lambda_{B}\big]^{2}+4(1-\alpha)\nu\beta\lambda_{B}}\bigg].\label{eq:lamJ_of_lamB}
\end{align}
From this relation, one can figure out the principal eigenvalue of
$\mathcal{J}$. Noticing that $\lambda_{\mathcal{J}}^{\max}=\rho(\mathcal{J})\geq 0,\lambda_{B}^{\max}=\rho(B)\geq 0$
due to the Perron-Frobenius theorem, we can focus on $\lambda_{\mathcal{J}}^{+}$
and $\lambda_{B}\in\mathbb{R}$ since the eigenmode with a fastest growth rate will not
be realized by negative or complex eigenvalue of $\mathcal{J}$. We also assume that $\rho(\mathcal{J})$ and $\rho(B)$ are non-zero. To
simplify the notation, denote $a=(1-\alpha)(1-\nu),b=(1-\alpha)\nu,c=(1-\beta)(1-\mu)$,
all of which are nonnegative, such that $\lambda_{\mathcal{J}}^{+}(\lambda_{B})=\frac{1}{2}\bigg[a+c+\alpha\lambda_{B}+\sqrt{(a-c+\alpha\lambda_{B})^{2}+4b\beta\lambda_{B}}\bigg]$.
We first consider the case $\lambda_{B}>0$
\begin{align}
\frac{\mathrm{d}\lambda_{\mathcal{J}}^{+}}{\mathrm{d}\lambda_{B}} & =\frac{\alpha}{2}\bigg[1+\frac{(a-c+\alpha\lambda_{B})+\frac{2b\beta}{\alpha}}{\sqrt{(a-c+\alpha\lambda_{B})^{2}+4b\beta\lambda_{B}}}\bigg]\nonumber \\
& >\frac{\alpha}{2}\bigg[1+\frac{(a-c+\alpha\lambda_{B})}{\sqrt{(a-c+\alpha\lambda_{B})^{2}+4b\beta\lambda_{B}}}\bigg]\nonumber \\
& >0,
\end{align}
where we have made use of the fact that $\frac{2b\beta}{\alpha}>0$
and $\left|\frac{(a-c+\alpha\lambda_{B})}{\sqrt{(a-c+\alpha\lambda_{B})^{2}+4b\beta\lambda_{B}}}\right|<1$
by assuming $\lambda_{B}>0$. It implies that $\max_{\lambda_{B}>0}\lambda_{\mathcal{J}}^{+}(\lambda_{B})=\lambda_{\mathcal{J}}^{+}(\lambda_{B}^{\text{max}})$.
Furthermore, it can be easily shown that $\lambda_{\mathcal{J}}^{+}(x)\geq\lambda_{\mathcal{J}}^{+}(-x)$
for $x\geq0$, which leads to the fact that $\forall\lambda_{B}<0,\lambda_{\mathcal{J}}^{+}(\lambda_{B})<\lambda_{\mathcal{J}}^{+}(|\lambda_{B}|)<\lambda_{\mathcal{J}}^{+}(\lambda_{B}^{\text{max}})$.
Hence, we can conclude that the maximal eigenvalue of $\mathcal{J}$
is given by $\lambda_{\mathcal{J}}^{\text{max}}=\lambda_{\mathcal{J}}^{+}(\lambda_{B}^{\text{max}})$,
i.e.,

\begin{align}
\lambda_{\mathcal{J}}^{\max}= & \frac{1}{2}\big[(1-\alpha)(1-\nu)+(1-\beta)(1-\mu)+\alpha\lambda_{B}^{\max}\big]\nonumber \\
& +\frac{1}{2}\sqrt{\big((1-\alpha)(1-\nu)-(1-\beta)(1-\mu)+\alpha\lambda_{B}^{\max}\big)^{2}+4(1-\alpha)\nu\beta\lambda_{B}^{\max}}.
\end{align}
The epidemic threshold is obtained by solving $\lambda_{\mathcal{J}}^{\max}(\beta,\alpha,\nu,\mu,\lambda_{B}^{\max})=1$. The resulting phase boundaries are in general nonlinear, which is more prominent in networks with sparse structures (especially relevant when contact and travel restrictions are enforced) and for disease with a long incubation period. We illustrate this in an example in Fig.~\ref{fig:nonlinear_phase_boundary}.

\begin{figure}
	\includegraphics[scale=0.4]{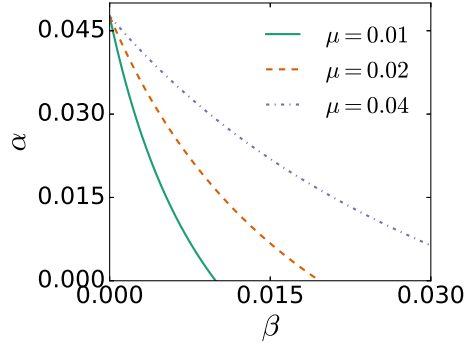}	
	\caption{Phase boundary of the SEIR model in the $(\beta,\alpha)$-plane. The parameters used are $\nu=0.05, \lambda_{B}^{\max}=2$. \label{fig:nonlinear_phase_boundary}}
\end{figure}

\subsection{Epidemic Threshold by the IBMF Approach}

Similar to the DMP approach, we can also derive the epidemic thresholds
through the individual-based mean-field (IBMF) approach. The initial
disease-free state is perturbed infinitesimally as $P_{S}^{i}(0)=1-\epsilon^{i}$,
in which case the probabilities $P_{E}^{i}(t)$ and $P_{I}^{i}(t)$
are also of order $\epsilon^{i}$ in the initial stage. We expand
Eq.~(\ref{eq:Emarginal}) and neglect terms of higher order of
$\epsilon^{i}$, leading to
\begin{equation}
P_{E}^{i}(t+1)\approx(1-\nu_{i})P_{E}^{i}(t)+P_{S}^{i}(t)\sum_{k\in\partial i}\big[\alpha_{ki}P_{E}^{k}(t)+\beta_{ki}P_{I}^{k}(t)\big].\label{eq:PEi_IBMF_linear}
\end{equation}
Equations~(\ref{eq:PEi_IBMF_linear}) and~(\ref{eq:Imarginal}) constitute a linear dynamical system of the probabilities $\{P_{E}^{i}(t),P_{I}^{i}(t)\}$.
They can be written in the matrix form as

\begin{equation}
\left(\begin{array}{c}
P_{E}(t+1)\\
P_{I}(t+1)
\end{array}\right)=\mathcal{J}^{\text{MF}}\left(\begin{array}{c}
P_{E}(t)\\
P_{I}(t)
\end{array}\right),\label{eq:IBMF_linear_dyn_matrix_form}
\end{equation}
where the $2N\times2N$ Jacobian matrix $\mathcal{J}^{\text{MF}}$
is defined as

\begin{equation}
\mathcal{J}^{\text{MF}}=\left(\begin{array}{cc}
(1-\nu)I+\alpha A & \beta A\\
\nu I & (1-\mu)I
\end{array}\right),
\end{equation}
where $I$ is the identity matrix and $A$ is the adjacency matrix
of the graph. The spectral radius $\rho(\mathcal{J}^{\text{MF}})$
determines the growth rate of the fastest mode of Eq.~(\ref{eq:IBMF_linear_dyn_matrix_form}).
Due to the Perron-Frobenius theorem, $\rho(\mathcal{J}^{\text{MF}})$
equals the leading eigenvalue $\lambda_{\mathcal{J}^{\text{MF}}}^{\max}$
of $\mathcal{J}^{\text{MF}}$, which is related to the leading eigenvalue
$\lambda_{A}^{\max}$ of the adjacency matrix of the graph. A similar
argument as in the DMP approach results in

\begin{align}
\lambda_{\mathcal{J}^{\text{MF}}}^{\max}= & \frac{1}{2}\big[(1-\nu)+(1-\mu)+\alpha\lambda_{A}^{\max}\big]\nonumber \\
& +\frac{1}{2}\sqrt{\big((1-\nu)-(1-\mu)+\alpha\lambda_{A}^{\max}\big)^{2}+4\nu\beta\lambda_{A}^{\max}}.
\end{align}
The epidemic threshold is obtained by solving $\lambda_{\mathcal{J}^{\text{MF}}}^{\max}(\beta,\alpha,\nu,\mu,\lambda_{A}^{\max})=1$.

\section{Basic Reproduction Number of The SEIR Model}
\label{app:ReproductionNo}
In this appendix, we provide a simple estimation of the basic reproduction
number $R_{0}$ for the SEIR model under investigation, which is defined
as the expected number of secondary infections from a single infection
in a population where all subjects are susceptible. We assume at time
$t=0$, a node randomly chosen from the network is exposed to the
virus. The exposed node has $\langle d\rangle$ susceptible neighbors
on average; the average number of neighbors it infects is computed
as
\begin{equation}
R_{0}=\langle d\rangle\sum_{t=1}^{\infty}\bigg[(1-\nu)^{t-1}\alpha+\sum_{\omega=2}^{t}(1-\nu)^{\omega-2}\nu(1-\mu)^{t-\omega}\beta\bigg],
\end{equation}
where $\omega$ is the time that the initial exposed node turns into
the infectious state and we have assumed that at the same step, the
process $E\to I$ or $I\to R$ occurs after the infection being transmitted.
The above expression can be simplified as 
\begin{align}
R_{0}= & \langle d\rangle\sum_{t=1}^{\infty}\bigg\{(1-\nu)^{t-1}\alpha+\frac{(1-\mu)^{t-2}\bigg[1-\big(\frac{1-\nu}{1-\mu}\big)^{t-1}\bigg]}{1-\frac{1-\nu}{1-\mu}}\nu\beta\bigg\}\nonumber \\
= & \langle d\rangle\sum_{t=1}^{\infty}\bigg\{(1-\nu)^{t-1}\alpha+\frac{(1-\mu)^{t-1}-(1-\nu)^{t-1}}{\nu-\mu}\nu\beta\bigg\}\nonumber \\
= & \langle d\rangle\bigg[\frac{\alpha}{\nu}+\frac{\nu\beta}{\nu-\mu}\left(\frac{1}{\mu}-\frac{1}{\nu}\right)\bigg]\nonumber \\
= & \langle d\rangle\left(\frac{\alpha}{\nu}+\frac{\beta}{\mu}\right).
\end{align}
The $R_{0}$ defined in this way only captures the average number
of contacts through $\langle d\rangle$, but neglects higher order
structures of the contact networks which could be very heterogeneous.

\section{The Leading Eigenvalue and Eigenvector of Matrix $B$}
\label{app:Eigenvalue}

\subsection{Reduction from Matrix $B$ to Matrix $M$}

In the main text, we claimed that the spectrum of the $2|\mathcal{E}|\times2|\mathcal{E}|$
nonbacktracking matrix can be obtained from a much smaller $2N\times2N$
matrix $M$ defined as

\begin{equation}
M=\begin{pmatrix}0 & D-I_{N}\\
-I_{N} & A
\end{pmatrix},\label{eq:M_matrix_def_sm}
\end{equation}
where $I_{N}$ is the $N$-dimensional identity matrix. The reduction
to Eq.~(\ref{eq:M_matrix_def_sm}) is a manifest of the Ihara-Bass
formula~\cite{Bass1992}, where an intuitive derivation of the reduction
can be found in Ref.~\cite{Krzakala2013}. The Ihara-Bass formula
has also been generalized to weight graphs and linked to Bethe free
energy in belief propagation~\cite{Watanabe2009}, which is relevant
in our study when heterogeneous transmission probabilities $\{\beta_{ij},\alpha_{ij}\}$
are considered. 

For completeness, we provide a derivation using the matrix notation
in this section. To this end, we define the $N\times2|\mathcal{E}|$
source matrix $\mathcal{S}$ and target matrix $\mathcal{T}$
\begin{equation}
\mathcal{S}_{ie}=\begin{cases}
1 & \text{if node \ensuremath{i} is the source of directed edge \ensuremath{e}, i.e., \ensuremath{\exists j,} \ensuremath{e=i\to j}},\\
0 & \text{\text{otherwise},}
\end{cases}
\end{equation}

\begin{equation}
\mathcal{T}_{ie}=\begin{cases}
1 & \text{if node \ensuremath{i} is the target of directed edge \ensuremath{e}, i.e., \ensuremath{\exists j,} \ensuremath{e=j\to i}},\\
0 & \text{\text{otherwise},}
\end{cases}
\end{equation}
and a $2|\mathcal{E}|\times2|\mathcal{E}|$ auxiliary matrix

\begin{equation}
J=\left(\begin{array}{cc}
0 & I_{2|\mathcal{E}|}\\
I_{2|\mathcal{E}|} & 0
\end{array}\right),
\end{equation}
where $I_{2|\mathcal{E}|}$ is the $2|\mathcal{E}|$-dimensional identity
matrix. We further index the set of directed edges according to the
order $(e_{1},...e_{|\mathcal{E}|},e_{|\mathcal{E}|+1},...e_{2|\mathcal{E}|})$
such that for $n\leq|\mathcal{E}|,e_{n}=i\to j$, one has $e_{n+|\mathcal{E}|}=j\to i$.
Under this notation, the nonbacktracking matrix $B$ can be written
as
\begin{equation}
B=\mathcal{S}^{\top}\mathcal{T}-J.
\end{equation}
It can also be easily verified that 
\begin{align}
\mathcal{S}\mathcal{T}^{\top} & =\mathcal{T}\mathcal{S}^{\top}=A,\qquad\mathcal{S}\mathcal{S}^{\top}=\mathcal{T}\mathcal{T}^{\top}=D,\label{eq:relation_ST_matrices}\\
\mathcal{S}J & =\mathcal{T},\qquad\mathcal{T}J=\mathcal{S},\label{eq:relation_ST_and_J_matrices}
\end{align}

The key element of the derivation in Ref.~\cite{Krzakala2013} is
that for a given $2|\mathcal{E}|$-dimensional $\boldsymbol{u}$,
one defines the corresponding $N$-dimensional incoming and outgoing
vectors

\begin{equation}
u_{i}^{\text{in}}=\sum_{j\in\partial i}u_{j\to i},\qquad u_{i}^{\text{out}}=\sum_{j\in\partial i}u_{i\to j},
\end{equation}
which can be expressed in the matrix form as
\begin{equation}
\boldsymbol{u}^{\text{in}}=\mathcal{T}\boldsymbol{u},\qquad\boldsymbol{u}^{\text{out}}=\mathcal{S}\boldsymbol{u}.
\end{equation}
Consider the vectors $(B\boldsymbol{u})^{\text{out}}$ and $(B\boldsymbol{u})^{\text{in}}$,
written in the matrix form as 

\begin{align}
\mathcal{S}B\boldsymbol{u} & =\mathcal{S}(\mathcal{S}^{\top}\mathcal{T}-J)\boldsymbol{u}=\mathcal{S}\mathcal{S}^{\top}\mathcal{T}\boldsymbol{u}-\mathcal{S}J\boldsymbol{u}\nonumber \\
& =D\mathcal{T}\boldsymbol{u}-\mathcal{T}\boldsymbol{u}=(D-I_{N})(\mathcal{T}\boldsymbol{u}),\\
\mathcal{T}B\boldsymbol{u} & =\mathcal{T}(\mathcal{S}^{\top}\mathcal{T}-J)\boldsymbol{u}=\mathcal{T}\mathcal{S}^{\top}\mathcal{T}\boldsymbol{u}-\mathcal{T}J\boldsymbol{u}\nonumber \\
& =A(\mathcal{T}\boldsymbol{u})-(\mathcal{S}\boldsymbol{u}),
\end{align}
where we have made used of the relations in Equations~(\ref{eq:relation_ST_matrices})
and (\ref{eq:relation_ST_and_J_matrices}). The above expressions
can be written as
\begin{equation}
\begin{pmatrix}\mathcal{S} & 0\\
0 & \mathcal{T}
\end{pmatrix}B\boldsymbol{u}=\begin{pmatrix}0 & D-I_{N}\\
-I_{N} & A
\end{pmatrix}\left(\begin{array}{c}
\mathcal{S}\boldsymbol{u}\\
\mathcal{T}\boldsymbol{u}
\end{array}\right)=:M\left(\begin{array}{c}
\mathcal{S}\boldsymbol{u}\\
\mathcal{T}\boldsymbol{u}
\end{array}\right),\label{eq:M_matrix_derived}
\end{equation}
where we identify the matrix $M$ defined in Eq.~(\ref{eq:M_matrix_def_sm}).
Now suppose $\boldsymbol{u}$ is an eigenvector of $B$ with eigenvalue
$\lambda_{B}$, then Eq.~(\ref{eq:M_matrix_derived}) leads to
\begin{equation}
M\left(\begin{array}{c}
\mathcal{S}\boldsymbol{u}\\
\mathcal{T}\boldsymbol{u}
\end{array}\right)=\lambda_{B}\left(\begin{array}{c}
\mathcal{S}\boldsymbol{u}\\
\mathcal{T}\boldsymbol{u}
\end{array}\right).
\end{equation}
which implies that $\left(\begin{array}{c}
\mathcal{S}\boldsymbol{u}\\
\mathcal{T}\boldsymbol{u}
\end{array}\right)$ or $\left(\begin{array}{c}
\boldsymbol{u}^{\text{out}}\\
\boldsymbol{u}^{\text{in}}
\end{array}\right)$ is the eigenvector of $M$ with eigenvalue $\lambda_{B}$. Therefore,
we can work with the much smaller matrix $M$ for computing the spectrum
of the nonbacktracking matrix $B$. 

Physically, this reduction comes from compressing the edge-based data
$\{u_{i\to j}\}$ to node-based data $\{u_{i}^{\text{out}},u_{i}^{\text{in}}\}$.
In the context of spreading processes, they correspond to edge-based
messages and node-based marginal probabilities, e.g., the cavity probability
$P_{S}^{i\to j}$ in the linearized dynamics satisfies
\begin{equation}
P_{S}^{i\to j}(t+1)\approx P_{S}^{i\to j}(t)-\sum_{k\in\partial i\backslash j}\big[\alpha\psi^{k\to i}(t)+\beta\phi^{k\to i}(t)\big],
\end{equation}
while the marginal probability satisfies

\begin{equation}
P_{S}^{i}(t+1)\approx P_{S}^{i}(t)-\sum_{k\in\partial i}\big[\alpha\psi^{k\to i}(t)+\beta\phi^{k\to i}(t)\big],
\end{equation}
where the probability of newly infection $\sum_{k\in\partial i}\big[\alpha\psi^{k\to i}(t)+\beta\phi^{k\to i}(t)\big]$
corresponds to the incoming vectors of the messages as $\alpha\boldsymbol{\psi}^{\text{in}}(t)+\beta\boldsymbol{\phi}^{\text{in}}(t)$.

\subsection{Eigenvalues and Eigenvectors of Matrix $M$}

Consider the eigenvalue equation of $M$

\begin{equation}
\begin{pmatrix}0 & D-I_{N}\\
-I_{N} & A
\end{pmatrix}\left(\begin{array}{c}
\boldsymbol{u}^{\text{out}}\\
\boldsymbol{u}^{\text{in}}
\end{array}\right)=\lambda_{M}\left(\begin{array}{c}
\boldsymbol{u}^{\text{out}}\\
\boldsymbol{u}^{\text{in}}
\end{array}\right),
\end{equation}
or explicitly

\begin{equation}
\begin{cases}
(D-I_{N})\boldsymbol{u}^{\text{in}}=\lambda_{M}\boldsymbol{u}^{\text{out}},\\
-\boldsymbol{u}^{\text{out}}+A\boldsymbol{u}^{\text{in}}=\lambda_{M}\boldsymbol{u}^{\text{in}}.
\end{cases}
\end{equation}
The above equations can be reduced to a nonlinear eigenvalue problem
\begin{equation}
\big[\lambda_{M}^{2}I-\lambda_{M}A+(D-I)\big]\boldsymbol{u}^{\text{in}}=0,\label{eq:IB_formula_nonlinear}
\end{equation}
and implies a relation between the outgoing-component $\boldsymbol{u}^{\text{out}}$
and the incoming-component $\boldsymbol{u}^{\text{in}}$ of the eigenmode
as

\begin{equation}
u_{i}^{\text{out}}=\frac{d_{i}-1}{\lambda_{M}}u_{i}^{\text{in}}.
\end{equation}

In the main text, it is argued that the leading eigenvector of the
matrix $B$ (denoted as $\tilde{\boldsymbol{u}}$), and the corresponding
incoming vector $\tilde{\boldsymbol{u}}^{\text{in}}=\mathcal{T}\tilde{\boldsymbol{u}}$
(i.e., the nonbacktracking centrality), are useful in predicting
the outcome of the epidemics. From the analysis in this section, both
the leading eigenvalue $\lambda_{B}^{\max}$ and the nonbacktracking
centrality $\tilde{\boldsymbol{u}}^{\text{in}}$ can be obtained through
the matrix $M$, which significantly reduces the computational complexity.
The Perron-Frobenius theorem guarantees that the leading eigenvector
$\tilde{\boldsymbol{u}}$ of the matrix $B$ can be chosen to be non-negative,
so does the nonbacktracking centrality $\tilde{\boldsymbol{u}}^{\text{in}}$.

\subsection{Exact Expression in Random Regular Graphs}

The leading eigenvalue of the matrix $M$ can be computed exactly
for random regular graphs. We first notice that the smallest eigenvalue
of the Laplacian matrix $L=D-A$ of a graph is $\lambda_{L}^{\text{min}}=0$~\cite{Godsil2001}. For regular graphs with degree $d$, we have
$D=dI$, and therefore $L$ commutes with $A$. It suggests that the
largest eigenvalue of $A$ is 
\begin{equation}
\lambda_{A}^{\max}=d.
\end{equation} 

Equation~(\ref{eq:IB_formula_nonlinear}) can be expressed as
\begin{equation}
A\boldsymbol{u}^{\text{in}}=\bigg(\lambda_{M}+\frac{d-1}{\lambda_{M}}\bigg)\boldsymbol{u}^{\text{in}},
\end{equation}
which implies that $\bigg(\lambda_{M}+\frac{d-1}{\lambda_{M}}\bigg)$
is an eigenvalue of the adjacency matrix $A$ with eigenvalue $\boldsymbol{u}^{\text{in}}$,
denoted as $\lambda_{A}$. The eigenvalue $\lambda_{M}$ of $M$ is
related to $\lambda_{A}$ as
\begin{equation}
\lambda_{M}=\frac{1}{2}\bigg[\lambda_{A}\pm\sqrt{\lambda_{A}^{2}-4(d-1)}\bigg].
\end{equation}
Therefore, the leading eigenvalue $\lambda_{M}^{\max}$ of $M$, which
is equal to the leading eigenvalue $\lambda_{B}^{\max}$, is identified
as
\begin{align}
\lambda_{M}^{\text{max}}=\lambda_{B}^{\max} & =\frac{1}{2}\bigg[\lambda_{A}^{\text{max}}+\sqrt{\big(\lambda_{A}^{\text{max}}\big)^{2}-4(d-1)}\bigg]\nonumber \\
& =d-1.
\end{align}

\subsection{Approximations of the $\lambda_{B}^{\max}$ in Uncorrelated Random Networks}
\label{app:approximation_accuracy_lambdaB}

For uncorrelated random networks, approximate expressions of the leading
eigenvalue $\lambda_{B}^{\max}$ can be derived, which can be useful
for estimating epidemics properties in large networks. In the annealed approximation
where only the information of the degree distribution $P(d)$ is retained,
$\lambda_{B}^{\max}$ is approximated as~\cite{Krzakala2013,Martin2014}
\begin{equation}
\lambda_{B}^{\max,\text{an}}\approx\frac{\langle d^{2}\rangle-\langle d\rangle}{\langle d\rangle}.\label{eq:lamB_approx_an}
\end{equation}
In a more refined approximation assuming uncorrelated networks but
taking into neighborhood information, $\lambda_{B}^{\max}$ is approximated
as~\cite{Satorras2020}
\begin{equation}
\lambda_{B}^{\max,\text{un}}\approx\frac{\sum_{ij}(d_{i}-1)A_{ij}(d_{j}-1)}{\sum_{i}d_{i}(d_{i}-1)}.\label{eq:lamB_approx_un}
\end{equation}

To examine the validity of the uncorrelated random network assumption,
we consider the degree correlation coefficient~\cite{Barabasi2016}
\begin{equation}
r_{d}=\frac{1}{\sigma^{2}}\sum_{k,l=1}^{d^{\max}}kl(e_{kl}-q_{k}q_{l}),
\end{equation}
where $e_{kl}$ is the element of the degree correlation matrix (i.e.,
the probability of finding an edge connecting two nodes of degree
$k$ and degree $l$), $q_{k}=\frac{kP(k)}{\langle d\rangle}$ (with
$P(k)$ being the degree distribution) is the probability of finding
an edge whose one end has degree $k$, and $\sigma^{2}=\sum_{k=1}^{d^{\max}}k^{2}q_{k}-\big(\sum_{k=1}^{d^{\max}}k^{2}q_{k}\big)^{2}$
is the variance of the measure $q$. The neutrality condition $r_{d}=0$
needs to be (at least roughly) satisfied for a network to be uncorrelated. 

In Table.~\ref{table:lamB_approx}, we examine the approximations offered
by Eqs~(\ref{eq:lamB_approx_an}) and (\ref{eq:lamB_approx_un})
for the networks considered in this paper. In general, $\lambda_{B}^{\max,\text{un}}$
provides a better approximation than $\lambda_{B}^{\max,\text{an}}$,
both of which predict $\lambda_{B}^{\max}$ quite well when $r_{d}$
is low. The two cases with a relatively poor approximation (especially
for $\lambda_{B}^{\max,\text{an}}$) are the SBM network with $c_{11}=30$
and the ER network with a clique. These two networks exhibit high
values of the degree correlation coefficient $r_{d}$, violating the
assumption of uncorrelated random network. On the other hand, the
existence of dense subgraph structures can cause localization of the
NB centrality, which also makes the approximation inaccurate. In light
of this, it has been proposed in Ref.~\cite{Satorras2020} to identify
some characteristic subgraph structures for an improved approximation
of $\lambda_{B}^{\max}$. 

\begin{table}[] 
	\setlength{\tabcolsep}{12pt}
	\begin{tabular}{ccccccc} \hline network                                    & $N$ & $\langle d \rangle$ & $\lambda_B^{\max}$ & $\lambda_B^{\max, \text{an}}$ & $\lambda_B^{\max, \text{un}}$ & $r_d$   \\ \hline ER($\langle d \rangle \approx 20$)         & 200 & 20.09               & 20                 & 20.02                         & 20                            & -0.0182 \\ \hline SBM, $c_{11}=10$                           & 200 & 21.94               & 21.79              & 21.8                          & 21.8                          & -0.0041 \\ \hline SBM, $c_{11}=20$                           & 200 & 24.64               & 26.19              & 25.38                         & 25.8                          & 0.2419  \\ \hline SBM, $c_{11}=30$                           & 200 & 27.15               & 33.26              & 29.9                          & 31.59                         & 0.4274  \\ \hline WP2013                                     & 91  & 8.55                & 10.17              & 10.26                         & 10.08                         & -0.0525 \\ \hline WP2015                                     & 213 & 21.22               & 24.49              & 24.2                          & 24.34                         & 0.0432  \\ \hline PS2014                                     & 242 & 36.65               & 42.18              & 40.64                         & 41.57                         & 0.1993  \\ \hline HS2013                                     & 326 & 20.47               & 25.19              & 23.38                         & 23.64                         & 0.0774  \\ \hline SF, $\gamma=2$                             & 400 & 13.29               & 24.43              & 25.9                          & 24.48                         & -0.0773 \\ \hline SF, $\gamma=2.5$                           & 400 & 10.18               & 16.75              & 17.87                         & 16.83                         & -0.06   \\ \hline SF, $\gamma=3$                             & 400 & 8.4                 & 11.27              & 12.14                         & 11.36                         & -0.0653 \\ \hline SF, $\gamma=3.5$                           & 400 & 7.48                & 8.55               & 8.88                          & 8.67                          & -0.0347 \\ \hline ER($\langle d \rangle \approx 5$)          & 200 & 5.2                 & 5.02               & 5.1                           & 5.04                          & -0.0582 \\ \hline ER($\langle d \rangle \approx 5$) + hub    & 200 & 5.54                & 6.05               & 6.48                          & 6.08                          & -0.0575 \\ \hline ER($\langle d \rangle \approx 5$) + clique & 200 & 7.05                & 18.15              & 11.09                         & 15.31                         & 0.6146  \\ \hline \end{tabular} 
	\caption{Approximation of the leading eigenvalue $\lambda_B^{\max}$ of the NB matrix. \label{table:lamB_approx}}
\end{table}

\end{widetext}

\bibliography{reference}

\begin{thebibliography}{58}%
\makeatletter
\providecommand \@ifxundefined [1]{%
 \@ifx{#1\undefined}
}%
\providecommand \@ifnum [1]{%
 \ifnum #1\expandafter \@firstoftwo
 \else \expandafter \@secondoftwo
 \fi
}%
\providecommand \@ifx [1]{%
 \ifx #1\expandafter \@firstoftwo
 \else \expandafter \@secondoftwo
 \fi
}%
\providecommand \natexlab [1]{#1}%
\providecommand \enquote  [1]{``#1''}%
\providecommand \bibnamefont  [1]{#1}%
\providecommand \bibfnamefont [1]{#1}%
\providecommand \citenamefont [1]{#1}%
\providecommand \href@noop [0]{\@secondoftwo}%
\providecommand \href [0]{\begingroup \@sanitize@url \@href}%
\providecommand \@href[1]{\@@startlink{#1}\@@href}%
\providecommand \@@href[1]{\endgroup#1\@@endlink}%
\providecommand \@sanitize@url [0]{\catcode `\\12\catcode `\$12\catcode
  `\&12\catcode `\#12\catcode `\^12\catcode `\_12\catcode `\%12\relax}%
\providecommand \@@startlink[1]{}%
\providecommand \@@endlink[0]{}%
\providecommand \url  [0]{\begingroup\@sanitize@url \@url }%
\providecommand \@url [1]{\endgroup\@href {#1}{\urlprefix }}%
\providecommand \urlprefix  [0]{URL }%
\providecommand \Eprint [0]{\href }%
\providecommand \doibase [0]{http://dx.doi.org/}%
\providecommand \selectlanguage [0]{\@gobble}%
\providecommand \bibinfo  [0]{\@secondoftwo}%
\providecommand \bibfield  [0]{\@secondoftwo}%
\providecommand \translation [1]{[#1]}%
\providecommand \BibitemOpen [0]{}%
\providecommand \bibitemStop [0]{}%
\providecommand \bibitemNoStop [0]{.\EOS\space}%
\providecommand \EOS [0]{\spacefactor3000\relax}%
\providecommand \BibitemShut  [1]{\csname bibitem#1\endcsname}%
\let\auto@bib@innerbib\@empty
\bibitem [{\citenamefont {Wang}\ \emph {et~al.}(2020)\citenamefont {Wang},
  \citenamefont {Wang}, \citenamefont {Chen},\ and\ \citenamefont
  {Qin}}]{WangYixuan2020}%
  \BibitemOpen
  \bibfield  {author} {\bibinfo {author} {\bibfnamefont {Yixuan}\ \bibnamefont
  {Wang}}, \bibinfo {author} {\bibfnamefont {Yuyi}\ \bibnamefont {Wang}},
  \bibinfo {author} {\bibfnamefont {Yan}\ \bibnamefont {Chen}}, \ and\ \bibinfo
  {author} {\bibfnamefont {Qingsong}\ \bibnamefont {Qin}},\ }\bibfield  {title}
  {\enquote {\bibinfo {title} {Unique epidemiological and clinical features of
  the emerging 2019 novel coronavirus pneumonia ({COVID}-19) implicate special
  control measures},}\ }\href {\doibase 10.1002/jmv.25748} {\bibfield
  {journal} {\bibinfo  {journal} {Journal of Medical Virology}\ }\textbf
  {\bibinfo {volume} {92}},\ \bibinfo {pages} {568--576} (\bibinfo {year}
  {2020})}\BibitemShut {NoStop}%
\bibitem [{\citenamefont {Li}\ \emph {et~al.}(2020{\natexlab{a}})\citenamefont
  {Li}, \citenamefont {Pei}, \citenamefont {Chen}, \citenamefont {Song},
  \citenamefont {Zhang}, \citenamefont {Yang},\ and\ \citenamefont
  {Shaman}}]{RuiyunLi2020}%
  \BibitemOpen
  \bibfield  {author} {\bibinfo {author} {\bibfnamefont {Ruiyun}\ \bibnamefont
  {Li}}, \bibinfo {author} {\bibfnamefont {Sen}\ \bibnamefont {Pei}}, \bibinfo
  {author} {\bibfnamefont {Bin}\ \bibnamefont {Chen}}, \bibinfo {author}
  {\bibfnamefont {Yimeng}\ \bibnamefont {Song}}, \bibinfo {author}
  {\bibfnamefont {Tao}\ \bibnamefont {Zhang}}, \bibinfo {author} {\bibfnamefont
  {Wan}\ \bibnamefont {Yang}}, \ and\ \bibinfo {author} {\bibfnamefont
  {Jeffrey}\ \bibnamefont {Shaman}},\ }\bibfield  {title} {\enquote {\bibinfo
  {title} {Substantial undocumented infection facilitates the rapid
  dissemination of novel coronavirus (sars-cov-2)},}\ }\href {\doibase
  10.1126/science.abb3221} {\bibfield  {journal} {\bibinfo  {journal}
  {Science}\ }\textbf {\bibinfo {volume} {368}},\ \bibinfo {pages} {489--493}
  (\bibinfo {year} {2020}{\natexlab{a}})}\BibitemShut {NoStop}%
\bibitem [{\citenamefont {He}\ \emph {et~al.}(2020)\citenamefont {He},
  \citenamefont {Lau}, \citenamefont {Wu}, \citenamefont {Deng}, \citenamefont
  {Wang}, \citenamefont {Hao}, \citenamefont {Lau}, \citenamefont {Wong},
  \citenamefont {Guan}, \citenamefont {Tan} \emph {et~al.}}]{He2020}%
  \BibitemOpen
  \bibfield  {author} {\bibinfo {author} {\bibfnamefont {Xi}~\bibnamefont
  {He}}, \bibinfo {author} {\bibfnamefont {Eric H.~Y.}\ \bibnamefont {Lau}},
  \bibinfo {author} {\bibfnamefont {Peng}\ \bibnamefont {Wu}}, \bibinfo
  {author} {\bibfnamefont {Xilong}\ \bibnamefont {Deng}}, \bibinfo {author}
  {\bibfnamefont {Jian}\ \bibnamefont {Wang}}, \bibinfo {author} {\bibfnamefont
  {Xinxin}\ \bibnamefont {Hao}}, \bibinfo {author} {\bibfnamefont {Yiu~Chung}\
  \bibnamefont {Lau}}, \bibinfo {author} {\bibfnamefont {Jessica~Y.}\
  \bibnamefont {Wong}}, \bibinfo {author} {\bibfnamefont {Yujuan}\ \bibnamefont
  {Guan}}, \bibinfo {author} {\bibfnamefont {Xinghua}\ \bibnamefont {Tan}},
  \emph {et~al.},\ }\bibfield  {title} {\enquote {\bibinfo {title} {Temporal
  dynamics in viral shedding and transmissibility of {COVID}-19},}\ }\href
  {\doibase 10.1038/s41591-020-0869-5} {\bibfield  {journal} {\bibinfo
  {journal} {Nature Medicine}\ }\textbf {\bibinfo {volume} {26}},\ \bibinfo
  {pages} {672--675} (\bibinfo {year} {2020})}\BibitemShut {NoStop}%
\bibitem [{\citenamefont {Moghadas}\ \emph
  {et~al.}(2020{\natexlab{a}})\citenamefont {Moghadas}, \citenamefont
  {Fitzpatrick}, \citenamefont {Sah}, \citenamefont {Pandey}, \citenamefont
  {Shoukat}, \citenamefont {Singer},\ and\ \citenamefont
  {Galvani}}]{Moghadas2020}%
  \BibitemOpen
  \bibfield  {author} {\bibinfo {author} {\bibfnamefont {Seyed~M.}\
  \bibnamefont {Moghadas}}, \bibinfo {author} {\bibfnamefont {Meagan~C.}\
  \bibnamefont {Fitzpatrick}}, \bibinfo {author} {\bibfnamefont {Pratha}\
  \bibnamefont {Sah}}, \bibinfo {author} {\bibfnamefont {Abhishek}\
  \bibnamefont {Pandey}}, \bibinfo {author} {\bibfnamefont {Affan}\
  \bibnamefont {Shoukat}}, \bibinfo {author} {\bibfnamefont {Burton~H.}\
  \bibnamefont {Singer}}, \ and\ \bibinfo {author} {\bibfnamefont {Alison~P.}\
  \bibnamefont {Galvani}},\ }\bibfield  {title} {\enquote {\bibinfo {title}
  {The implications of silent transmission for the control of {COVID}-19
  outbreaks},}\ }\href {\doibase 10.1073/pnas.2008373117} {\bibfield  {journal}
  {\bibinfo  {journal} {Proceedings of the National Academy of Sciences}\
  }\textbf {\bibinfo {volume} {117}},\ \bibinfo {pages} {17513--17515}
  (\bibinfo {year} {2020}{\natexlab{a}})}\BibitemShut {NoStop}%
\bibitem [{\citenamefont {Gatto}\ \emph {et~al.}(2020)\citenamefont {Gatto},
  \citenamefont {Bertuzzo}, \citenamefont {Mari}, \citenamefont {Miccoli},
  \citenamefont {Carraro}, \citenamefont {Casagrandi},\ and\ \citenamefont
  {Rinaldo}}]{Gatto2020}%
  \BibitemOpen
  \bibfield  {author} {\bibinfo {author} {\bibfnamefont {Marino}\ \bibnamefont
  {Gatto}}, \bibinfo {author} {\bibfnamefont {Enrico}\ \bibnamefont
  {Bertuzzo}}, \bibinfo {author} {\bibfnamefont {Lorenzo}\ \bibnamefont
  {Mari}}, \bibinfo {author} {\bibfnamefont {Stefano}\ \bibnamefont {Miccoli}},
  \bibinfo {author} {\bibfnamefont {Luca}\ \bibnamefont {Carraro}}, \bibinfo
  {author} {\bibfnamefont {Renato}\ \bibnamefont {Casagrandi}}, \ and\ \bibinfo
  {author} {\bibfnamefont {Andrea}\ \bibnamefont {Rinaldo}},\ }\bibfield
  {title} {\enquote {\bibinfo {title} {Spread and dynamics of the {COVID}-19
  epidemic in italy: Effects of emergency containment measures},}\ }\href
  {\doibase 10.1073/pnas.2004978117} {\bibfield  {journal} {\bibinfo  {journal}
  {Proceedings of the National Academy of Sciences}\ }\textbf {\bibinfo
  {volume} {117}},\ \bibinfo {pages} {10484--10491} (\bibinfo {year}
  {2020})}\BibitemShut {NoStop}%
\bibitem [{\citenamefont {Chinazzi}\ \emph {et~al.}(2020)\citenamefont
  {Chinazzi}, \citenamefont {Davis}, \citenamefont {Ajelli}, \citenamefont
  {Gioannini}, \citenamefont {Litvinova}, \citenamefont {Merler}, \citenamefont
  {Pastore~y Piontti}, \citenamefont {Mu}, \citenamefont {Rossi}, \citenamefont
  {Sun} \emph {et~al.}}]{Chinazzi2020}%
  \BibitemOpen
  \bibfield  {author} {\bibinfo {author} {\bibfnamefont {Matteo}\ \bibnamefont
  {Chinazzi}}, \bibinfo {author} {\bibfnamefont {Jessica~T.}\ \bibnamefont
  {Davis}}, \bibinfo {author} {\bibfnamefont {Marco}\ \bibnamefont {Ajelli}},
  \bibinfo {author} {\bibfnamefont {Corrado}\ \bibnamefont {Gioannini}},
  \bibinfo {author} {\bibfnamefont {Maria}\ \bibnamefont {Litvinova}}, \bibinfo
  {author} {\bibfnamefont {Stefano}\ \bibnamefont {Merler}}, \bibinfo {author}
  {\bibfnamefont {Ana}\ \bibnamefont {Pastore~y Piontti}}, \bibinfo {author}
  {\bibfnamefont {Kunpeng}\ \bibnamefont {Mu}}, \bibinfo {author}
  {\bibfnamefont {Luca}\ \bibnamefont {Rossi}}, \bibinfo {author}
  {\bibfnamefont {Kaiyuan}\ \bibnamefont {Sun}},  \emph {et~al.},\ }\bibfield
  {title} {\enquote {\bibinfo {title} {The effect of travel restrictions on the
  spread of the 2019 novel coronavirus ({COVID}-19) outbreak},}\ }\href
  {\doibase 10.1126/science.aba9757} {\bibfield  {journal} {\bibinfo  {journal}
  {Science}\ }\textbf {\bibinfo {volume} {368}},\ \bibinfo {pages} {395--400}
  (\bibinfo {year} {2020})}\BibitemShut {NoStop}%
\bibitem [{\citenamefont {Hao}\ \emph {et~al.}(2020)\citenamefont {Hao},
  \citenamefont {Cheng}, \citenamefont {Wu}, \citenamefont {Wu}, \citenamefont
  {Lin},\ and\ \citenamefont {Wang}}]{Hao2020}%
  \BibitemOpen
  \bibfield  {author} {\bibinfo {author} {\bibfnamefont {Xingjie}\ \bibnamefont
  {Hao}}, \bibinfo {author} {\bibfnamefont {Shanshan}\ \bibnamefont {Cheng}},
  \bibinfo {author} {\bibfnamefont {Degang}\ \bibnamefont {Wu}}, \bibinfo
  {author} {\bibfnamefont {Tangchun}\ \bibnamefont {Wu}}, \bibinfo {author}
  {\bibfnamefont {Xihong}\ \bibnamefont {Lin}}, \ and\ \bibinfo {author}
  {\bibfnamefont {Chaolong}\ \bibnamefont {Wang}},\ }\bibfield  {title}
  {\enquote {\bibinfo {title} {Reconstruction of the full transmission dynamics
  of {COVID}-19 in wuhan},}\ }\href {\doibase 10.1038/s41586-020-2554-8}
  {\bibfield  {journal} {\bibinfo  {journal} {Nature}\ }\textbf {\bibinfo
  {volume} {584}},\ \bibinfo {pages} {420--424} (\bibinfo {year}
  {2020})}\BibitemShut {NoStop}%
\bibitem [{\citenamefont {Moghadas}\ \emph
  {et~al.}(2020{\natexlab{b}})\citenamefont {Moghadas}, \citenamefont
  {Shoukat}, \citenamefont {Fitzpatrick}, \citenamefont {Wells}, \citenamefont
  {Sah}, \citenamefont {Pandey}, \citenamefont {Sachs}, \citenamefont {Wang},
  \citenamefont {Meyers}, \citenamefont {Singer} \emph
  {et~al.}}]{Moghadas2020Apr}%
  \BibitemOpen
  \bibfield  {author} {\bibinfo {author} {\bibfnamefont {Seyed~M.}\
  \bibnamefont {Moghadas}}, \bibinfo {author} {\bibfnamefont {Affan}\
  \bibnamefont {Shoukat}}, \bibinfo {author} {\bibfnamefont {Meagan~C.}\
  \bibnamefont {Fitzpatrick}}, \bibinfo {author} {\bibfnamefont {Chad~R.}\
  \bibnamefont {Wells}}, \bibinfo {author} {\bibfnamefont {Pratha}\
  \bibnamefont {Sah}}, \bibinfo {author} {\bibfnamefont {Abhishek}\
  \bibnamefont {Pandey}}, \bibinfo {author} {\bibfnamefont {Jeffrey~D.}\
  \bibnamefont {Sachs}}, \bibinfo {author} {\bibfnamefont {Zheng}\ \bibnamefont
  {Wang}}, \bibinfo {author} {\bibfnamefont {Lauren~A.}\ \bibnamefont
  {Meyers}}, \bibinfo {author} {\bibfnamefont {Burton~H.}\ \bibnamefont
  {Singer}},  \emph {et~al.},\ }\bibfield  {title} {\enquote {\bibinfo {title}
  {Projecting hospital utilization during the {COVID}-19 outbreaks in the
  united states},}\ }\href {\doibase 10.1073/pnas.2004064117} {\bibfield
  {journal} {\bibinfo  {journal} {Proceedings of the National Academy of
  Sciences}\ }\textbf {\bibinfo {volume} {117}},\ \bibinfo {pages} {9122--9126}
  (\bibinfo {year} {2020}{\natexlab{b}})}\BibitemShut {NoStop}%
\bibitem [{\citenamefont {Pastor-Satorras}\ \emph {et~al.}(2015)\citenamefont
  {Pastor-Satorras}, \citenamefont {Castellano}, \citenamefont {Van~Mieghem},\
  and\ \citenamefont {Vespignani}}]{Satorras2015}%
  \BibitemOpen
  \bibfield  {author} {\bibinfo {author} {\bibfnamefont {Romualdo}\
  \bibnamefont {Pastor-Satorras}}, \bibinfo {author} {\bibfnamefont {Claudio}\
  \bibnamefont {Castellano}}, \bibinfo {author} {\bibfnamefont {Piet}\
  \bibnamefont {Van~Mieghem}}, \ and\ \bibinfo {author} {\bibfnamefont
  {Alessandro}\ \bibnamefont {Vespignani}},\ }\bibfield  {title} {\enquote
  {\bibinfo {title} {Epidemic processes in complex networks},}\ }\href
  {\doibase 10.1103/RevModPhys.87.925} {\bibfield  {journal} {\bibinfo
  {journal} {Rev. Mod. Phys.}\ }\textbf {\bibinfo {volume} {87}},\ \bibinfo
  {pages} {925--979} (\bibinfo {year} {2015})}\BibitemShut {NoStop}%
\bibitem [{\citenamefont {Ferguson}\ \emph {et~al.}()\citenamefont {Ferguson},
  \citenamefont {Laydon}, \citenamefont {Nedjati~Gilani}, \citenamefont {Imai},
  \citenamefont {Ainslie}, \citenamefont {Baguelin}, \citenamefont {Bhatia},
  \citenamefont {Boonyasiri}, \citenamefont {Cucunuba~Perez}, \citenamefont
  {Cuomo-Dannenburg} \emph {et~al.}}]{Ferguson2020}%
  \BibitemOpen
  \bibfield  {author} {\bibinfo {author} {\bibfnamefont {N}~\bibnamefont
  {Ferguson}}, \bibinfo {author} {\bibfnamefont {D}~\bibnamefont {Laydon}},
  \bibinfo {author} {\bibfnamefont {G}~\bibnamefont {Nedjati~Gilani}}, \bibinfo
  {author} {\bibfnamefont {N}~\bibnamefont {Imai}}, \bibinfo {author}
  {\bibfnamefont {K}~\bibnamefont {Ainslie}}, \bibinfo {author} {\bibfnamefont
  {M}~\bibnamefont {Baguelin}}, \bibinfo {author} {\bibfnamefont
  {S}~\bibnamefont {Bhatia}}, \bibinfo {author} {\bibfnamefont {A}~\bibnamefont
  {Boonyasiri}}, \bibinfo {author} {\bibfnamefont {ZULMA}\ \bibnamefont
  {Cucunuba~Perez}}, \bibinfo {author} {\bibfnamefont {G}~\bibnamefont
  {Cuomo-Dannenburg}},  \emph {et~al.},\ }\bibfield  {title} {\enquote
  {\bibinfo {title} {Report 9: Impact of non-pharmaceutical interventions
  ({NPI}s) to reduce covid19 mortality and healthcare demand},}\ }\href
  {\doibase 10.25561/77482} {\bibfield  {journal} {\bibinfo  {journal}
  {https://spiral.imperial.ac.uk:8443/handle/10044/1/77482}\
  }10.25561/77482}\BibitemShut {NoStop}%
\bibitem [{\citenamefont {Adam}(2020)}]{Adam2020}%
  \BibitemOpen
  \bibfield  {author} {\bibinfo {author} {\bibfnamefont {David}\ \bibnamefont
  {Adam}},\ }\bibfield  {title} {\enquote {\bibinfo {title} {Special report:
  The simulations driving the world's response to {COVID}-19},}\ }\href
  {\doibase 10.1038/d41586-020-01003-6} {\bibfield  {journal} {\bibinfo
  {journal} {Nature}\ }\textbf {\bibinfo {volume} {580}},\ \bibinfo {pages}
  {316--318} (\bibinfo {year} {2020})}\BibitemShut {NoStop}%
\bibitem [{\citenamefont {Reich}\ \emph {et~al.}(2020)\citenamefont {Reich},
  \citenamefont {Shalev},\ and\ \citenamefont {Kalvari}}]{Reich2020}%
  \BibitemOpen
  \bibfield  {author} {\bibinfo {author} {\bibfnamefont {Ofir}\ \bibnamefont
  {Reich}}, \bibinfo {author} {\bibfnamefont {Guy}\ \bibnamefont {Shalev}}, \
  and\ \bibinfo {author} {\bibfnamefont {Tom}\ \bibnamefont {Kalvari}},\
  }\bibfield  {title} {\enquote {\bibinfo {title} {Modeling {COVID}-19 on a
  network: super-spreaders, testing and containment},}\ }\href {\doibase
  10.1101/2020.04.30.20081828} {\bibfield  {journal} {\bibinfo  {journal}
  {medRxiv}\ } (\bibinfo {year} {2020}),\
  10.1101/2020.04.30.20081828}\BibitemShut {NoStop}%
\bibitem [{\citenamefont {Davies}\ \emph {et~al.}(2020)\citenamefont {Davies},
  \citenamefont {Kucharski}, \citenamefont {Eggo}, \citenamefont {Gimma},
  \citenamefont {Edmunds}, \citenamefont {Jombart}, \citenamefont {O'Reilly},
  \citenamefont {Endo}, \citenamefont {Hellewell}, \citenamefont {Nightingale}
  \emph {et~al.}}]{Davies2020}%
  \BibitemOpen
  \bibfield  {author} {\bibinfo {author} {\bibfnamefont {Nicholas~G.}\
  \bibnamefont {Davies}}, \bibinfo {author} {\bibfnamefont {Adam~J.}\
  \bibnamefont {Kucharski}}, \bibinfo {author} {\bibfnamefont {Rosalind~M.}\
  \bibnamefont {Eggo}}, \bibinfo {author} {\bibfnamefont {Amy}\ \bibnamefont
  {Gimma}}, \bibinfo {author} {\bibfnamefont {W.~John}\ \bibnamefont
  {Edmunds}}, \bibinfo {author} {\bibfnamefont {Thibaut}\ \bibnamefont
  {Jombart}}, \bibinfo {author} {\bibfnamefont {Kathleen}\ \bibnamefont
  {O'Reilly}}, \bibinfo {author} {\bibfnamefont {Akira}\ \bibnamefont {Endo}},
  \bibinfo {author} {\bibfnamefont {Joel}\ \bibnamefont {Hellewell}}, \bibinfo
  {author} {\bibfnamefont {Emily~S.}\ \bibnamefont {Nightingale}},  \emph
  {et~al.},\ }\bibfield  {title} {\enquote {\bibinfo {title} {Effects of
  non-pharmaceutical interventions on {COVID}-19 cases, deaths, and demand for
  hospital services in the uk: a modelling study},}\ }\href {\doibase
  10.1016/S2468-2667(20)30133-X} {\bibfield  {journal} {\bibinfo  {journal}
  {The Lancet Public Health}\ }\textbf {\bibinfo {volume} {5}},\ \bibinfo
  {pages} {e375--e385} (\bibinfo {year} {2020})}\BibitemShut {NoStop}%
\bibitem [{\citenamefont {Karrer}\ and\ \citenamefont
  {Newman}(2010)}]{Karrer2010}%
  \BibitemOpen
  \bibfield  {author} {\bibinfo {author} {\bibfnamefont {Brian}\ \bibnamefont
  {Karrer}}\ and\ \bibinfo {author} {\bibfnamefont {M.~E.~J.}\ \bibnamefont
  {Newman}},\ }\bibfield  {title} {\enquote {\bibinfo {title} {Message passing
  approach for general epidemic models},}\ }\href {\doibase
  10.1103/PhysRevE.82.016101} {\bibfield  {journal} {\bibinfo  {journal} {Phys.
  Rev. E}\ }\textbf {\bibinfo {volume} {82}},\ \bibinfo {pages} {016101}
  (\bibinfo {year} {2010})}\BibitemShut {NoStop}%
\bibitem [{\citenamefont {Lokhov}\ \emph {et~al.}(2014)\citenamefont {Lokhov},
  \citenamefont {M\'ezard}, \citenamefont {Ohta},\ and\ \citenamefont
  {Zdeborov\'a}}]{Lokhov2014}%
  \BibitemOpen
  \bibfield  {author} {\bibinfo {author} {\bibfnamefont {Andrey~Y.}\
  \bibnamefont {Lokhov}}, \bibinfo {author} {\bibfnamefont {Marc}\ \bibnamefont
  {M\'ezard}}, \bibinfo {author} {\bibfnamefont {Hiroki}\ \bibnamefont {Ohta}},
  \ and\ \bibinfo {author} {\bibfnamefont {Lenka}\ \bibnamefont
  {Zdeborov\'a}},\ }\bibfield  {title} {\enquote {\bibinfo {title} {Inferring
  the origin of an epidemic with a dynamic message-passing algorithm},}\ }\href
  {\doibase 10.1103/PhysRevE.90.012801} {\bibfield  {journal} {\bibinfo
  {journal} {Phys. Rev. E}\ }\textbf {\bibinfo {volume} {90}},\ \bibinfo
  {pages} {012801} (\bibinfo {year} {2014})}\BibitemShut {NoStop}%
\bibitem [{\citenamefont {Lokhov}\ \emph {et~al.}(2015)\citenamefont {Lokhov},
  \citenamefont {M\'ezard},\ and\ \citenamefont {Zdeborov\'a}}]{Lokhov2015}%
  \BibitemOpen
  \bibfield  {author} {\bibinfo {author} {\bibfnamefont {Andrey~Y.}\
  \bibnamefont {Lokhov}}, \bibinfo {author} {\bibfnamefont {Marc}\ \bibnamefont
  {M\'ezard}}, \ and\ \bibinfo {author} {\bibfnamefont {Lenka}\ \bibnamefont
  {Zdeborov\'a}},\ }\bibfield  {title} {\enquote {\bibinfo {title} {Dynamic
  message-passing equations for models with unidirectional dynamics},}\ }\href
  {\doibase 10.1103/PhysRevE.91.012811} {\bibfield  {journal} {\bibinfo
  {journal} {Phys. Rev. E}\ }\textbf {\bibinfo {volume} {91}},\ \bibinfo
  {pages} {012811} (\bibinfo {year} {2015})}\BibitemShut {NoStop}%
\bibitem [{\citenamefont {Shrestha}\ \emph {et~al.}(2015)\citenamefont
  {Shrestha}, \citenamefont {Scarpino},\ and\ \citenamefont
  {Moore}}]{Shrestha2015}%
  \BibitemOpen
  \bibfield  {author} {\bibinfo {author} {\bibfnamefont {Munik}\ \bibnamefont
  {Shrestha}}, \bibinfo {author} {\bibfnamefont {Samuel~V.}\ \bibnamefont
  {Scarpino}}, \ and\ \bibinfo {author} {\bibfnamefont {Cristopher}\
  \bibnamefont {Moore}},\ }\bibfield  {title} {\enquote {\bibinfo {title}
  {Message-passing approach for recurrent-state epidemic models on networks},}\
  }\href {\doibase 10.1103/PhysRevE.92.022821} {\bibfield  {journal} {\bibinfo
  {journal} {Phys. Rev. E}\ }\textbf {\bibinfo {volume} {92}},\ \bibinfo
  {pages} {022821} (\bibinfo {year} {2015})}\BibitemShut {NoStop}%
\bibitem [{\citenamefont {Wang}\ \emph {et~al.}(2017)\citenamefont {Wang},
  \citenamefont {Tang}, \citenamefont {Stanley},\ and\ \citenamefont
  {Braunstein}}]{Wang2017}%
  \BibitemOpen
  \bibfield  {author} {\bibinfo {author} {\bibfnamefont {Wei}\ \bibnamefont
  {Wang}}, \bibinfo {author} {\bibfnamefont {Ming}\ \bibnamefont {Tang}},
  \bibinfo {author} {\bibfnamefont {H~Eugene}\ \bibnamefont {Stanley}}, \ and\
  \bibinfo {author} {\bibfnamefont {Lidia~A}\ \bibnamefont {Braunstein}},\
  }\bibfield  {title} {\enquote {\bibinfo {title} {Unification of theoretical
  approaches for epidemic spreading on complex networks},}\ }\href {\doibase
  10.1088/1361-6633/aa5398} {\bibfield  {journal} {\bibinfo  {journal} {Reports
  on Progress in Physics}\ }\textbf {\bibinfo {volume} {80}},\ \bibinfo {pages}
  {036603} (\bibinfo {year} {2017})}\BibitemShut {NoStop}%
\bibitem [{\citenamefont {Keeling}\ and\ \citenamefont
  {Rohani}(2008)}]{Keeling2008}%
  \BibitemOpen
  \bibfield  {author} {\bibinfo {author} {\bibfnamefont {Matt~J.}\ \bibnamefont
  {Keeling}}\ and\ \bibinfo {author} {\bibfnamefont {Pejman}\ \bibnamefont
  {Rohani}},\ }\href {http://www.jstor.org/stable/j.ctvcm4gk0} {\emph {\bibinfo
  {title} {Modeling Infectious Diseases in Humans and Animals}}}\ (\bibinfo
  {publisher} {Princeton University Press},\ \bibinfo {year}
  {2008})\BibitemShut {NoStop}%
\bibitem [{\citenamefont {Tian}\ \emph {et~al.}(2021)\citenamefont {Tian},
  \citenamefont {Li}, \citenamefont {Qi}, \citenamefont {Tang}, \citenamefont
  {Tang}, \citenamefont {Liu}, \citenamefont {Li}, \citenamefont {Cheng},
  \citenamefont {Li}, \citenamefont {Shi}, \citenamefont {Liu},\ and\
  \citenamefont {Tang}}]{Tian2020}%
  \BibitemOpen
  \bibfield  {author} {\bibinfo {author} {\bibfnamefont {Liang}\ \bibnamefont
  {Tian}}, \bibinfo {author} {\bibfnamefont {Xuefei}\ \bibnamefont {Li}},
  \bibinfo {author} {\bibfnamefont {Fei}\ \bibnamefont {Qi}}, \bibinfo {author}
  {\bibfnamefont {Qian-Yuan}\ \bibnamefont {Tang}}, \bibinfo {author}
  {\bibfnamefont {Viola}\ \bibnamefont {Tang}}, \bibinfo {author}
  {\bibfnamefont {Jiang}\ \bibnamefont {Liu}}, \bibinfo {author} {\bibfnamefont
  {Zhiyuan}\ \bibnamefont {Li}}, \bibinfo {author} {\bibfnamefont {Xingye}\
  \bibnamefont {Cheng}}, \bibinfo {author} {\bibfnamefont {Xuanxuan}\
  \bibnamefont {Li}}, \bibinfo {author} {\bibfnamefont {Yingchen}\ \bibnamefont
  {Shi}}, \bibinfo {author} {\bibfnamefont {Haiguang}\ \bibnamefont {Liu}}, \
  and\ \bibinfo {author} {\bibfnamefont {Lei-Han}\ \bibnamefont {Tang}},\
  }\bibfield  {title} {\enquote {\bibinfo {title} {Harnessing peak transmission
  around symptom onset for non-pharmaceutical intervention and containment of
  the {COVID}-19 pandemic},}\ }\href {\doibase 10.1038/s41467-021-21385-z}
  {\bibfield  {journal} {\bibinfo  {journal} {Nature Communications}\ }\textbf
  {\bibinfo {volume} {12}},\ \bibinfo {pages} {1147} (\bibinfo {year}
  {2021})}\BibitemShut {NoStop}%
\bibitem [{\citenamefont {Li}\ \emph {et~al.}(2020{\natexlab{b}})\citenamefont
  {Li}, \citenamefont {Guan}, \citenamefont {Wu}, \citenamefont {Wang},
  \citenamefont {Zhou}, \citenamefont {Tong}, \citenamefont {Ren},
  \citenamefont {Leung}, \citenamefont {Lau}, \citenamefont {Wong} \emph
  {et~al.}}]{QunLi2020}%
  \BibitemOpen
  \bibfield  {author} {\bibinfo {author} {\bibfnamefont {Qun}\ \bibnamefont
  {Li}}, \bibinfo {author} {\bibfnamefont {Xuhua}\ \bibnamefont {Guan}},
  \bibinfo {author} {\bibfnamefont {Peng}\ \bibnamefont {Wu}}, \bibinfo
  {author} {\bibfnamefont {Xiaoye}\ \bibnamefont {Wang}}, \bibinfo {author}
  {\bibfnamefont {Lei}\ \bibnamefont {Zhou}}, \bibinfo {author} {\bibfnamefont
  {Yeqing}\ \bibnamefont {Tong}}, \bibinfo {author} {\bibfnamefont {Ruiqi}\
  \bibnamefont {Ren}}, \bibinfo {author} {\bibfnamefont {Kathy~S.M.}\
  \bibnamefont {Leung}}, \bibinfo {author} {\bibfnamefont {Eric~H.Y.}\
  \bibnamefont {Lau}}, \bibinfo {author} {\bibfnamefont {Jessica~Y.}\
  \bibnamefont {Wong}},  \emph {et~al.},\ }\bibfield  {title} {\enquote
  {\bibinfo {title} {Early transmission dynamics in wuhan, china, of novel
  coronavirus-infected pneumonia},}\ }\href {\doibase 10.1056/NEJMoa2001316}
  {\bibfield  {journal} {\bibinfo  {journal} {New England Journal of Medicine}\
  }\textbf {\bibinfo {volume} {382}},\ \bibinfo {pages} {1199--1207} (\bibinfo
  {year} {2020}{\natexlab{b}})},\ \bibinfo {note} {pMID: 31995857}\BibitemShut
  {NoStop}%
\bibitem [{\citenamefont {Ashcroft}\ \emph {et~al.}(2020)\citenamefont
  {Ashcroft}, \citenamefont {Huisman}, \citenamefont {Lehtinen}, \citenamefont
  {Bouman}, \citenamefont {Althaus}, \citenamefont {Regoes},\ and\
  \citenamefont {Bonhoeffer}}]{Ashcroft2020}%
  \BibitemOpen
  \bibfield  {author} {\bibinfo {author} {\bibfnamefont {Peter}\ \bibnamefont
  {Ashcroft}}, \bibinfo {author} {\bibfnamefont {Jana~S}\ \bibnamefont
  {Huisman}}, \bibinfo {author} {\bibfnamefont {Sonja}\ \bibnamefont
  {Lehtinen}}, \bibinfo {author} {\bibfnamefont {Judith~A}\ \bibnamefont
  {Bouman}}, \bibinfo {author} {\bibfnamefont {Christian~L}\ \bibnamefont
  {Althaus}}, \bibinfo {author} {\bibfnamefont {Roland~R}\ \bibnamefont
  {Regoes}}, \ and\ \bibinfo {author} {\bibfnamefont {Sebastian}\ \bibnamefont
  {Bonhoeffer}},\ }\bibfield  {title} {\enquote {\bibinfo {title} {{COVID}-19
  infectivity profile correction},}\ }\href {\doibase 10.4414/smw.2020.20336}
  {\bibfield  {journal} {\bibinfo  {journal} {Swiss Med Wkly. 2020;150:w20336}\
  } (\bibinfo {year} {2020}),\ 10.4414/smw.2020.20336}\BibitemShut {NoStop}%
\bibitem [{\citenamefont {Perera}\ \emph {et~al.}(2020)\citenamefont {Perera},
  \citenamefont {Tso}, \citenamefont {Tsang}, \citenamefont {Tsang},
  \citenamefont {Fung}, \citenamefont {Leung}, \citenamefont {Chin},
  \citenamefont {Chu}, \citenamefont {Cheng}, \citenamefont {Poon},
  \citenamefont {Chuang},\ and\ \citenamefont {Peiris}}]{Perera2020}%
  \BibitemOpen
  \bibfield  {author} {\bibinfo {author} {\bibfnamefont {Ranawaka~A.P.M.}\
  \bibnamefont {Perera}}, \bibinfo {author} {\bibfnamefont {Eugene}\
  \bibnamefont {Tso}}, \bibinfo {author} {\bibfnamefont {Owen~T.Y.}\
  \bibnamefont {Tsang}}, \bibinfo {author} {\bibfnamefont {Dominic~N.C.}\
  \bibnamefont {Tsang}}, \bibinfo {author} {\bibfnamefont {Kitty}\ \bibnamefont
  {Fung}}, \bibinfo {author} {\bibfnamefont {Yonna~W.Y.}\ \bibnamefont
  {Leung}}, \bibinfo {author} {\bibfnamefont {Alex~W.H.}\ \bibnamefont {Chin}},
  \bibinfo {author} {\bibfnamefont {Daniel~K.W.}\ \bibnamefont {Chu}}, \bibinfo
  {author} {\bibfnamefont {Samuel~M.S.}\ \bibnamefont {Cheng}}, \bibinfo
  {author} {\bibfnamefont {Leo~L.M.}\ \bibnamefont {Poon}}, \bibinfo {author}
  {\bibfnamefont {Vivien~W.M.}\ \bibnamefont {Chuang}}, \ and\ \bibinfo
  {author} {\bibfnamefont {Malik}\ \bibnamefont {Peiris}},\ }\bibfield  {title}
  {\enquote {\bibinfo {title} {{SARS}-{CoV}-2 virus culture and subgenomic
  {RNA} for respiratory specimens from patients with mild coronavirus
  disease},}\ }\href {\doibase 10.3201/eid2611.203219} {\bibfield  {journal}
  {\bibinfo  {journal} {Emerging Infectious Diseases}\ }\textbf {\bibinfo
  {volume} {26}},\ \bibinfo {pages} {11} (\bibinfo {year} {2020})}\BibitemShut
  {NoStop}%
\bibitem [{\citenamefont {Koh}\ \emph {et~al.}(2020)\citenamefont {Koh},
  \citenamefont {Naing}, \citenamefont {Chaw}, \citenamefont {Rosledzana},
  \citenamefont {Alikhan}, \citenamefont {Jamaludin}, \citenamefont {Amin},
  \citenamefont {Omar}, \citenamefont {Shazli}, \citenamefont {Griffith},
  \citenamefont {Pastore},\ and\ \citenamefont {Wong}}]{Koh2020}%
  \BibitemOpen
  \bibfield  {author} {\bibinfo {author} {\bibfnamefont {Wee~Chian}\
  \bibnamefont {Koh}}, \bibinfo {author} {\bibfnamefont {Lin}\ \bibnamefont
  {Naing}}, \bibinfo {author} {\bibfnamefont {Liling}\ \bibnamefont {Chaw}},
  \bibinfo {author} {\bibfnamefont {Muhammad~Ali}\ \bibnamefont {Rosledzana}},
  \bibinfo {author} {\bibfnamefont {Mohammad~Fathi}\ \bibnamefont {Alikhan}},
  \bibinfo {author} {\bibfnamefont {Sirajul~Adli}\ \bibnamefont {Jamaludin}},
  \bibinfo {author} {\bibfnamefont {Faezah}\ \bibnamefont {Amin}}, \bibinfo
  {author} {\bibfnamefont {Asiah}\ \bibnamefont {Omar}}, \bibinfo {author}
  {\bibfnamefont {Alia}\ \bibnamefont {Shazli}}, \bibinfo {author}
  {\bibfnamefont {Matthew}\ \bibnamefont {Griffith}}, \bibinfo {author}
  {\bibfnamefont {Roberta}\ \bibnamefont {Pastore}}, \ and\ \bibinfo {author}
  {\bibfnamefont {Justin}\ \bibnamefont {Wong}},\ }\bibfield  {title} {\enquote
  {\bibinfo {title} {What do we know about {SARS}-{CoV}-2 transmission? a
  systematic review and meta-analysis of the secondary attack rate and
  associated risk factors},}\ }\href {\doibase 10.1371/journal.pone.0240205}
  {\bibfield  {journal} {\bibinfo  {journal} {{PLOS} {ONE}}\ }\textbf {\bibinfo
  {volume} {15}},\ \bibinfo {pages} {e0240205} (\bibinfo {year}
  {2020})}\BibitemShut {NoStop}%
\bibitem [{\citenamefont {Byambasuren}\ \emph {et~al.}(2020)\citenamefont
  {Byambasuren}, \citenamefont {Cardona}, \citenamefont {Bell}, \citenamefont
  {Clark}, \citenamefont {McLaws},\ and\ \citenamefont
  {Glasziou}}]{Byambasuren2020}%
  \BibitemOpen
  \bibfield  {author} {\bibinfo {author} {\bibfnamefont {Oyungerel}\
  \bibnamefont {Byambasuren}}, \bibinfo {author} {\bibfnamefont {Magnolia}\
  \bibnamefont {Cardona}}, \bibinfo {author} {\bibfnamefont {Katy}\
  \bibnamefont {Bell}}, \bibinfo {author} {\bibfnamefont {Justin}\ \bibnamefont
  {Clark}}, \bibinfo {author} {\bibfnamefont {Mary-Louise}\ \bibnamefont
  {McLaws}}, \ and\ \bibinfo {author} {\bibfnamefont {Paul}\ \bibnamefont
  {Glasziou}},\ }\bibfield  {title} {\enquote {\bibinfo {title} {Estimating the
  extent of asymptomatic {COVID}-19 and its potential for community
  transmission: Systematic review and meta-analysis},}\ }\href {\doibase
  10.3138/jammi-2020-0030} {\bibfield  {journal} {\bibinfo  {journal} {Official
  Journal of the Association of Medical Microbiology and Infectious Disease
  Canada}\ }\textbf {\bibinfo {volume} {5}},\ \bibinfo {pages} {223--234}
  (\bibinfo {year} {2020})}\BibitemShut {NoStop}%
\bibitem [{\citenamefont {Youssef}\ and\ \citenamefont
  {Scoglio}(2011)}]{Youssef2011}%
  \BibitemOpen
  \bibfield  {author} {\bibinfo {author} {\bibfnamefont {Mina}\ \bibnamefont
  {Youssef}}\ and\ \bibinfo {author} {\bibfnamefont {Caterina}\ \bibnamefont
  {Scoglio}},\ }\bibfield  {title} {\enquote {\bibinfo {title} {An
  individual-based approach to {SIR} epidemics in contact networks},}\ }\href
  {\doibase https://doi.org/10.1016/j.jtbi.2011.05.029} {\bibfield  {journal}
  {\bibinfo  {journal} {Journal of Theoretical Biology}\ }\textbf {\bibinfo
  {volume} {283}},\ \bibinfo {pages} {136 -- 144} (\bibinfo {year}
  {2011})}\BibitemShut {NoStop}%
\bibitem [{\citenamefont {Mo}\ \emph {et~al.}(2020)\citenamefont {Mo},
  \citenamefont {Feng}, \citenamefont {Shen}, \citenamefont {Tam},
  \citenamefont {Li}, \citenamefont {Yin},\ and\ \citenamefont
  {Zhao}}]{MoBaichuan2020}%
  \BibitemOpen
  \bibfield  {author} {\bibinfo {author} {\bibfnamefont {Baichuan}\
  \bibnamefont {Mo}}, \bibinfo {author} {\bibfnamefont {Kairui}\ \bibnamefont
  {Feng}}, \bibinfo {author} {\bibfnamefont {Yu}~\bibnamefont {Shen}}, \bibinfo
  {author} {\bibfnamefont {Clarence}\ \bibnamefont {Tam}}, \bibinfo {author}
  {\bibfnamefont {Daqing}\ \bibnamefont {Li}}, \bibinfo {author} {\bibfnamefont
  {Yafeng}\ \bibnamefont {Yin}}, \ and\ \bibinfo {author} {\bibfnamefont
  {Jinhua}\ \bibnamefont {Zhao}},\ }\href@noop {} {\enquote {\bibinfo {title}
  {Modeling epidemic spreading through public transit using time-varying
  encounter network},}\ }\bibinfo {howpublished} {arXiv:2004.04602} (\bibinfo
  {year} {2020})\BibitemShut {NoStop}%
\bibitem [{\citenamefont {Basnarkov}(2021)}]{Basnarkov2020}%
  \BibitemOpen
  \bibfield  {author} {\bibinfo {author} {\bibfnamefont {Lasko}\ \bibnamefont
  {Basnarkov}},\ }\bibfield  {title} {\enquote {\bibinfo {title} {Seair
  epidemic spreading model of covid-19},}\ }\href {\doibase
  https://doi.org/10.1016/j.chaos.2020.110394} {\bibfield  {journal} {\bibinfo
  {journal} {Chaos, Solitons \& Fractals}\ }\textbf {\bibinfo {volume} {142}},\
  \bibinfo {pages} {110394} (\bibinfo {year} {2021})}\BibitemShut {NoStop}%
\bibitem [{\citenamefont {Koher}\ \emph {et~al.}(2019)\citenamefont {Koher},
  \citenamefont {Lentz}, \citenamefont {Gleeson},\ and\ \citenamefont
  {H\"ovel}}]{Koher2019}%
  \BibitemOpen
  \bibfield  {author} {\bibinfo {author} {\bibfnamefont {Andreas}\ \bibnamefont
  {Koher}}, \bibinfo {author} {\bibfnamefont {Hartmut H.~K.}\ \bibnamefont
  {Lentz}}, \bibinfo {author} {\bibfnamefont {James~P.}\ \bibnamefont
  {Gleeson}}, \ and\ \bibinfo {author} {\bibfnamefont {Philipp}\ \bibnamefont
  {H\"ovel}},\ }\bibfield  {title} {\enquote {\bibinfo {title} {Contact-based
  model for epidemic spreading on temporal networks},}\ }\href {\doibase
  10.1103/PhysRevX.9.031017} {\bibfield  {journal} {\bibinfo  {journal} {Phys.
  Rev. X}\ }\textbf {\bibinfo {volume} {9}},\ \bibinfo {pages} {031017}
  (\bibinfo {year} {2019})}\BibitemShut {NoStop}%
\bibitem [{\citenamefont {Sun}\ \emph {et~al.}(2021)\citenamefont {Sun},
  \citenamefont {Saad},\ and\ \citenamefont {Lokhov}}]{HanlinSun2019}%
  \BibitemOpen
  \bibfield  {author} {\bibinfo {author} {\bibfnamefont {Hanlin}\ \bibnamefont
  {Sun}}, \bibinfo {author} {\bibfnamefont {David}\ \bibnamefont {Saad}}, \
  and\ \bibinfo {author} {\bibfnamefont {Andrey~Y.}\ \bibnamefont {Lokhov}},\
  }\bibfield  {title} {\enquote {\bibinfo {title} {Competition, collaboration,
  and optimization in multiple interacting spreading processes},}\ }\href
  {\doibase 10.1103/PhysRevX.11.011048} {\bibfield  {journal} {\bibinfo
  {journal} {Phys. Rev. X}\ }\textbf {\bibinfo {volume} {11}},\ \bibinfo
  {pages} {011048} (\bibinfo {year} {2021})}\BibitemShut {NoStop}%
\bibitem [{\citenamefont {Mossong}\ \emph {et~al.}(2008)\citenamefont
  {Mossong}, \citenamefont {Hens}, \citenamefont {Jit}, \citenamefont
  {Beutels}, \citenamefont {Auranen}, \citenamefont {Mikolajczyk},
  \citenamefont {Massari}, \citenamefont {Salmaso}, \citenamefont {Tomba},
  \citenamefont {Wallinga} \emph {et~al.}}]{Mossong2008}%
  \BibitemOpen
  \bibfield  {author} {\bibinfo {author} {\bibfnamefont {Jo{\"e}l}\
  \bibnamefont {Mossong}}, \bibinfo {author} {\bibfnamefont {Niel}\
  \bibnamefont {Hens}}, \bibinfo {author} {\bibfnamefont {Mark}\ \bibnamefont
  {Jit}}, \bibinfo {author} {\bibfnamefont {Philippe}\ \bibnamefont {Beutels}},
  \bibinfo {author} {\bibfnamefont {Kari}\ \bibnamefont {Auranen}}, \bibinfo
  {author} {\bibfnamefont {Rafael}\ \bibnamefont {Mikolajczyk}}, \bibinfo
  {author} {\bibfnamefont {Marco}\ \bibnamefont {Massari}}, \bibinfo {author}
  {\bibfnamefont {Stefania}\ \bibnamefont {Salmaso}}, \bibinfo {author}
  {\bibfnamefont {Gianpaolo~Scalia}\ \bibnamefont {Tomba}}, \bibinfo {author}
  {\bibfnamefont {Jacco}\ \bibnamefont {Wallinga}},  \emph {et~al.},\
  }\bibfield  {title} {\enquote {\bibinfo {title} {Social contacts and mixing
  patterns relevant to the spread of infectious diseases},}\ }\href {\doibase
  10.1371/journal.pmed.0050074} {\bibfield  {journal} {\bibinfo  {journal}
  {{PLoS} Medicine}\ }\textbf {\bibinfo {volume} {5}},\ \bibinfo {pages} {e74}
  (\bibinfo {year} {2008})}\BibitemShut {NoStop}%
\bibitem [{soc()}]{sociopatterns}%
  \BibitemOpen
  \href@noop {} {}\bibinfo {howpublished}
  {http://www.sociopatterns.org/}\BibitemShut {NoStop}%
\bibitem [{\citenamefont {Altarelli}\ \emph {et~al.}(2014)\citenamefont
  {Altarelli}, \citenamefont {Braunstein}, \citenamefont {Dall'Asta},
  \citenamefont {Wakeling},\ and\ \citenamefont {Zecchina}}]{Altarelli2014}%
  \BibitemOpen
  \bibfield  {author} {\bibinfo {author} {\bibfnamefont {F.}~\bibnamefont
  {Altarelli}}, \bibinfo {author} {\bibfnamefont {A.}~\bibnamefont
  {Braunstein}}, \bibinfo {author} {\bibfnamefont {L.}~\bibnamefont
  {Dall'Asta}}, \bibinfo {author} {\bibfnamefont {J.~R.}\ \bibnamefont
  {Wakeling}}, \ and\ \bibinfo {author} {\bibfnamefont {R.}~\bibnamefont
  {Zecchina}},\ }\bibfield  {title} {\enquote {\bibinfo {title} {Containing
  epidemic outbreaks by message-passing techniques},}\ }\href {\doibase
  10.1103/PhysRevX.4.021024} {\bibfield  {journal} {\bibinfo  {journal} {Phys.
  Rev. X}\ }\textbf {\bibinfo {volume} {4}},\ \bibinfo {pages} {021024}
  (\bibinfo {year} {2014})}\BibitemShut {NoStop}%
\bibitem [{\citenamefont {Bogu\~n\'a}\ and\ \citenamefont
  {Pastor-Satorras}(2002)}]{Boguna2002}%
  \BibitemOpen
  \bibfield  {author} {\bibinfo {author} {\bibfnamefont {Mari\'an}\
  \bibnamefont {Bogu\~n\'a}}\ and\ \bibinfo {author} {\bibfnamefont {Romualdo}\
  \bibnamefont {Pastor-Satorras}},\ }\bibfield  {title} {\enquote {\bibinfo
  {title} {Epidemic spreading in correlated complex networks},}\ }\href
  {\doibase 10.1103/PhysRevE.66.047104} {\bibfield  {journal} {\bibinfo
  {journal} {Phys. Rev. E}\ }\textbf {\bibinfo {volume} {66}},\ \bibinfo
  {pages} {047104} (\bibinfo {year} {2002})}\BibitemShut {NoStop}%
\bibitem [{Note1()}]{Note1}%
  \BibitemOpen
  \bibinfo {note} {We remark that there are other conventions used when
  defining the NB matrix in the literature; a common convention corresponds to
  the transpose of $B$ defined in Eq.~(\ref {eq:NB_matrix_def})~\cite
  {Krzakala2013}.}\BibitemShut {Stop}%
\bibitem [{\citenamefont {Krzakala}\ \emph {et~al.}(2013)\citenamefont
  {Krzakala}, \citenamefont {Moore}, \citenamefont {Mossel}, \citenamefont
  {Neeman}, \citenamefont {Sly}, \citenamefont {Zdeborov{\'a}},\ and\
  \citenamefont {Zhang}}]{Krzakala2013}%
  \BibitemOpen
  \bibfield  {author} {\bibinfo {author} {\bibfnamefont {Florent}\ \bibnamefont
  {Krzakala}}, \bibinfo {author} {\bibfnamefont {Cristopher}\ \bibnamefont
  {Moore}}, \bibinfo {author} {\bibfnamefont {Elchanan}\ \bibnamefont
  {Mossel}}, \bibinfo {author} {\bibfnamefont {Joe}\ \bibnamefont {Neeman}},
  \bibinfo {author} {\bibfnamefont {Allan}\ \bibnamefont {Sly}}, \bibinfo
  {author} {\bibfnamefont {Lenka}\ \bibnamefont {Zdeborov{\'a}}}, \ and\
  \bibinfo {author} {\bibfnamefont {Pan}\ \bibnamefont {Zhang}},\ }\bibfield
  {title} {\enquote {\bibinfo {title} {Spectral redemption in clustering sparse
  networks},}\ }\href {\doibase 10.1073/pnas.1312486110} {\bibfield  {journal}
  {\bibinfo  {journal} {Proceedings of the National Academy of Sciences}\
  }\textbf {\bibinfo {volume} {110}},\ \bibinfo {pages} {20935--20940}
  (\bibinfo {year} {2013})}\BibitemShut {NoStop}%
\bibitem [{\citenamefont {Karrer}\ \emph {et~al.}(2014)\citenamefont {Karrer},
  \citenamefont {Newman},\ and\ \citenamefont {Zdeborov\'a}}]{Karrer2014}%
  \BibitemOpen
  \bibfield  {author} {\bibinfo {author} {\bibfnamefont {Brian}\ \bibnamefont
  {Karrer}}, \bibinfo {author} {\bibfnamefont {M.~E.~J.}\ \bibnamefont
  {Newman}}, \ and\ \bibinfo {author} {\bibfnamefont {Lenka}\ \bibnamefont
  {Zdeborov\'a}},\ }\bibfield  {title} {\enquote {\bibinfo {title} {Percolation
  on sparse networks},}\ }\href {\doibase 10.1103/PhysRevLett.113.208702}
  {\bibfield  {journal} {\bibinfo  {journal} {Phys. Rev. Lett.}\ }\textbf
  {\bibinfo {volume} {113}},\ \bibinfo {pages} {208702} (\bibinfo {year}
  {2014})}\BibitemShut {NoStop}%
\bibitem [{\citenamefont {Horn}\ and\ \citenamefont
  {Johnson}(2012)}]{Horn2012}%
  \BibitemOpen
  \bibfield  {author} {\bibinfo {author} {\bibfnamefont {Roger~A.}\
  \bibnamefont {Horn}}\ and\ \bibinfo {author} {\bibfnamefont {Charles~R.}\
  \bibnamefont {Johnson}},\ }\href@noop {} {\emph {\bibinfo {title} {Matrix
  Analysis}}},\ \bibinfo {edition} {2nd}\ ed.\ (\bibinfo  {publisher}
  {Cambridge University Press},\ \bibinfo {address} {New York},\ \bibinfo
  {year} {2012})\BibitemShut {NoStop}%
\bibitem [{\citenamefont {Reyna-Lara}\ \emph {et~al.}(2021)\citenamefont
  {Reyna-Lara}, \citenamefont {Soriano-Pa\~nos}, \citenamefont {G\'omez},
  \citenamefont {Granell}, \citenamefont {Matamalas}, \citenamefont
  {Steinegger}, \citenamefont {Arenas},\ and\ \citenamefont
  {G\'omez-Garde\~nes}}]{Adriana2021}%
  \BibitemOpen
  \bibfield  {author} {\bibinfo {author} {\bibfnamefont {Adriana}\ \bibnamefont
  {Reyna-Lara}}, \bibinfo {author} {\bibfnamefont {David}\ \bibnamefont
  {Soriano-Pa\~nos}}, \bibinfo {author} {\bibfnamefont {Sergio}\ \bibnamefont
  {G\'omez}}, \bibinfo {author} {\bibfnamefont {Clara}\ \bibnamefont
  {Granell}}, \bibinfo {author} {\bibfnamefont {Joan~T.}\ \bibnamefont
  {Matamalas}}, \bibinfo {author} {\bibfnamefont {Benjamin}\ \bibnamefont
  {Steinegger}}, \bibinfo {author} {\bibfnamefont {Alex}\ \bibnamefont
  {Arenas}}, \ and\ \bibinfo {author} {\bibfnamefont {Jes\'us}\ \bibnamefont
  {G\'omez-Garde\~nes}},\ }\bibfield  {title} {\enquote {\bibinfo {title}
  {Virus spread versus contact tracing: Two competing contagion processes},}\
  }\href {\doibase 10.1103/PhysRevResearch.3.013163} {\bibfield  {journal}
  {\bibinfo  {journal} {Phys. Rev. Research}\ }\textbf {\bibinfo {volume}
  {3}},\ \bibinfo {pages} {013163} (\bibinfo {year} {2021})}\BibitemShut
  {NoStop}%
\bibitem [{\citenamefont {Watanabe}\ and\ \citenamefont
  {Fukumizu}(2009)}]{Watanabe2009}%
  \BibitemOpen
  \bibfield  {author} {\bibinfo {author} {\bibfnamefont {Yusuke}\ \bibnamefont
  {Watanabe}}\ and\ \bibinfo {author} {\bibfnamefont {Kenji}\ \bibnamefont
  {Fukumizu}},\ }\bibfield  {title} {\enquote {\bibinfo {title} {Graph zeta
  function in the bethe free energy and loopy belief propagation},}\ }in\ \href
  {http://papers.nips.cc/paper/3779-graph-zeta-function-in-the-bethe-free-energy-and-loopy-belief-propagation.pdf}
  {\emph {\bibinfo {booktitle} {Advances in Neural Information Processing
  Systems 22}}},\ \bibinfo {editor} {edited by\ \bibinfo {editor}
  {\bibfnamefont {Y.}~\bibnamefont {Bengio}}, \bibinfo {editor} {\bibfnamefont
  {D.}~\bibnamefont {Schuurmans}}, \bibinfo {editor} {\bibfnamefont {J.~D.}\
  \bibnamefont {Lafferty}}, \bibinfo {editor} {\bibfnamefont {C.~K.~I.}\
  \bibnamefont {Williams}}, \ and\ \bibinfo {editor} {\bibfnamefont
  {A.}~\bibnamefont {Culotta}}}\ (\bibinfo  {publisher} {Curran Associates,
  Inc.},\ \bibinfo {year} {2009})\ pp.\ \bibinfo {pages}
  {2017--2025}\BibitemShut {NoStop}%
\bibitem [{\citenamefont {Shu}\ \emph {et~al.}(2015)\citenamefont {Shu},
  \citenamefont {Wang}, \citenamefont {Tang},\ and\ \citenamefont
  {Do}}]{Shu2015}%
  \BibitemOpen
  \bibfield  {author} {\bibinfo {author} {\bibfnamefont {Panpan}\ \bibnamefont
  {Shu}}, \bibinfo {author} {\bibfnamefont {Wei}\ \bibnamefont {Wang}},
  \bibinfo {author} {\bibfnamefont {Ming}\ \bibnamefont {Tang}}, \ and\
  \bibinfo {author} {\bibfnamefont {Younghae}\ \bibnamefont {Do}},\ }\bibfield
  {title} {\enquote {\bibinfo {title} {Numerical identification of epidemic
  thresholds for susceptible-infected-recovered model on finite-size
  networks},}\ }\href {\doibase 10.1063/1.4922153} {\bibfield  {journal}
  {\bibinfo  {journal} {Chaos: An Interdisciplinary Journal of Nonlinear
  Science}\ }\textbf {\bibinfo {volume} {25}},\ \bibinfo {pages} {063104}
  (\bibinfo {year} {2015})}\BibitemShut {NoStop}%
\bibitem [{\citenamefont {Bianconi}(2018)}]{Bianconi2018}%
  \BibitemOpen
  \bibfield  {author} {\bibinfo {author} {\bibfnamefont {Ginestra}\
  \bibnamefont {Bianconi}},\ }\bibfield  {title} {\enquote {\bibinfo {title}
  {Rare events and discontinuous percolation transitions},}\ }\href {\doibase
  10.1103/PhysRevE.97.022314} {\bibfield  {journal} {\bibinfo  {journal} {Phys.
  Rev. E}\ }\textbf {\bibinfo {volume} {97}},\ \bibinfo {pages} {022314}
  (\bibinfo {year} {2018})}\BibitemShut {NoStop}%
\bibitem [{\citenamefont {K{\"u}hn}\ and\ \citenamefont
  {Rogers}(2017)}]{Kuehn2017}%
  \BibitemOpen
  \bibfield  {author} {\bibinfo {author} {\bibfnamefont {Reimer}\ \bibnamefont
  {K{\"u}hn}}\ and\ \bibinfo {author} {\bibfnamefont {Tim}\ \bibnamefont
  {Rogers}},\ }\bibfield  {title} {\enquote {\bibinfo {title} {Heterogeneous
  micro-structure of percolation in sparse networks},}\ }\href {\doibase
  10.1209/0295-5075/118/68003} {\bibfield  {journal} {\bibinfo  {journal}
  {{EPL} (Europhysics Letters)}\ }\textbf {\bibinfo {volume} {118}},\ \bibinfo
  {pages} {68003} (\bibinfo {year} {2017})}\BibitemShut {NoStop}%
\bibitem [{\citenamefont {Rasmussen}\ and\ \citenamefont
  {Williams}(2006)}]{Rasmussen2006}%
  \BibitemOpen
  \bibfield  {author} {\bibinfo {author} {\bibfnamefont {Carl~Edward}\
  \bibnamefont {Rasmussen}}\ and\ \bibinfo {author} {\bibfnamefont {Christopher
  K.~I.}\ \bibnamefont {Williams}},\ }\href@noop {} {\emph {\bibinfo {title}
  {Gaussian Processes for Machine Learning}}},\ Adaptive computation and
  machine learning\ (\bibinfo  {publisher} {The MIT Press},\ \bibinfo {year}
  {2006})\BibitemShut {NoStop}%
\bibitem [{\citenamefont {Martin}\ \emph {et~al.}(2014)\citenamefont {Martin},
  \citenamefont {Zhang},\ and\ \citenamefont {Newman}}]{Martin2014}%
  \BibitemOpen
  \bibfield  {author} {\bibinfo {author} {\bibfnamefont {Travis}\ \bibnamefont
  {Martin}}, \bibinfo {author} {\bibfnamefont {Xiao}\ \bibnamefont {Zhang}}, \
  and\ \bibinfo {author} {\bibfnamefont {M.~E.~J.}\ \bibnamefont {Newman}},\
  }\bibfield  {title} {\enquote {\bibinfo {title} {Localization and centrality
  in networks},}\ }\href {\doibase 10.1103/PhysRevE.90.052808} {\bibfield
  {journal} {\bibinfo  {journal} {Phys. Rev. E}\ }\textbf {\bibinfo {volume}
  {90}},\ \bibinfo {pages} {052808} (\bibinfo {year} {2014})}\BibitemShut
  {NoStop}%
\bibitem [{\citenamefont {Rogers}(2015)}]{Rogers2015}%
  \BibitemOpen
  \bibfield  {author} {\bibinfo {author} {\bibfnamefont {T.}~\bibnamefont
  {Rogers}},\ }\bibfield  {title} {\enquote {\bibinfo {title} {Assessing node
  risk and vulnerability in epidemics on networks},}\ }\href {\doibase
  10.1209/0295-5075/109/28005} {\bibfield  {journal} {\bibinfo  {journal}
  {{EPL} (Europhysics Letters)}\ }\textbf {\bibinfo {volume} {109}},\ \bibinfo
  {pages} {28005} (\bibinfo {year} {2015})}\BibitemShut {NoStop}%
\bibitem [{\citenamefont {Pastor-Satorras}\ and\ \citenamefont
  {Castellano}(2020)}]{Satorras2020}%
  \BibitemOpen
  \bibfield  {author} {\bibinfo {author} {\bibfnamefont {Romualdo}\
  \bibnamefont {Pastor-Satorras}}\ and\ \bibinfo {author} {\bibfnamefont
  {Claudio}\ \bibnamefont {Castellano}},\ }\bibfield  {title} {\enquote
  {\bibinfo {title} {The localization of non-backtracking centrality in
  networks and its physical consequences},}\ }\href {\doibase
  10.1038/s41598-020-78582-x} {\bibfield  {journal} {\bibinfo  {journal}
  {Scientific Reports}\ }\textbf {\bibinfo {volume} {10}},\ \bibinfo {pages}
  {21639} (\bibinfo {year} {2020})}\BibitemShut {NoStop}%
\bibitem [{\citenamefont {Mastrandrea}\ \emph {et~al.}(2015)\citenamefont
  {Mastrandrea}, \citenamefont {Fournet},\ and\ \citenamefont
  {Barrat}}]{Mastrandrea2015}%
  \BibitemOpen
  \bibfield  {author} {\bibinfo {author} {\bibfnamefont {Rossana}\ \bibnamefont
  {Mastrandrea}}, \bibinfo {author} {\bibfnamefont {Julie}\ \bibnamefont
  {Fournet}}, \ and\ \bibinfo {author} {\bibfnamefont {Alain}\ \bibnamefont
  {Barrat}},\ }\bibfield  {title} {\enquote {\bibinfo {title} {Contact patterns
  in a high school: A comparison between data collected using wearable sensors,
  contact diaries and friendship surveys},}\ }\href {\doibase
  10.1371/journal.pone.0136497} {\bibfield  {journal} {\bibinfo  {journal}
  {{PLOS} {ONE}}\ }\textbf {\bibinfo {volume} {10}},\ \bibinfo {pages}
  {e0136497} (\bibinfo {year} {2015})}\BibitemShut {NoStop}%
\bibitem [{\citenamefont {Barab{\'a}si}\ and\ \citenamefont
  {P{\'o}sfai}(2016)}]{Barabasi2016}%
  \BibitemOpen
  \bibfield  {author} {\bibinfo {author} {\bibfnamefont {A.L.}\ \bibnamefont
  {Barab{\'a}si}}\ and\ \bibinfo {author} {\bibfnamefont {M.}~\bibnamefont
  {P{\'o}sfai}},\ }\href@noop {} {\emph {\bibinfo {title} {Network Science}}}\
  (\bibinfo  {publisher} {Cambridge University Press},\ \bibinfo {address}
  {Cambridge, United Kingdom},\ \bibinfo {year} {2016})\BibitemShut {NoStop}%
\bibitem [{\citenamefont {Gemmetto}\ \emph {et~al.}(2014)\citenamefont
  {Gemmetto}, \citenamefont {Barrat},\ and\ \citenamefont
  {Cattuto}}]{Gemmetto2014}%
  \BibitemOpen
  \bibfield  {author} {\bibinfo {author} {\bibfnamefont {Valerio}\ \bibnamefont
  {Gemmetto}}, \bibinfo {author} {\bibfnamefont {Alain}\ \bibnamefont
  {Barrat}}, \ and\ \bibinfo {author} {\bibfnamefont {Ciro}\ \bibnamefont
  {Cattuto}},\ }\bibfield  {title} {\enquote {\bibinfo {title} {Mitigation of
  infectious disease at school: targeted class closure vs school closure},}\
  }\href {\doibase 10.1186/s12879-014-0695-9} {\bibfield  {journal} {\bibinfo
  {journal} {{BMC} Infectious Diseases}\ }\textbf {\bibinfo {volume} {14}},\
  \bibinfo {pages} {695} (\bibinfo {year} {2014})}\BibitemShut {NoStop}%
\bibitem [{\citenamefont {G{\'{e}}nois}\ \emph {et~al.}(2015)\citenamefont
  {G{\'{e}}nois}, \citenamefont {Vestergaard}, \citenamefont {Fournet},
  \citenamefont {Panisson}, \citenamefont {Bonmarin},\ and\ \citenamefont
  {Barrat}}]{GNOIS2015}%
  \BibitemOpen
  \bibfield  {author} {\bibinfo {author} {\bibfnamefont {Mathieu}\ \bibnamefont
  {G{\'{e}}nois}}, \bibinfo {author} {\bibfnamefont {Christian~L.}\
  \bibnamefont {Vestergaard}}, \bibinfo {author} {\bibfnamefont {Julie}\
  \bibnamefont {Fournet}}, \bibinfo {author} {\bibfnamefont {Andr{\'{e}}}\
  \bibnamefont {Panisson}}, \bibinfo {author} {\bibfnamefont {Isabelle}\
  \bibnamefont {Bonmarin}}, \ and\ \bibinfo {author} {\bibfnamefont {Alain}\
  \bibnamefont {Barrat}},\ }\bibfield  {title} {\enquote {\bibinfo {title}
  {Data on face-to-face contacts in an office building suggest a low-cost
  vaccination strategy based on community linkers},}\ }\href {\doibase
  10.1017/nws.2015.10} {\bibfield  {journal} {\bibinfo  {journal} {Network
  Science}\ }\textbf {\bibinfo {volume} {3}},\ \bibinfo {pages} {326--347}
  (\bibinfo {year} {2015})}\BibitemShut {NoStop}%
\bibitem [{\citenamefont {G{\'e}nois}\ and\ \citenamefont
  {Barrat}(2018)}]{Genois2018}%
  \BibitemOpen
  \bibfield  {author} {\bibinfo {author} {\bibfnamefont {Mathieu}\ \bibnamefont
  {G{\'e}nois}}\ and\ \bibinfo {author} {\bibfnamefont {Alain}\ \bibnamefont
  {Barrat}},\ }\bibfield  {title} {\enquote {\bibinfo {title} {Can co-location
  be used as a proxy for face-to-face contacts?}}\ }\href {\doibase
  10.1140/epjds/s13688-018-0140-1} {\bibfield  {journal} {\bibinfo  {journal}
  {EPJ Data Science}\ }\textbf {\bibinfo {volume} {7}},\ \bibinfo {pages} {11}
  (\bibinfo {year} {2018})}\BibitemShut {NoStop}%
\bibitem [{\citenamefont {Prem}\ \emph {et~al.}(2017)\citenamefont {Prem},
  \citenamefont {Cook},\ and\ \citenamefont {Jit}}]{Prem2017}%
  \BibitemOpen
  \bibfield  {author} {\bibinfo {author} {\bibfnamefont {Kiesha}\ \bibnamefont
  {Prem}}, \bibinfo {author} {\bibfnamefont {Alex~R.}\ \bibnamefont {Cook}}, \
  and\ \bibinfo {author} {\bibfnamefont {Mark}\ \bibnamefont {Jit}},\
  }\bibfield  {title} {\enquote {\bibinfo {title} {Projecting social contact
  matrices in 152 countries using contact surveys and demographic data},}\
  }\href {\doibase 10.1371/journal.pcbi.1005697} {\bibfield  {journal}
  {\bibinfo  {journal} {PLOS Computational Biology}\ }\textbf {\bibinfo
  {volume} {13}},\ \bibinfo {pages} {e1005697} (\bibinfo {year}
  {2017})}\BibitemShut {NoStop}%
\bibitem [{\citenamefont {Bianconi}\ \emph {et~al.}(2021)\citenamefont
  {Bianconi}, \citenamefont {Sun}, \citenamefont {Rapisardi},\ and\
  \citenamefont {Arenas}}]{Bianconi2021}%
  \BibitemOpen
  \bibfield  {author} {\bibinfo {author} {\bibfnamefont {Ginestra}\
  \bibnamefont {Bianconi}}, \bibinfo {author} {\bibfnamefont {Hanlin}\
  \bibnamefont {Sun}}, \bibinfo {author} {\bibfnamefont {Giacomo}\ \bibnamefont
  {Rapisardi}}, \ and\ \bibinfo {author} {\bibfnamefont {Alex}\ \bibnamefont
  {Arenas}},\ }\bibfield  {title} {\enquote {\bibinfo {title} {Message-passing
  approach to epidemic tracing and mitigation with apps},}\ }\href {\doibase
  10.1103/PhysRevResearch.3.L012014} {\bibfield  {journal} {\bibinfo  {journal}
  {Phys. Rev. Research}\ }\textbf {\bibinfo {volume} {3}},\ \bibinfo {pages}
  {L012014} (\bibinfo {year} {2021})}\BibitemShut {NoStop}%
\bibitem [{\citenamefont {Baker}\ \emph {et~al.}(2020)\citenamefont {Baker},
  \citenamefont {Biazzo}, \citenamefont {Braunstein}, \citenamefont {Catania},
  \citenamefont {Dall'Asta}, \citenamefont {Ingrosso}, \citenamefont
  {Krzakala}, \citenamefont {Mazza}, \citenamefont {M\'ezard}, \citenamefont
  {Muntoni}, \citenamefont {Refinetti}, \citenamefont {Mannelli},\ and\
  \citenamefont {Zdeborov\'a}}]{Baker2020}%
  \BibitemOpen
  \bibfield  {author} {\bibinfo {author} {\bibfnamefont {Antoine}\ \bibnamefont
  {Baker}}, \bibinfo {author} {\bibfnamefont {Indaco}\ \bibnamefont {Biazzo}},
  \bibinfo {author} {\bibfnamefont {Alfredo}\ \bibnamefont {Braunstein}},
  \bibinfo {author} {\bibfnamefont {Giovanni}\ \bibnamefont {Catania}},
  \bibinfo {author} {\bibfnamefont {Luca}\ \bibnamefont {Dall'Asta}}, \bibinfo
  {author} {\bibfnamefont {Alessandro}\ \bibnamefont {Ingrosso}}, \bibinfo
  {author} {\bibfnamefont {Florent}\ \bibnamefont {Krzakala}}, \bibinfo
  {author} {\bibfnamefont {Fabio}\ \bibnamefont {Mazza}}, \bibinfo {author}
  {\bibfnamefont {Marc}\ \bibnamefont {M\'ezard}}, \bibinfo {author}
  {\bibfnamefont {Anna~Paola}\ \bibnamefont {Muntoni}}, \bibinfo {author}
  {\bibfnamefont {Maria}\ \bibnamefont {Refinetti}}, \bibinfo {author}
  {\bibfnamefont {Stefano~Sarao}\ \bibnamefont {Mannelli}}, \ and\ \bibinfo
  {author} {\bibfnamefont {Lenka}\ \bibnamefont {Zdeborov\'a}},\ }\href@noop {}
  {\enquote {\bibinfo {title} {Epidemic mitigation by statistical inference
  from contact tracing data},}\ } (\bibinfo {year} {2020}),\ \Eprint
  {http://arxiv.org/abs/arXiv:2009.09422} {arXiv:2009.09422} \BibitemShut
  {NoStop}%
\bibitem [{\citenamefont {Lokhov}\ and\ \citenamefont
  {Saad}(2017)}]{Lokhov2017}%
  \BibitemOpen
  \bibfield  {author} {\bibinfo {author} {\bibfnamefont {Andrey~Y.}\
  \bibnamefont {Lokhov}}\ and\ \bibinfo {author} {\bibfnamefont {David}\
  \bibnamefont {Saad}},\ }\bibfield  {title} {\enquote {\bibinfo {title}
  {Optimal deployment of resources for maximizing impact in spreading
  processes},}\ }\href {\doibase 10.1073/pnas.1614694114} {\bibfield  {journal}
  {\bibinfo  {journal} {Proceedings of the National Academy of Sciences}\
  }\textbf {\bibinfo {volume} {114}},\ \bibinfo {pages} {E8138--E8146}
  (\bibinfo {year} {2017})}\BibitemShut {NoStop}%
\bibitem [{\citenamefont {Bass}(1992)}]{Bass1992}%
  \BibitemOpen
  \bibfield  {author} {\bibinfo {author} {\bibfnamefont {Hyman}\ \bibnamefont
  {Bass}},\ }\bibfield  {title} {\enquote {\bibinfo {title} {The
  {Ihara-Selberg} zeta function of a tree lattice},}\ }\href {\doibase
  10.1142/S0129167X92000357} {\bibfield  {journal} {\bibinfo  {journal}
  {International Journal of Mathematics}\ }\textbf {\bibinfo {volume} {03}},\
  \bibinfo {pages} {717--797} (\bibinfo {year} {1992})}\BibitemShut {NoStop}%
\bibitem [{\citenamefont {Godsil}\ and\ \citenamefont
  {Royle}(2001)}]{Godsil2001}%
  \BibitemOpen
  \bibfield  {author} {\bibinfo {author} {\bibfnamefont {Chris}\ \bibnamefont
  {Godsil}}\ and\ \bibinfo {author} {\bibfnamefont {Gordon}\ \bibnamefont
  {Royle}},\ }\href {\doibase 10.1007/978-1-4613-0163-9} {\emph {\bibinfo
  {title} {Algebraic Graph Theory}}}\ (\bibinfo  {publisher} {Springer New
  York},\ \bibinfo {year} {2001})\BibitemShut {NoStop}%
\end{thebibliography}%

\end{document}